\documentclass[aps,reprint,twocolumn,pra,groupedaddress,floatfix]{revtex4-1}
\usepackage{amsmath,amssymb,amsfonts}
\usepackage{mathrsfs}
\usepackage{stackengine}
\usepackage{soul}
\usepackage{graphicx}
\usepackage{color}
\begin{document}
\title{Phase shifted $\mathcal{PT}$-symmetric periodic structures}
\author{S. Vignesh Raja}
\email{vickyneeshraja@gmail.com}
\author{A. Govindarajan}
\email{govin.nld@gmail.com}
\author{A. Mahalingam}
\email{drmaha@annauniv.edu}
\author{M. Lakshmanan}
\email{lakshman.cnld@gmail.com}
\affiliation{$^{*,\ddagger}$Department of Physics, Anna University, Chennai - 600 025, India}
\affiliation{$^{\dagger,\mathsection}$Centre for Nonlinear Dynamics, School of Physics, Bharathidasan University, Tiruchirappalli - 620 024, India}
\begin{abstract}
	We report the spectral features of a phase-shifted parity and time ($\mathcal{PT}$)-symmetric fiber Bragg grating (PPTFBG) and demonstrate its functionality as a demultiplexer in the unbroken $\mathcal{PT}$-symmetric regime. The length of the  proposed system is of the order of millimeters and the lasing spectra in the broken $\mathcal{PT}$-symmetric regime can be easily tuned in terms of intensity, bandwidth and wavelength by varying the magnitude of the phase shift in the middle of the structure. Surprisingly, the multi-modal lasing spectra are suppressed by virtue of judiciously selected phase and the gain-loss value. Also, it is possible to obtain sidelobe-less spectra in the broken $\mathcal{PT}$-symmetric regime, without a need for an apodization profile, which is  a traditional way to tame the unwanted sidelobes. The system is found to show narrow band single-mode lasing behavior for a wide range of phase shift values for given values of gain and loss. Moreover, we report the intensity tunable reflection and transmission characteristics in the unbroken regime via variation in gain and loss. At the exceptional point, the system shows unidirectional wave transport phenomenon  independent of the presence of phase shift in the middle of the grating. For the right light incidence direction, the system exhibits zero reflection wavelengths within the stopband at the exceptional point. We also investigate the role of multiple phase shifts placed at fixed locations along the length of the FBG and the variations in the spectra when the phase shift and gain-loss values are tuned. In the broken $\mathcal{PT}$-symmetric regime, the presence of multiple phase shifts aids in controlling the number of reflectivity peaks besides controlling their magnitude. The advantage of the proposed model is that it exhibits multi-functional capabilities like demultiplexing, filtering and lasing in a short length of the grating depending on the different operating regimes.  
\end{abstract}
\maketitle
\section{Introduction}

From the invention of mirrors, lens to optical fibers of today, the quest to alter light propagation in a medium seems to be unceasing one. At present, the field of refractive index engineering has witnessed one of its major milestones in the form of introducing non-Hermitian notions into the traditional optical structures \cite{el2007theory,kottos2010optical,ruter2010observation,lin2011unidirectional,feng2017non,el2018non,ozdemir2019parity}. Before familiarization of $\mathcal{PT}$-symmetric concepts, overcoming the inherent loss remained to be one of the challenging aspects in designing any physically realizable optical structure \cite{lupu2013switching, govindarajan2018tailoring}. When most of the researchers were investigating the methodologies  to nullify the intrinsic loss, non-Hermitian physicists and mathematicians opted for the manipulation of the inherent loss of the system with the aid of newly designed artifacts known as $\mathcal{PT}$-symmetric structures \cite{bender1998real,kottos2010optical,ruter2010observation}. It was figured out that by judiciously controlling the inherent loss and extrinsic gain,  it is possible to realize $\mathcal{PT}$-symmetric systems that can breed many surprising optical phenomena \cite{lin2011unidirectional,govindarajan2018tailoring,govindarajan2019nonlinear}. Mathematically, the $\mathcal{PT}$-symmetric condition, which is required to breed such optical behaviors, is defined in terms of the refractive index as $n (z) = n^* (-z)$. Physically, this kind of complex refractive index profile is achieved in any $\mathcal{PT}$-symmetric structure by means of devising regions of gain and loss built in a counter balanced architecture \cite{regensburger2012parity,phang2013ultrafast,govindarajan2020}.  Apart from  fundamental aspects, research in the field of  $\mathcal{PT}$-symmetric optics has strongly changed towards the possibility of employing them in a wide range of applications in next generation light wave communication systems with larger  tunability and reconfiguration \cite{PhysRevA.91.053825}. 

Fiber Bragg grating (FBG) is an indispensable optical element extensively employed across diverse fields of physics ranging from simple filters \cite{agrawal1994phase,erdogan1997fiber,giles1997lightwave,hill1997fiber,othonos1997fiber,winful1979theory}  to optical signal processing \cite{radic1995theory,radic1995theory,yousefi2015all}. Hill \emph{et al.} discovered that the fiber is vulnerable to intense UV radiation which results in an irreversible alteration of the core index, and thus constituting a periodic index variation in the core \cite{hill1978photosensitivity, hill1997fiber}. Such a repeated periodic pattern composed of small sections of constant period ($\Lambda$) and fixed index ($n$), is termed as a grating \cite{meltz1989formation,giles1997lightwave}. It is known that due to the mismatch  between the core 
and grating indexes, a fraction of the optical signal is reflected at each period \cite{hill1997fiber}. These reflected signals get added up constructively at one selective wavelength known as Bragg wavelength ($\lambda_b$). FBG is an extremely wavelength discriminative device which strongly reflects back the wavelengths satisfying the Bragg condition (where the photonic bandgap gets formed) while the rest of the optical signals are transmitted \cite{othonos1997fiber}. Such a wavelength selective reflection (transmission) phenomenon arises as a consequence of energy coupling between different counter-propagating modes of the FBG. 

The bands of wavelengths which are reflected strongly near the Bragg wavelength are designated as the stopband of the grating. The possibility to detune the stopband of the grating at ease stands out to be one of the distinct features of these distributed feedback structures \cite{karimi2012all,broderick1998bistable}. Without a doubt, this must be regarded as the dominant reason which promoted FBG to emerge as a separate research domain today. In addition, it also offers numerous other intriguing features like compactness, inexpensiveness, low insertion, high return loss, and so forth \cite{giles1997lightwave,erdogan1997fiber,hill1997fiber,othonos1997fiber}. The research field of fiber Bragg grating can be broadly categorized into two separate regimes,  namely linear and nonlinear domains. The nonlinear domain, which includes the  study of gap soliton formation, steering dynamics of the grating structures via optical bistability,  and multi-stability is a fascinating one in the perspective of conventional \cite{winful2000raman,mills1987gap,winful1979theory,de1990switching} and $\mathcal{PT}$-symmetric structures \cite{raja2019multifaceted, raja2020tailoring}. Nevertheless, we confine our investigation here to the study of  linear dynamics of phase shifted periodic structures alone by introducing gain and loss with the reason that the linear system itself is less understood so far from the perspective of $\mathcal{PT}$-symmetry.

Ever since the pioneering work of Agrawal \emph{et al.} \cite{agrawal1994phase}, the phase-shifted FBGs paved the way for the scientific community to unearth many intriguing optical behaviors from the perspective of both linear and nonlinear gratings \cite{agrawal1994phase,radic1994optical,radic1995theory,melloni2000all,suryanto2009numerical,liu2018proposal}.  With the introduction of phase-shift into the structure, a narrow range of wavelengths inside the stop-band of the FBG transmits the incoming optical field. This span of wavelengths can be altered by fine-tuning the amount of phase-shift \cite{agrawal1994phase}. Driven by these luxuries, these phase-shifted FBGs are used to build all-optical demultiplexers \cite{agrawal1994phase}, low power all-optical switches \cite{radic1994optical,radic1995theory}, and signal processing devices \cite{melloni2000all}. 

Initially, Kulishov \emph{et al}. \cite{kulishov2005nonreciprocal}, formulated the coupled mode theory of a Bragg grating with gain and loss in the linear regime and also demonstrated the direction dependent transmission, reflection, delay and dispersion characteristics of the same system. They further made evident that if the light launching condition is reversed, significant amplification of spectra occurs under  grating index modulation (real and imaginary) mismatch condition. Another significant contribution in that work includes demonstration of reflectionless transmission when the real and imaginary parts of modulation index are equal. Some of the notable contributions in the field of distributed feedback structures (DFB) with gain and loss were given by Longhi which include the demonstration of spectral singularities \cite{longhi2010optical}, simultaneous coherent lasing and absorption behavior \cite{longhi2010pt}. Subsequently, Lin \emph{et al.}  demonstrated the light propagation dynamics  in the context of linear $\mathcal{PT}$-symmetric fiber Bragg grating  (PTFBG) at the exact $\mathcal{PT}$-symmetric phase and coined the reflectionless  optical wave transmission mechanism as \textit{unidirectional invisibility} \cite{lin2011unidirectional}. Later, Huang \emph{et al.} extended this concept of $\mathcal{PT}$-symmetry to nonuniform chirped gratings in concatenation with active and passive grating structures \cite{huang2014type}. It is worthwhile to recall  that the remarkable design of frequency comb in supersymmetric (SUSY) DFB structure  was exhibited by  Longhi \cite{longhi2015supersymmetric}.  Lupu \emph{et al.} demonstrated the concept of $\mathcal{PT}$-symmetry in a apodized grating with the aid of duty cycle methods \cite{lupu2016tailoring}. Quite recently, Correa \emph{et al.} came up with the possibility of designing Bragg gratings whose optical dynamics can be described by a Dirac like equation in the presence of $\mathcal{PT}$-symmetric Hamiltonian \cite{correa2017confluent}. Following these works, we recently demonstrated that it is possible to construct direction dependent delay lines and dispersion compensator in a chirped and apodized PTFBG \cite{raja2020tailoring}. Furthermore, we would like to construct $\mathcal{PT}$-symmetric FBG structures which can provide narrow band tunable lasing spectra in terms of intensity, linewidth and wavelength. Consequently, it is favorable to control the spectra by simply fine tuning the amount of phase-shift in the middle section of the grating. This gives the overall motivation and the necessity to investigate the spectral features of PPTFBGs.

Before we investigate the spectral characteristics of our proposed model, we would to like to clearly point out the advantages of the system compared to the other configurations reported in the literature to realize a tunable all-optical narrow band lasing spectra \cite{sun2006single,cheng2008single,he2009tunable,feng2009switchable,pan2010wavelength}. In the context of the aforementioned conventional systems, the insertions of discrete components in the form of Fabry P{\'e}rot filters and saturable absorbers  to achieve narrow band wavelength selectivity (bandpass) into the system contribute to insertion loss \cite{cheng2008single,feng2009switchable}. The amplification at the sidelobes of the spectra need to be eliminated by employing an apodized FBG. However, it unnecessarily truncates the spectral response in the absence of chirping \cite{hill1994chirped,ennser1998optimization,pastor1996design}. These configurations demand the usage of an extra gain fiber ($Er^{3+}$) having larger length in the ring for stabilization which increases the overall length of the system to the order of meters \cite{cheng2008single,sun2006single}. 
It is important to remember that higher the number of discrete components in a system, higher will be the loss and this reduces the compactness and reconfigurability of the system. But a broken phase shifted $\mathcal{PT}$-symmetric FBG model, illustrated in this work, has an edge over the conventional systems to realize a narrow band lasing spectra as illustrated in the following sections. The same phase-shifted PTFBG itself will do the filtering, amplification, side-lobe suppression  without a need for additional discrete components. Moreover, the length of the phase-shifted PTFBG proposed in this work is of the order of millimeters. This has a definite role in increasing the compactness of the overall system. Moreover, the same device can be configured to function as a demultiplexer and phase-shifted modulator in the unbroken $\mathcal{PT}$-symmetric regime.  

To accomplish our above motivation, we segment the remaining part of this article as follows: Section \ref{Sec:2} illustrates the mathematical model of the proposed system. In Sec. \ref{Sec:3}, we investigate the grating characteristics of the proposed system in the unbroken $\mathcal{PT}$-symmetric regime followed by the demonstration of unidirectional wave transport phenomenon at the exceptional point in Sec. \ref{Sec:4}. The lasing behavior in the broken $\mathcal{PT}$-symmetric regime is elucidated in Sec. \ref{Sec:5}. Finally, the article is concluded in Sec.\ref{Sec:6}.

\section{mathematical model}
\label{Sec:2}

\begin{figure}
	\centering	\includegraphics[width=1\linewidth]{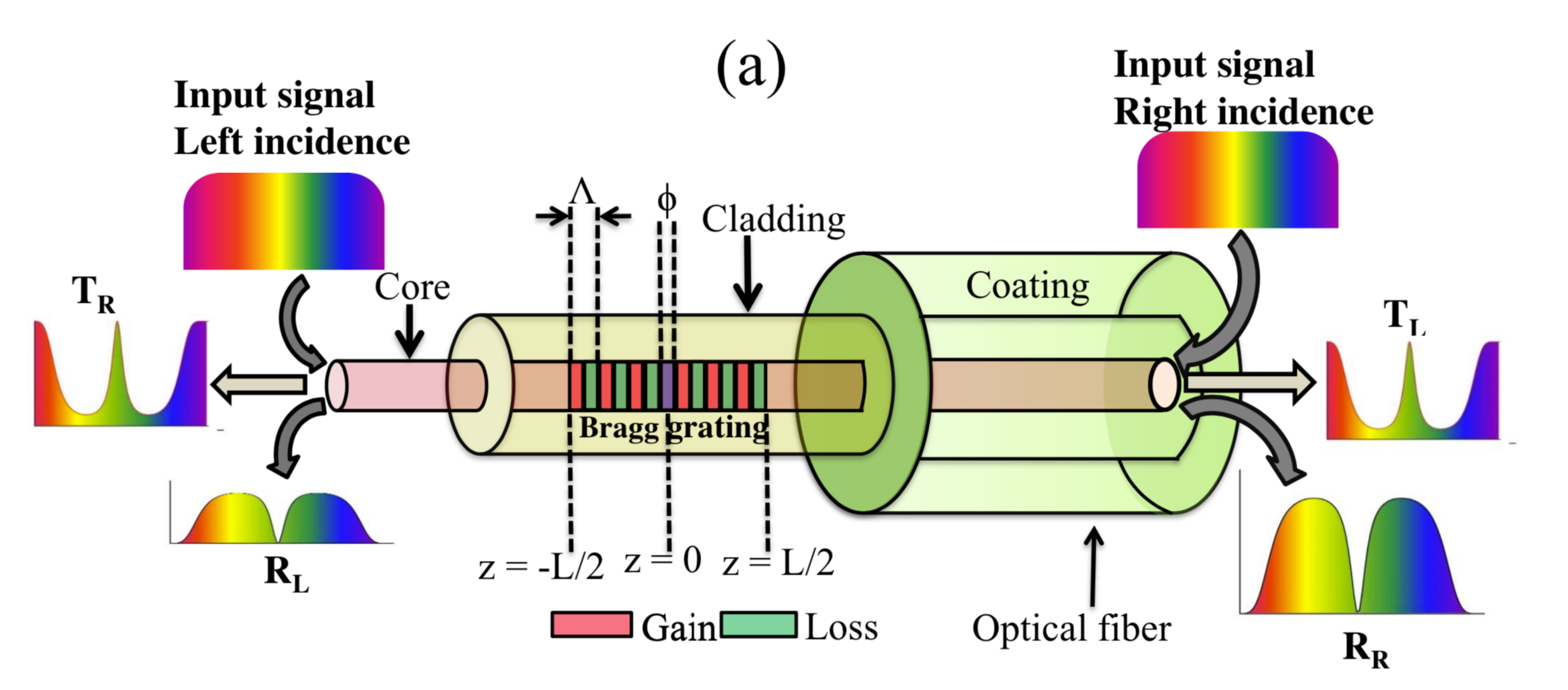}\\
	\centering	\includegraphics[width=1\linewidth]{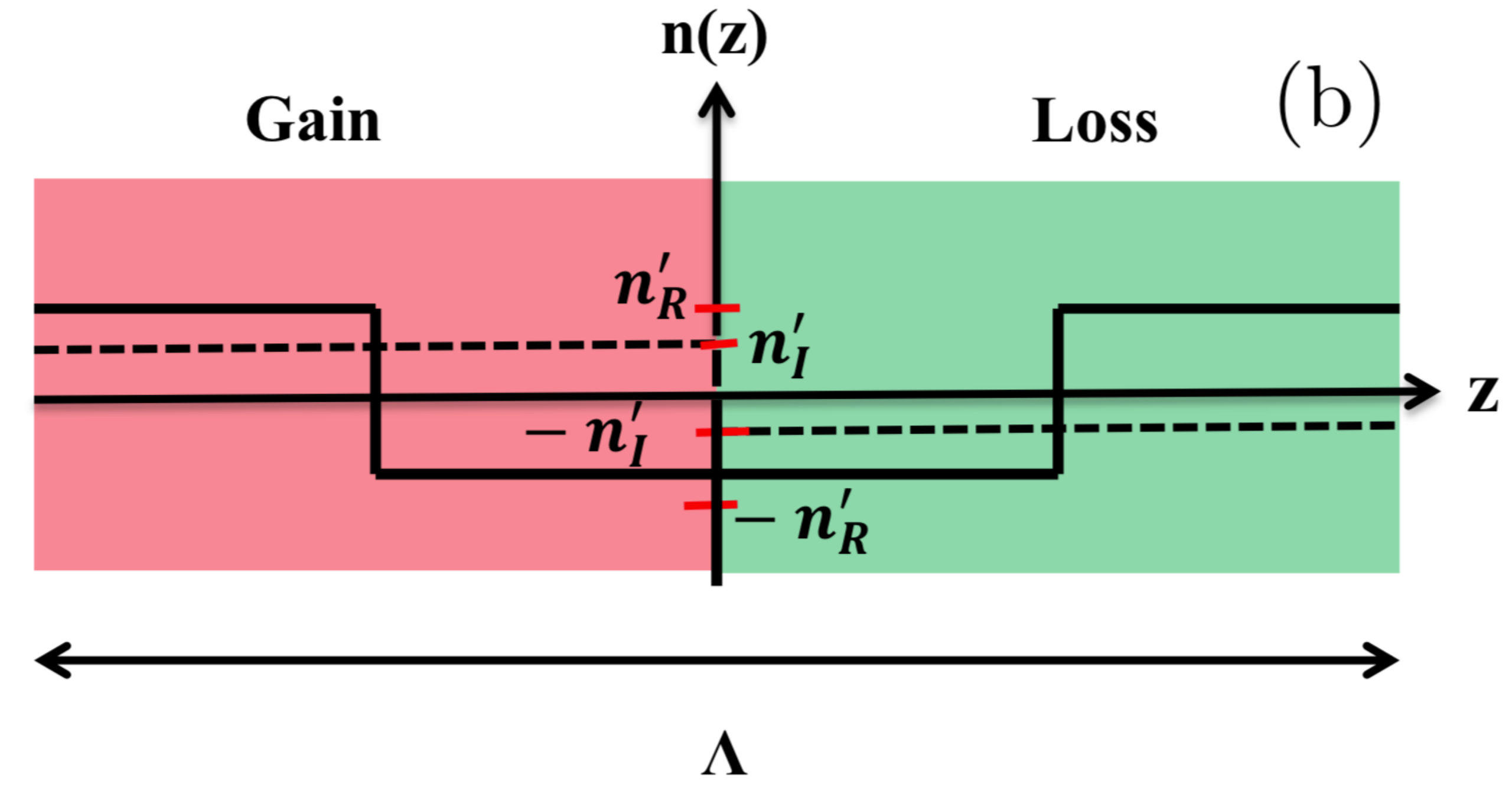}\\
	\caption{(a) Schematic of a phase-shifted $\mathcal{PT}$-symmetric fiber Bragg grating (PPTFBG) where two uniform PPTFBGs each having a length of $L/2$ which is separated with a phase-shift $\phi$ in the middle of the structures.  (b) Each unit cell of uniform structure with grating period $\Lambda$ consists of equal gain (green) and loss regions (red) to maintain the $\mathcal{PT}$-symmetric refractive index ($n(z)$). }
	\label{fig0}
\end{figure}
\begin{figure}
	\centering	\includegraphics[width=0.5\linewidth]{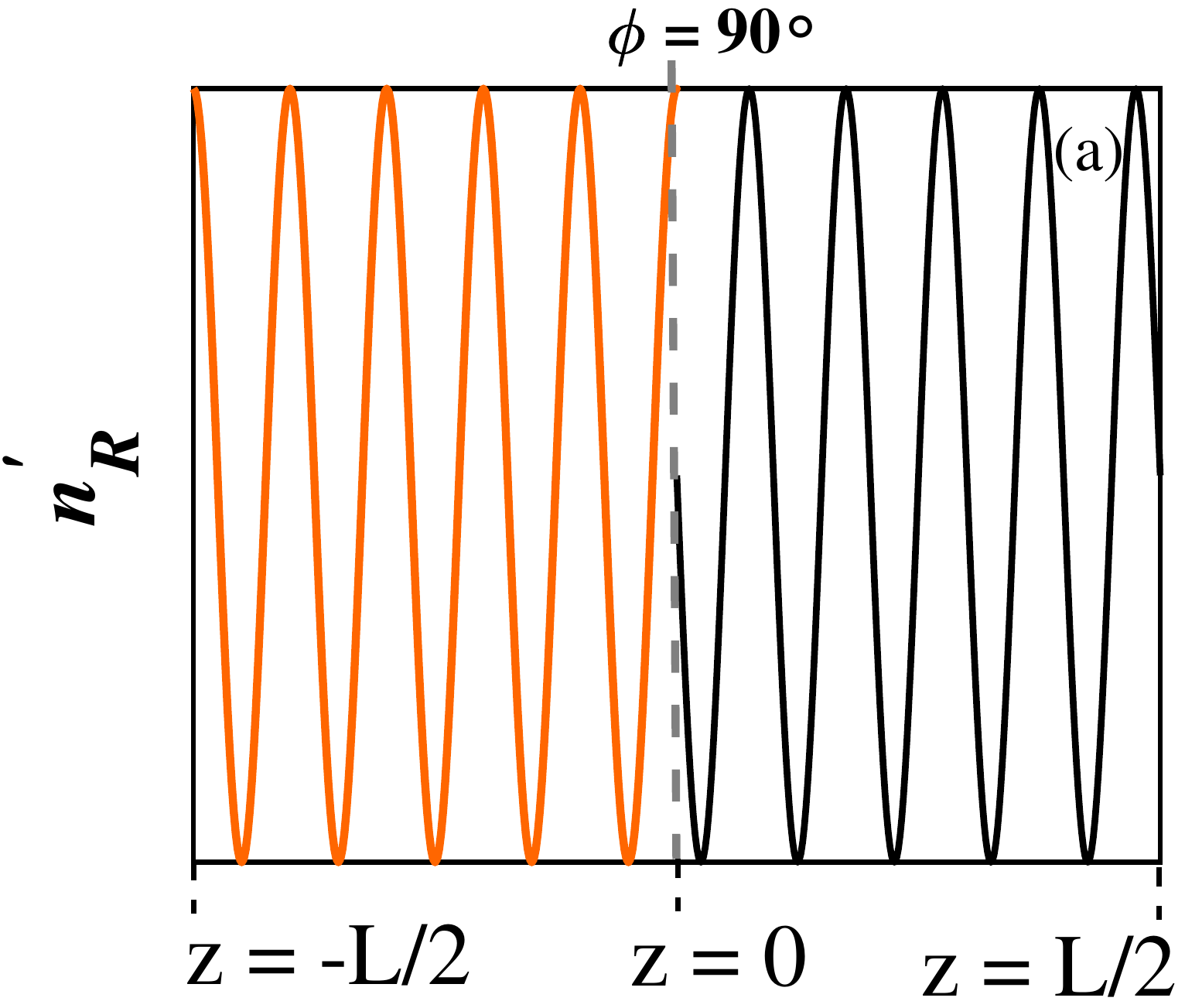}\centering	\includegraphics[width=0.5\linewidth]{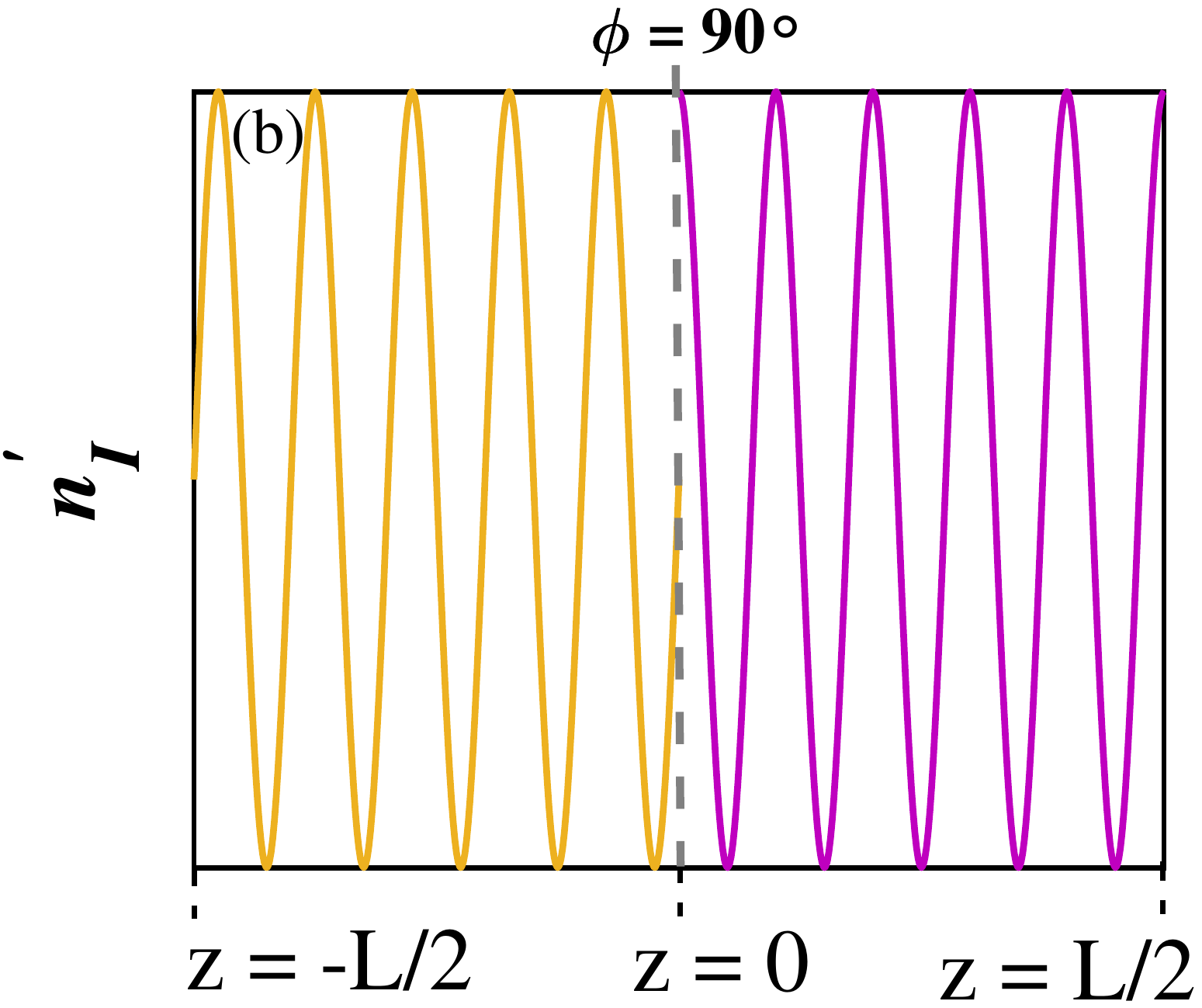}\\\centering	\includegraphics[width=0.5\linewidth]{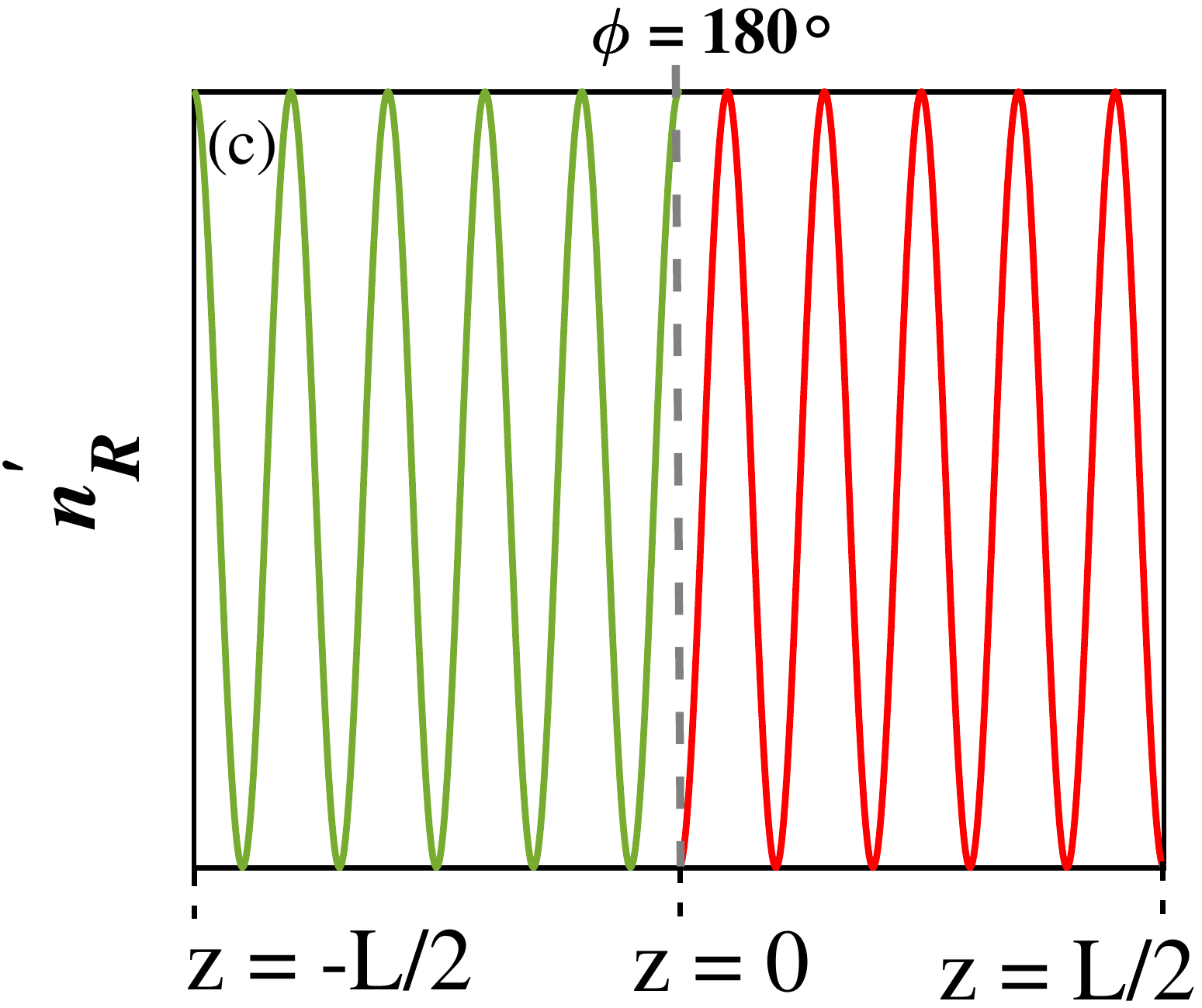}\centering	\includegraphics[width=0.5\linewidth]{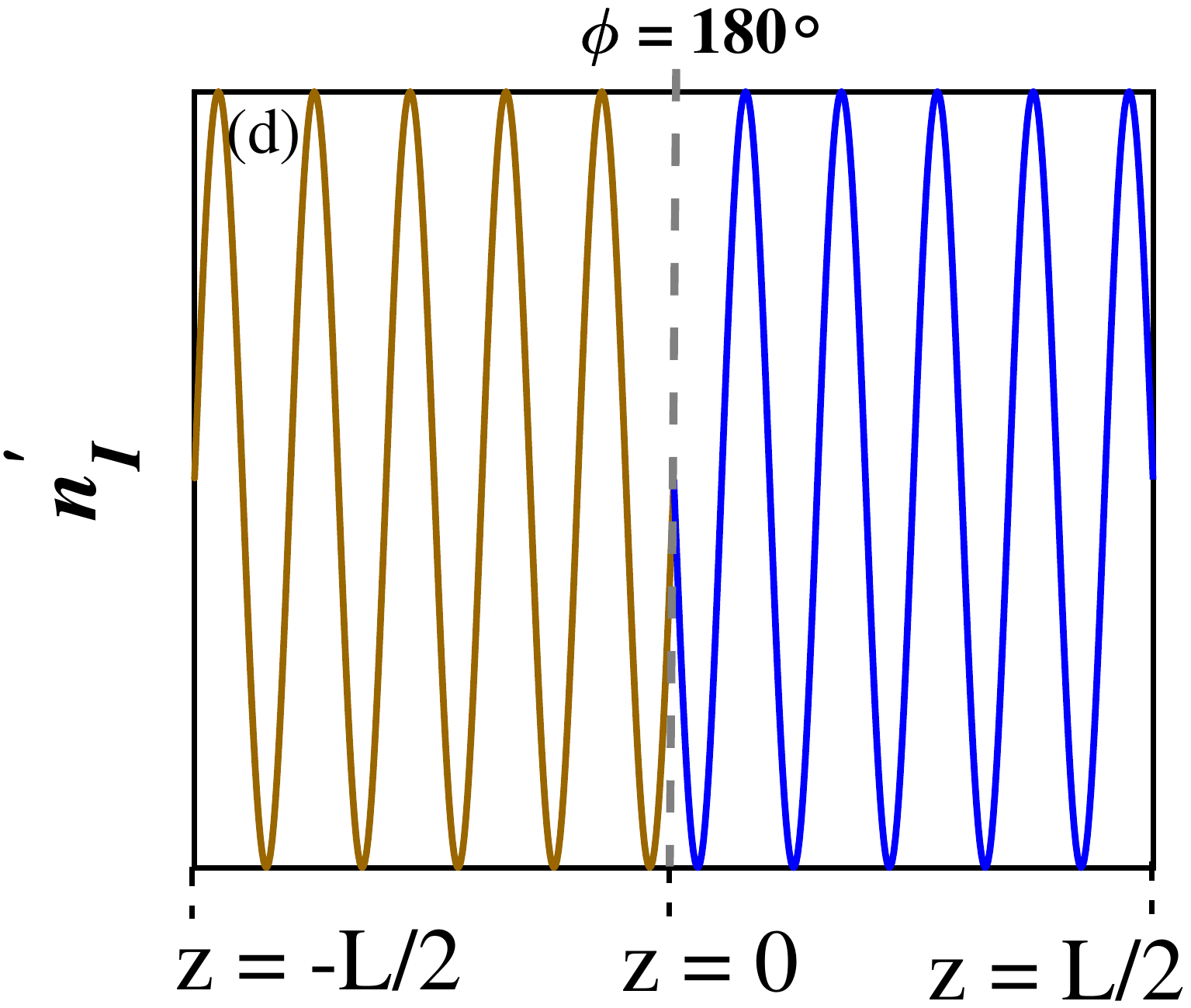}\\\centering	\includegraphics[width=0.5\linewidth]{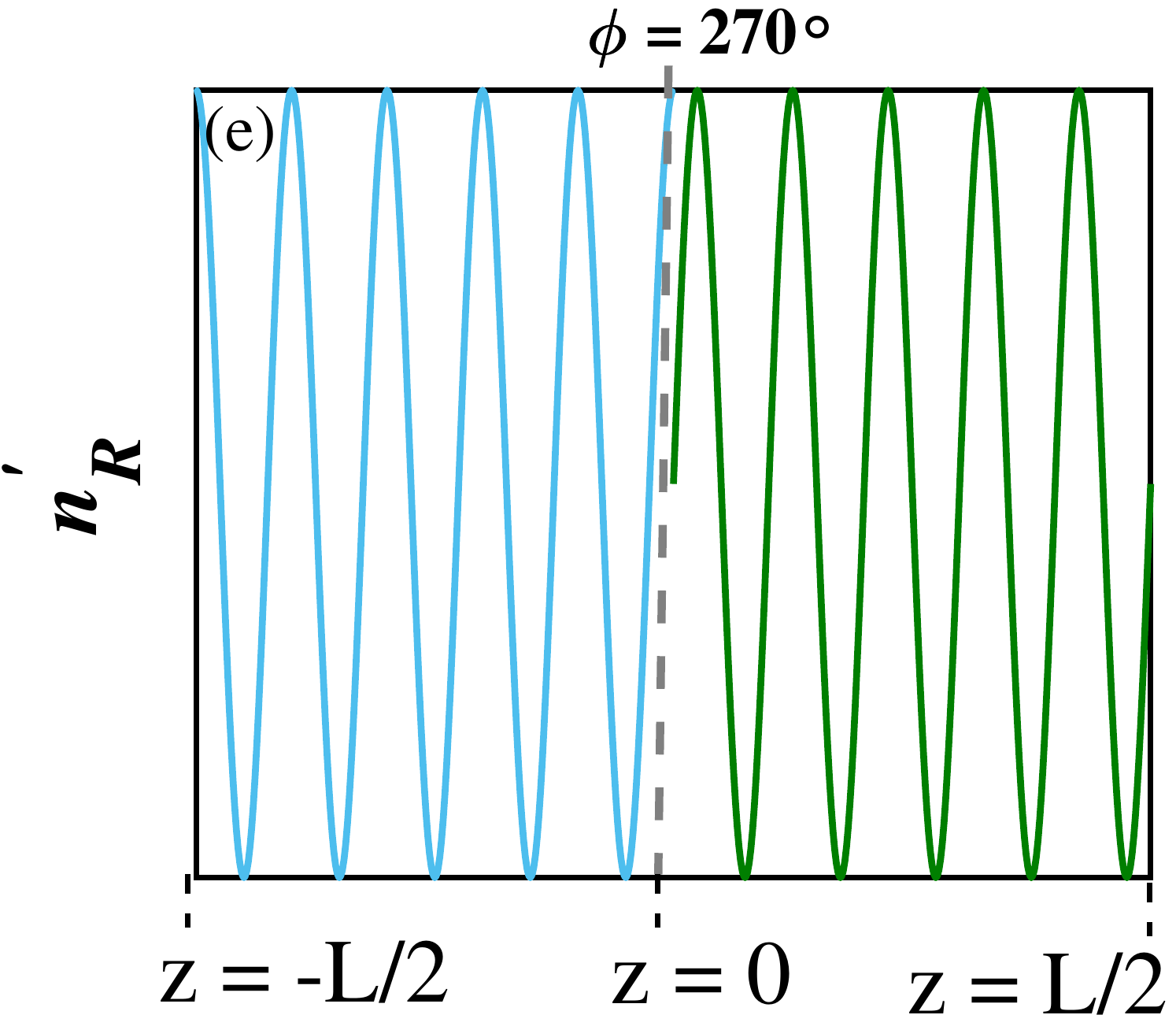}\centering	\includegraphics[width=0.5\linewidth]{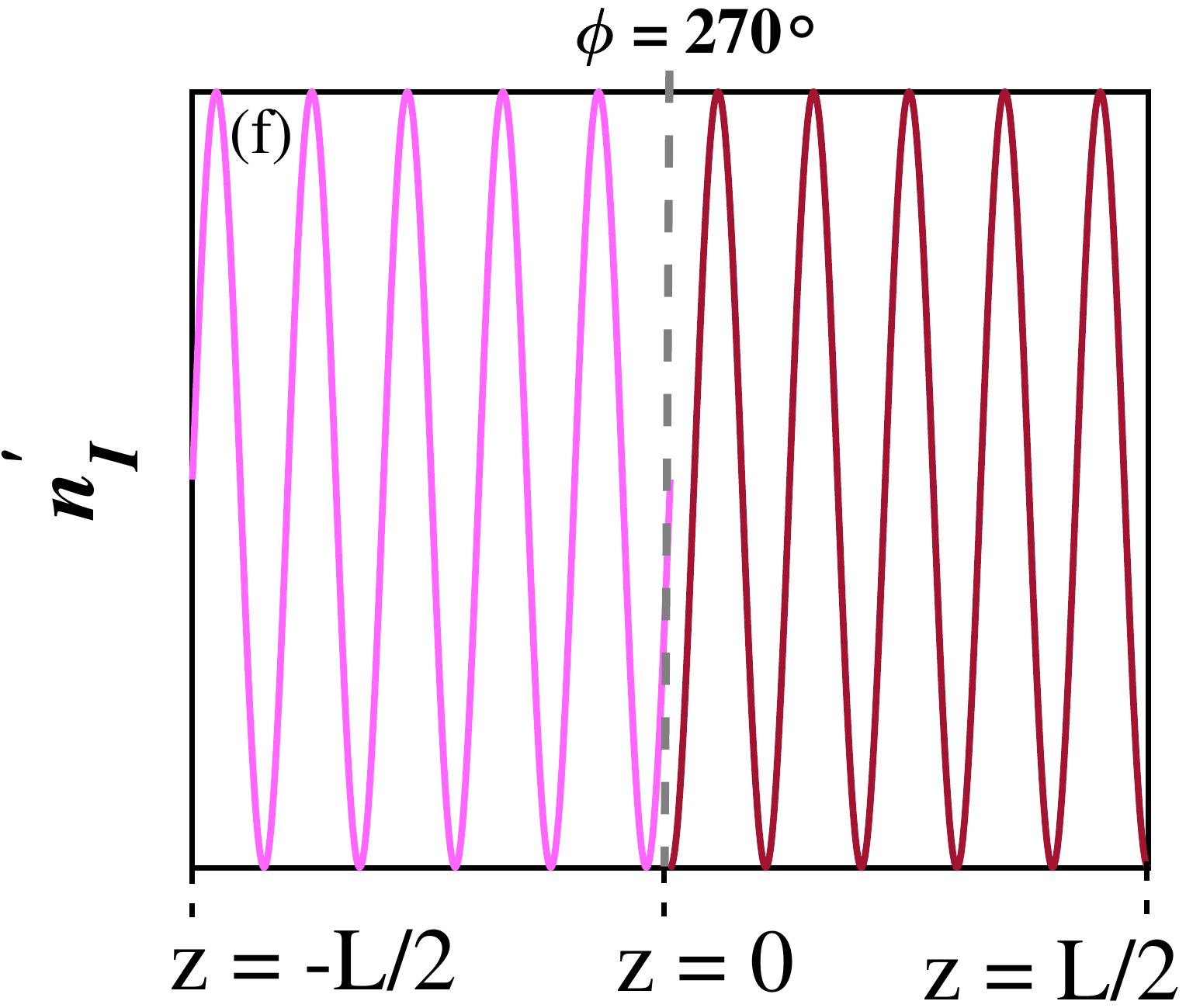}
	\caption{Schematics of the variation of real and imaginary parts of refractive index for various phase-shift values are drawn in the left and right panels, respectively. They are plotted at a phase of (a), (b) $\phi=90^\circ$, (c), (d) $\phi=180^\circ$, (e) and (f) $\phi = 270^\circ$. In all the plots, one can observe that the different phase values contribute to the various types of refractive index profiles which predominantly get changed in the middle of the device with respect to the phase values ranging from $90^\circ$ to $270^\circ$.}
	\label{fig01}
\end{figure}

 We consider a $\mathcal{PT}$-symmetric refractive index distribution [$n(z)$] that includes the effect of discrete phase shift ($\phi$), which is introduced in the middle of the periodic structure ($z=0$). The total distance $L$ is divided into many small unit cells with each having a grating period $\Lambda$ (see Fig. \ref{fig0}). The refractive index distribution is given by
\begin{equation}
n(z)=n_{0}+n_{1R}\cos\left(\frac{2\pi}{\Lambda}z+\phi\right)+in_{1I}\sin\left(\frac{2\pi}{\Lambda}z+\phi\right),
\label{Eq:norm1}
\end{equation}

where $n_0$, $n_{1R}$, and $n_{1I}$  represent the effective refractive index of the core, real and imaginary parts of the grating's modulation strength, respectively. It is to be noted that Eq. \eqref{Eq:norm1} refers to the  gradually varying sinusoidal profile that may be hard to fabricate in real situation. To overcome this, one can employ square wave forms instead of sinusoidal forms as shown in Fig. 1(b). This will enable one to study the entire structure consisting of each unit cell of length $l$ obeying sine and cosine modulations of the considered profile in Eq. \eqref{Eq:norm1}. Also, it is assumed that the real ($n_{1R}$) and imaginary ($n_{1I}$) parts of the  modulation strength are weak perturbations when compared to the uniform index of the background material ($n_0$), viz., $n_{1R}, n_{1I}\ll n_0$ \cite{lin2011unidirectional}. Hence, after neglecting the higher order terms, the equation resulting from squaring Eq. (\ref{Eq:norm1}) can be reduced to the final form,
\begin{gather}
\nonumber n^2(z)=n_0^2+2n_0n_{1R}cos(2\pi/\Lambda z+\phi)\\+2n_0n_{1I}sin(2\pi/\Lambda z+\phi)
\label{Eq:normm4}
\end{gather}
Equation (\ref{Eq:normm4}) can be rewritten as
\begin{gather}
\nonumber n^2(z)=n_0^2+n_0(n_{1R}+n_{1I})\exp\left(\frac{2i\pi}{\Lambda}z+\phi\right)\\+n_0(n_{1R}-n_{1I})\exp\left(\frac{-2i\pi}{\Lambda}z+\phi\right)
\label{Eq:normm5}
\end{gather}

The coupled mode equations that describe the system of interest can be found by substituting the refractive index distribution given in Eq. (\ref{Eq:norm1}) in the time-independent Helmholtz equation. It describes the propagation characteristics of the incoming optical field inside the $\mathcal{PT}$-symmetric optical structure \cite{lin2011unidirectional} which can be given by
\begin{equation}
\cfrac{d^2E}{dz^2}+k^2\left(\cfrac{n^2(z)}{n_0^2}\right)E=0.
\label{Eq:normm2}
\end{equation}
In Eq. (\ref{Eq:normm2}), $k$ stands for the wave vector and $E$ describes the optical field as a superposition of forward and backward fields traveling inside the grating and it reads as
\begin{equation}
 E=E_f(z) \exp(ik z)+E_b(z) \exp(-ik z),
\label{Eq:normm3}
\end{equation}
where $E_f$ and $E_b$ refer to the slowly varying amplitudes of forward and backward propagating fields, respectively. 
Note that based on slowly varying envelope approximation, the second derivatives of the forward ($E_f''$) and backward field ($E_b''$) envelopes can be neglected. On substituting Eqs. (\ref{Eq:normm3})  and \eqref{Eq:normm5} together  in Eq. \eqref{Eq:normm2} and averaging over the  rapidly oscillating terms $\exp\left[\pm i\left(\frac{2\pi}{\Lambda}z+\phi+k z\right)\right]$ by synchronous approximation, the resulting equation reads as
\begin{gather}
\nonumber 2i k E_f'\exp(i k z)-2i k E_b'\exp(-ik z)\\\nonumber +k^2\frac{(n_{1R}+n_{1I})}{n_0} E_b \exp\left[i\left(\frac{2\pi}{\Lambda}z+\phi-k z\right)\right]\\+k^2\frac{(n_{1R}-n_{1I})}{n_0} E_f \exp\left[-i\left(\frac{2\pi}{\Lambda}z+\phi-k z\right)\right]=0.
\label{Eq:normm6}
\end{gather}

 We can check that Eq. (6) implies the following system of first order coupled linear differential equations for the  forward and backward propagating fields $E_f$ and $E_b$, respectively,
 \begin{gather}
 \nonumber k \frac{(n_{1R}+n_{1I})}{n_0} E_b \exp\left[i\left(\frac{2\pi}{\Lambda}z+\phi-k z\right)\right]\\+2i E_f'\exp(i k z)=0,
 \label{Eq:normm7}
 \end{gather}
\begin{gather}
\nonumber k \frac{(n_{1R}-n_{1I})}{n_0} E_f \exp\left[-i\left(\frac{2\pi}{\Lambda}z+\phi-k z\right)\right]\\-2i E_b'\exp(-i k z)=0.
\label{Eq:normm8}
\end{gather}

From the fundamental definition of wave vector, the standard forms of coupling ($\kappa$) and gain-loss coefficients ($g$) read as \cite{lin2011unidirectional,sarma2014modulation,miri2012bragg,raja2019multifaceted,raja2020tailoring} 
\begin{gather}
k n_{1R}/2 n_0 = \pi n_{1R}/\lambda = \kappa, \quad k n_{1I}/2 n_0 = \pi n_{1I}/\lambda = g
\label{Eq:normm9}
\end{gather}
 where $\lambda$ is the operating wavelength. Also, the detuning parameter is given by
 \begin{gather}
 \delta=2\pi n_0 \left(\cfrac{1}{\lambda}-\cfrac{1}{\lambda_b}\right), 
 \label{Eq:normm10}
 \end{gather}
 where $\lambda_b = 2n_{0} \Lambda$. Substituting these standard notations in Eq. (\ref{Eq:normm7}) and Eq. (\ref{Eq:normm8}), Eqs. (7) and (8) can be rewritten as
 \begin{gather}
 \frac{dE_f}{dz}=i\left({\kappa}+{g}\right)e^{i\phi} e^{-2i\delta z} E_b, 
 \label{Eq:normm11}\\
 \frac{dE_b}{dz}=-i\left({\kappa}-{g}\right)e^{-i\phi} e^{2i\delta z} E_f,
 \label{Eq:normm12}
 \end{gather}
 
The resulting coupled mode equations that describe the proposed system can be found out by substituting transformations, $u,v=E_{f,b}\exp(\mp i\delta z)$ \cite{raja2020tailoring} in Eqs. (\ref{Eq:normm11}) and (\ref{Eq:normm12}), which read as \cite{radic1995theory,raja2020tailoring},
\begin{gather}
+i\frac{du}{dz}+\delta u+\left({\kappa}+{g}\right)e^{i\phi} v=0, 
\label{Eq:norm2}\\
-i\frac{dv}{dz}+\delta v+\left({\kappa}-{g}\right)e^{-i\phi} u=0,
\label{Eq:norm3}
\end{gather}
where $u$, $v$, $z$, $\delta$, $\kappa$, and $g$ correspond to forward, backward field amplitudes, the spatial coordinate, detuning, coupling, and gain/loss parameter,  respectively.

Since the central theme of the article is to demonstrate tunable spectral characteristics of a PPTFBG by varying the magnitude of the phase-shift (PS) in the middle of the grating, it is important to comment on the existing techniques for the fabrication of phase-shifted fiber Bragg gratings (PFBGs). This includes the UV
post-processing \cite{cusano2009microstructured},  moving fiber-scanning beam method \cite{loh1995complex}, CO$_2$ laser irradiation \cite{xia2005phase}, shielded phase mask method \cite{rota2013dual}, Moire method \cite{legoubin1991formation}, and so on. A detailed comparative study of various fabrication techniques to realize a PFBG was outlined by Chehura \emph{et al.} and they concluded that the scanning beam method  is comparatively advantageous over other existing methods \cite{chehura2010simple}. Even though this technique is simple and more reliable, the magnitude of PS etched on the structure cannot be varied, once it is fabricated. For some specific applications (as enumerated previously), the magnitude of the PS needs to be tunable rather than remaining static. Under such circumstances, one should opt for a fabrication process that would allow dynamical variation in the PS value, post the formulation of the device. Some of the widely used techniques in this context includes in-grating bubble technique \cite{liao2013tunable},  mechanical tuning with piezoelectric transducers \cite{chen2010phase}, and heating element method \cite{janos1995permanent}. Each one of the aforementioned techniques possess some disadvantages and the cost ineffectiveness is one of the major concerns among all. Falah \emph{et al.} came up with an idea of \textit{4-point beam bending arrangement}, which would allow tuning the PS value via variation in the optical path by spatially varying the strain within the periodic structure through the micrometer screw adjustment \cite{falah2016reconfigurable}. This method was proven to posses set and forget capability, consume less power and tunable over the full scale range of PS from 0 to $2\pi$. More importantly, the degree of accuracy of PS value achieved with respect to the micrometer screw variation was found to show good agreement with theoretical results. Recently, variable-velocity scanning method was proposed by Zhou \emph{et.al}., which allows fabrication of PFBGs with asymmetric uniform grating section on either side of the PS region \cite{zhou2018novel}. Even though it is experimentally established that it is possible to obtain tunable PFBGs, we confine our investigation to the symmetric PFBGs (consisting of two uniform grating structures with each length of $L/2$ with PS placed in their middle) with equal gain and loss as shown in Fig. \ref{fig0}. With this brief note, we affirm that our proposed model is experimentally realizable and applications such as tunable lasing can be established with ease, thanks to the ability of the structure to provide tunable phase shift. 

Although the mathematical model (\ref{Eq:norm2}) and (\ref{Eq:norm3}) of the system can be solved by direct-integration technique with standard FBG boundary condition,  the phase-shifted FBG with gain and loss shown in Fig. \ref{fig0} can also be  investigated   with the aid of transfer matrix method (piece-wise uniform technique) \cite{agrawal1994phase,raja2020tailoring}.  This is a most commonly used technique to model a linear and nonuniform FBG, and it also gives a clear picture on the practical realization of these structures. Moreover, the investigation of the spectra of the grating becomes a relatively simple task as this method is faster than the direct-integration approach in the case of more complicated grating physical structures. The modeling of phase-shifted FBG system  relies on the fact that the overall structure is formed by concatenating two uniform and symmetric FBGs with a phase-shift value of $\phi$ in between them \cite{erdogan1997fiber}. The phase shift $\phi$ at  $z=0$ can be inserted by simply multiplying the matrix corresponding to the first uniform FBG ($M_1$) with the diagonal matrix ($M_{ph}$) having elements $\exp(\pm i\phi/2)$ followed by a second uniform FBG ($M_2$). Thus, the resultant second order matrix ($F$) that describes the overall system reads as
\begin{gather}
F=M_1 \times M_{ph} \times M_2,
\label{Eq:noorm1}
\end{gather}
Hence, the input and output fields are related by,
\begin{gather}
\left[\begin{array}{c}
u_n\\
v_n
\end{array}\right]=M_1 \times M_{ph} \times M_2\left[\begin{array}{c}
u_0\\
v_0
\end{array}\right].
\label{Eq:Norm9}
\end{gather}
Here $u_n$ and $v_n$ represent the output amplitudes  as a function of input amplitudes $u_0$ and $v_0$ and $M_1$, $M_2$ denote the matrices which relate the input and output fields of two uniform FBGs separated by the discrete phase-shift ($\phi$).  
Modeling the uniform $\mathcal{PT}$-symmetric grating ($-L/2\leq z<0$ and $0<z\leq L/2$) requires dividing the length of the uniform grating ($\hat{L} = L/2$) into $n$ number of small sections of each of length $l$  \cite{raja2020tailoring}. Each piecewise section of length $l$ is modeled by a corresponding matrix $m_1$, $m_2$, $\dots$ $m_n$ and hence the uniform gratings of length $\hat{L}$ are represented by the product matrices 
\begin{gather}
M_1= M_2 = m_1  m_2 m_3 \dots m_{n-1} m_n
\end{gather}
\begin{gather}
m_1=\left[\begin{array}{cc}
m_{11} & m_{12}\\
m_{21} & m_{22}
\end{array}\right],\\\nonumber
\end{gather}
where
\begin{gather}
\nonumber  m_{11} = m_{22}^* =\cosh(\hat{\sigma} l  )+i\left(\cfrac{\delta}{\hat{\sigma}}\right)\sinh(\hat{\sigma} l),
\\\nonumber m_{12}=i\left(\cfrac{\kappa+g}{\hat{\sigma}}\right)\sinh(\hat{\sigma} l),
\\ {m_{21}=-i\left(\cfrac{\kappa-g}{\hat{\sigma}}\right)\sinh(\hat{\sigma} l).}
\label{Eq:Norm}
\end{gather}

 The procedure in which the TMM elements in Eq. (\ref{Eq:Norm}) are calculated is  straightforward. By considering the harmonic solutions in the form of $u,v=A_f,A_b \exp(i\hat{\sigma}z)$, one arrives at the dispersion relation for the common propagation constant for $m_1$, $m_2$, $\dots$, $m_n$ as, $\hat{\sigma}=\sqrt{\left(\kappa^2-g^2-\delta^2\right)}$.  Once the propagation constant is found, the appropriate boundary values are applied to the Bragg gratings to arrive at the matrix elements.

Also, the matrix that represents the phase-shift region is given by \cite{erdogan1997fiber}
\begin{gather}
\nonumber \\M_{ph}=\left[\begin{array}{cc}
exp(i\phi/2) & 0\\
0 & exp(-i\phi/2)
\end{array}\right]
\end{gather}

Note that the phase-shift ($\phi$) is introduced in the middle of the grating  such that on either sides of the phase shift region, we have two uniform FBGs as indicated  in Fig. \ref{fig0}(a). Each uniform PTFBG features a number of alternating regions of gain and loss. A single unit cell of period $\Lambda$ is constituted by having a real [$n_{R}^{'}$ $=$ $n_{1R}$ $cos(2 \pi z/\Lambda +\phi)$] and imaginary [$ n_{I}^{'}$ $=$ $i$ $n_{1I}$ $cos(2 \pi z/\Lambda +\phi)$] modulation of refractive index as shown in Fig. \ref{fig1}(b). Figures \ref{fig01}(a) -- \ref{fig01}(f) show the corresponding variation in $n_{R}^{'}$ and  $n_{I}^{'}$ when a particular value of phase shift is included into the system. To elucidate further, when $z<0$, the modulations of refractive index profile are unperturbed by the phase shift ($\phi=0$). At $z = 0$, the phase term ($\phi$) is added to the modulation profile in the middle of the grating and thus we visualize discontinuities (exactly in the middle) in $n_{R}^{'}$ and  $n_{I}^{'}$ profiles.  For $z>0$, these profiles get changed with the modified phase and thus the two uniform PTFBGs feature a difference in the phases of $\phi$. These variations in $n_{R}^{'}$ and  $n_{I}^{'}$ are responsible for the presence of narrow transmission band within the stopband of the grating as indicated later on in Fig. \ref{fig1}(a). The location of this transmission band is dictated by the amount of phase-shift ($\phi$) in the middle.

The resultant matrix $F$ reads as,
\begin{gather}
F=\left[\begin{array}{cc}
F_{11}& F_{12} \\
F_{21}  & F_{22} 
\end{array}\right],\\\nonumber
\end{gather}
where
\begin{gather}
\nonumber F_{11}=M_{11}^2 e^{i\phi/2}+M_{12} M_{21} e^{-i\phi/2}, \\\nonumber F_{12}= M_{11} M_{12} e^{i\phi/2}+M_{12} M_{22} e^{-i\phi/2},\\
\nonumber F_{21}=M_{21} M_{11} e^{i\phi/2}+M_{22} M_{21} e^{-i\phi/2},\\ F_{22}= M_{12} M_{21} e^{i\phi/2}+M_{22}^2 e^{-i\phi/2}.
\\\nonumber
\end{gather}
Thus the reflected and the transmission amplitudes of full PPTFBGs can be found from the final matrix $F$ as,
\begin{gather}
 r_L=- F_{21}/F_{22}=-\cfrac{M_{11}M_{21} \exp{(i\phi)}+M_{21}M_{22}}{M_{12}M_{21}\exp{(i\phi)}+M_{22}^2},\\ r_R=F_{12}/F_{22}=\cfrac{M_{11}M_{12} \exp{(i\phi)}+M_{12}M_{22}}{M_{12}M_{21}\exp{(i\phi)}+M_{22}^2},\\
t_L =t_R = t = |F_{11}F_{22}-F_{12}F_{21}|/F_{22}=1/F_{22}.
\label{Eq:Norm10}
\end{gather}
After some mathematical manipulations, using Eq. (\ref{Eq:Norm8}), the explicit relation for the reflection and transmission amplitudes can be found as,
\begin{gather}
r_L=\cfrac {i(\kappa-g)[\hat{\sigma} r_1(1+e^{i\phi})-i \delta r_1^2(1-e^{i \phi})]}{\kappa^2-g^2 (1+r_1^2 e^{i\phi})-\delta^2(1+r_1^2)-2i\delta \hat{\sigma}r_1},\label{Eq:Norm10a}\\
r_R=\cfrac {i(\kappa+g)[\hat{\sigma} r_1(1+e^{i\phi})-i \delta r_1^2(1-e^{i \phi})]}{\kappa^2-g^2(1+r_1^2 e^{i\phi})-\delta^2(1+r_1^2)-2i\delta \hat{\sigma}r_1} \label{Eq:Norm10b},\\
t=\cfrac {e^{i\phi/2} \hat{\sigma}^2 sech^2(\hat{\sigma}z)}{\kappa^2-g^2(1+r_1^2 e^{i\phi})-\delta^2(1+r_1^2)-2i\delta \hat{\sigma}r_1}.
\label{Eq:Noorm10c}
\end{gather}
In Eqs. (\ref{Eq:Norm10a}) --(\ref{Eq:Noorm10c}), $r_1=tanh(\hat{\sigma}L)$.

Further, if the applications pertaining to optical network communications demand more than a few transmission windows within the stop band of the spectra, one can make use of the concept of introducing multiple phase shifts as discussed below \cite{agrawal1994phase}.
\begin{figure}
	\centering
	\includegraphics[width=1\linewidth]{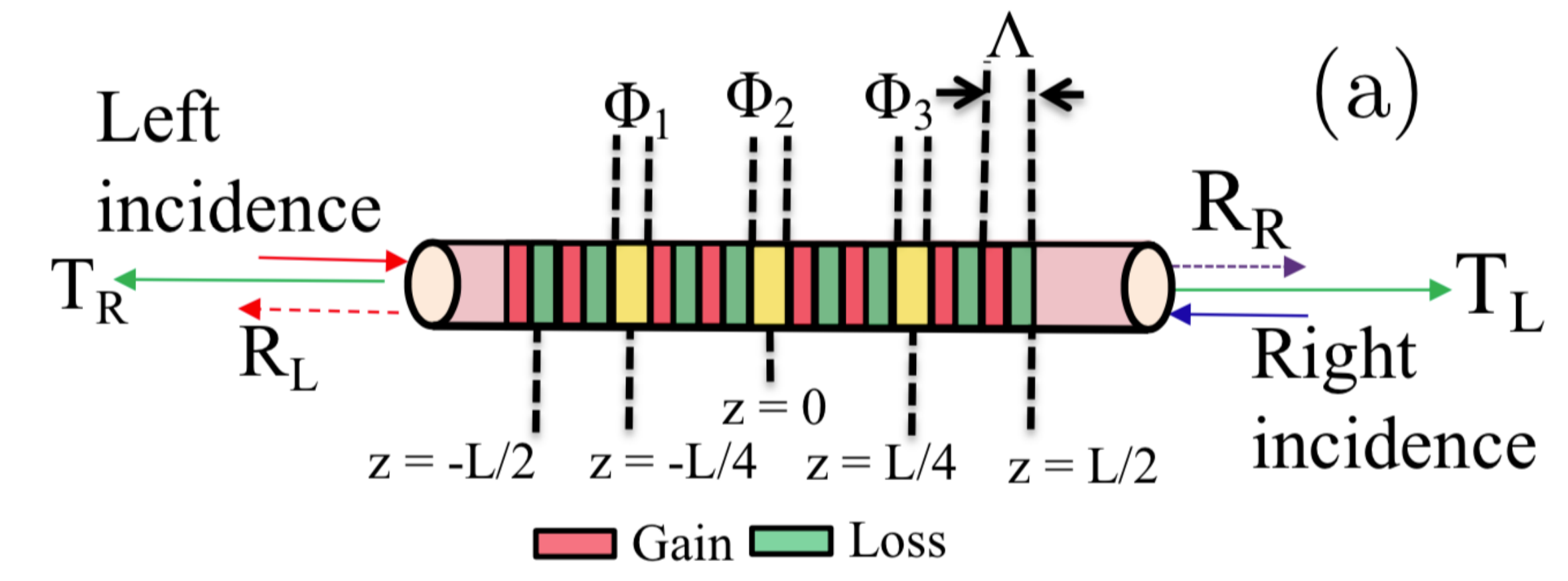}\\	\includegraphics[width=0.95\linewidth]{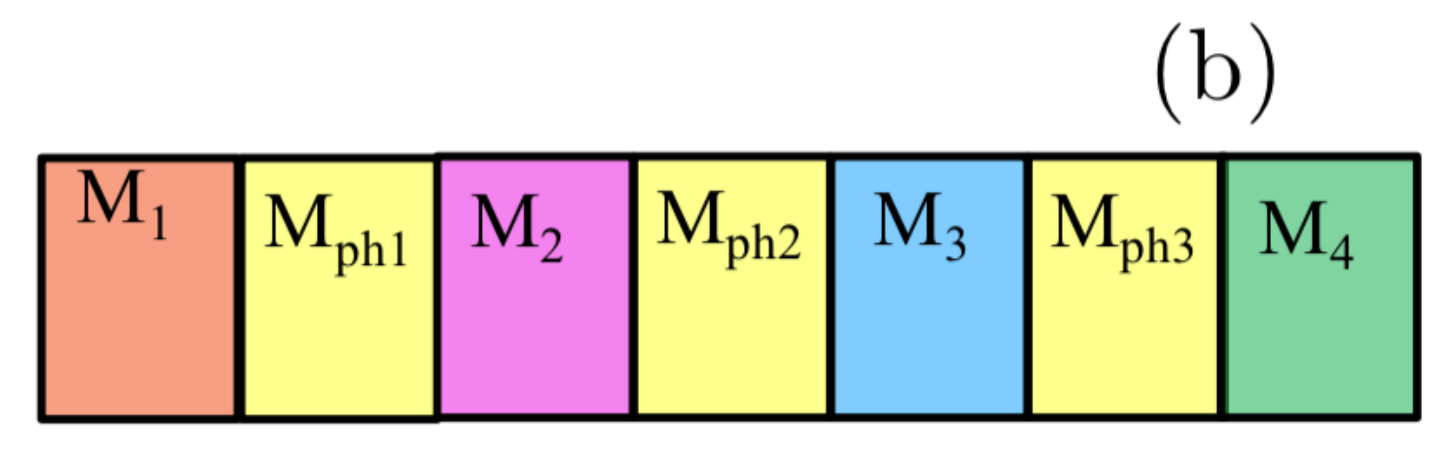}
	\caption{(a) Schematic of multiple phase shifted PTFBG with phase shift regions located at $z =-L/4$, $0$, $L/4$. (b) Portrays the implementation of transfer matrix method for the multi phase shift scheme.}
	\label{figm0}  
\end{figure}

The multiple phase shift regions can be modeled mathematically by including the phase matrix at the respective locations in the combined transfer matrix. As an example, first, second and third phase shift regions $z_1$, $z_2$, $z_3$ be located at $z = -L/4$, $0$, and $L/4$, respectively and the corresponding phase matrices are given by $M_{ph1}$, $M_{ph2}$ and $M_{ph3}$. Then we can identify a matrix, $M_1$ which represents the matrix that accounts for the resultant of all the small grating sections from $z = -L/2$ to $z = -L/4$. Similarly, $M_2$ represents the combined matrix obtained by cascading all the individual grating sections from $z = -L/4$ to $z = 0$. Likewise, $M_3$, $M_4$ represent the matrix acquired by cascading the individual grating  sections between $z = 0$ to $z = L/4$ and $z = L/4$ to $z = L/2$, respectively. The resultant matrix $M$ which stands for the overall structure from $z = -L/2$ to $z = L/2$ is given by
\begin{gather}\label{Eq: mul1}
M_{multi} = M_1 \times M_{ph1} \times M_2  \times M_{ph2} \times M_3\times  M_{ph3} \times M_4.
\end{gather}
It is to be noted that the explicit mathematical form in the case of multiple phase shift is cumbersome to find and hence the reflection and transmission amplitudes are computed from the transfer matrix routine and are given by,
\begin{gather}\label{Eq: mul2}
\nonumber r_L=-M_{21_{multi}}/M_{22_{multi}},\\
\nonumber r_R=M_{12_{multi}}/M_{22_{multi}},\\
t=det(M_{multi})/M_{22_{multi}}.
\end{gather}

Finally, the reflection and transmission coefficients can be expressed as
\begin{gather}\label{Eq: mul3}
R_{L}=|r_{L}|^2, \quad R_{R} = |r_{R}|^2, \quad T=|t|^{2}. 
\end{gather}
Throughout this paper, the length ($L$) and the coupling coefficient of the device ($\kappa$) are taken to be $4$ mm and $10$ cm$^{-1}$, respectively (unless specified).
\section{spectral characteristics of PPTFBG with a single phase shifted region}
\subsection{Unbroken $\mathcal{PT}$-symmetric regime}
\label{Sec:3}
\begin{figure}
	\centering
 	\includegraphics[width=0.5\linewidth]{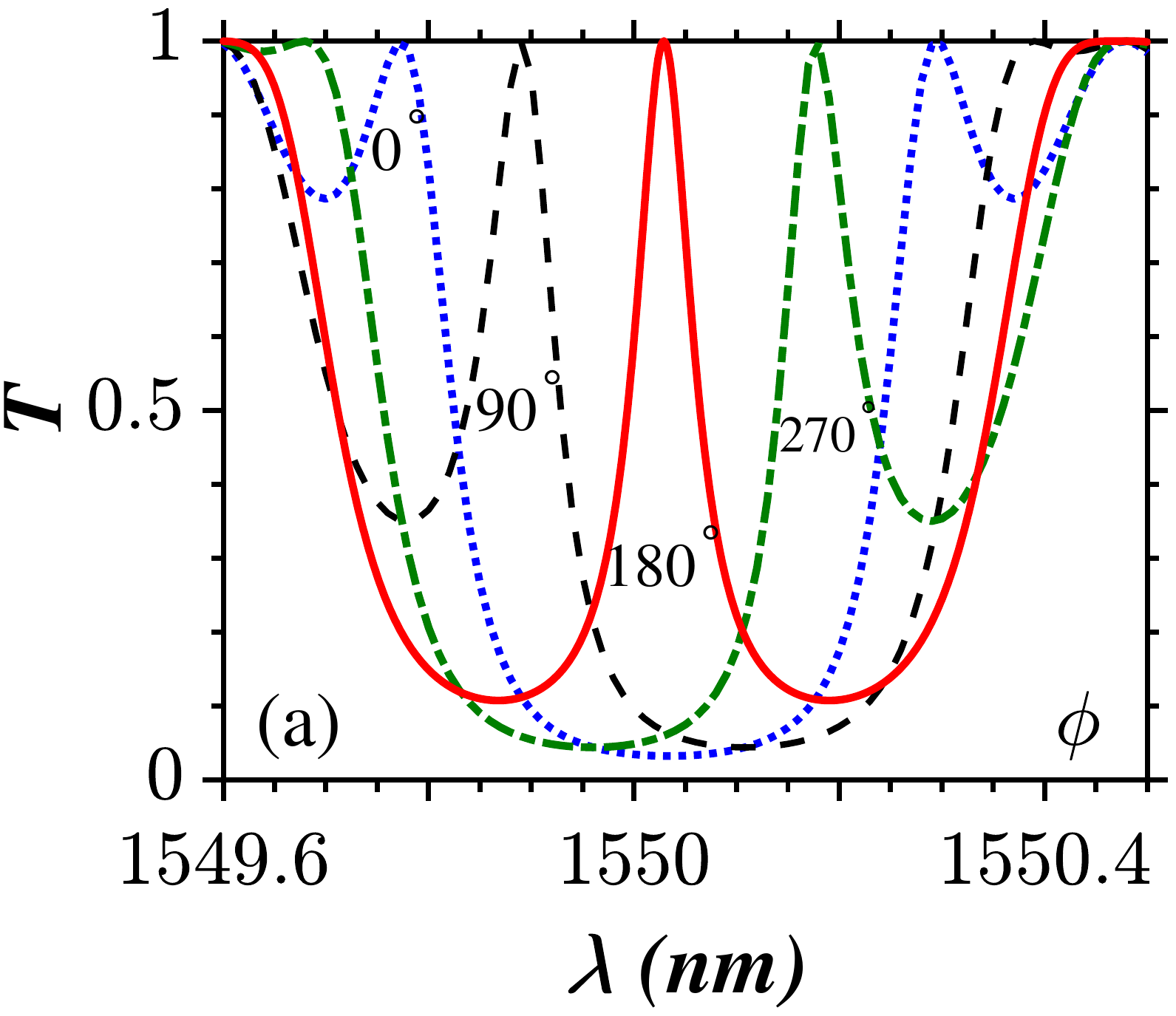}\includegraphics[width=0.5\linewidth]{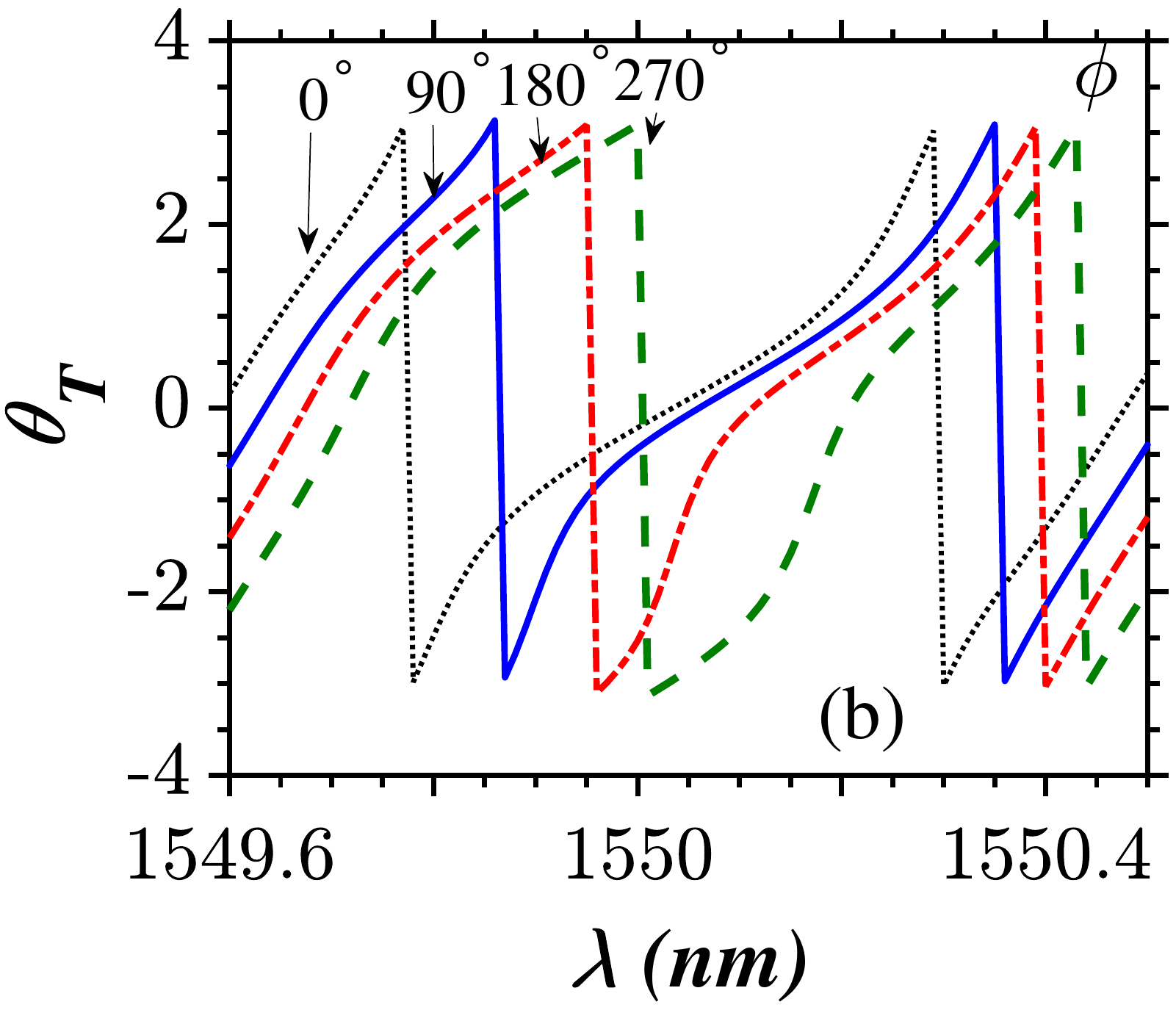}\\\includegraphics[width=0.5\linewidth]{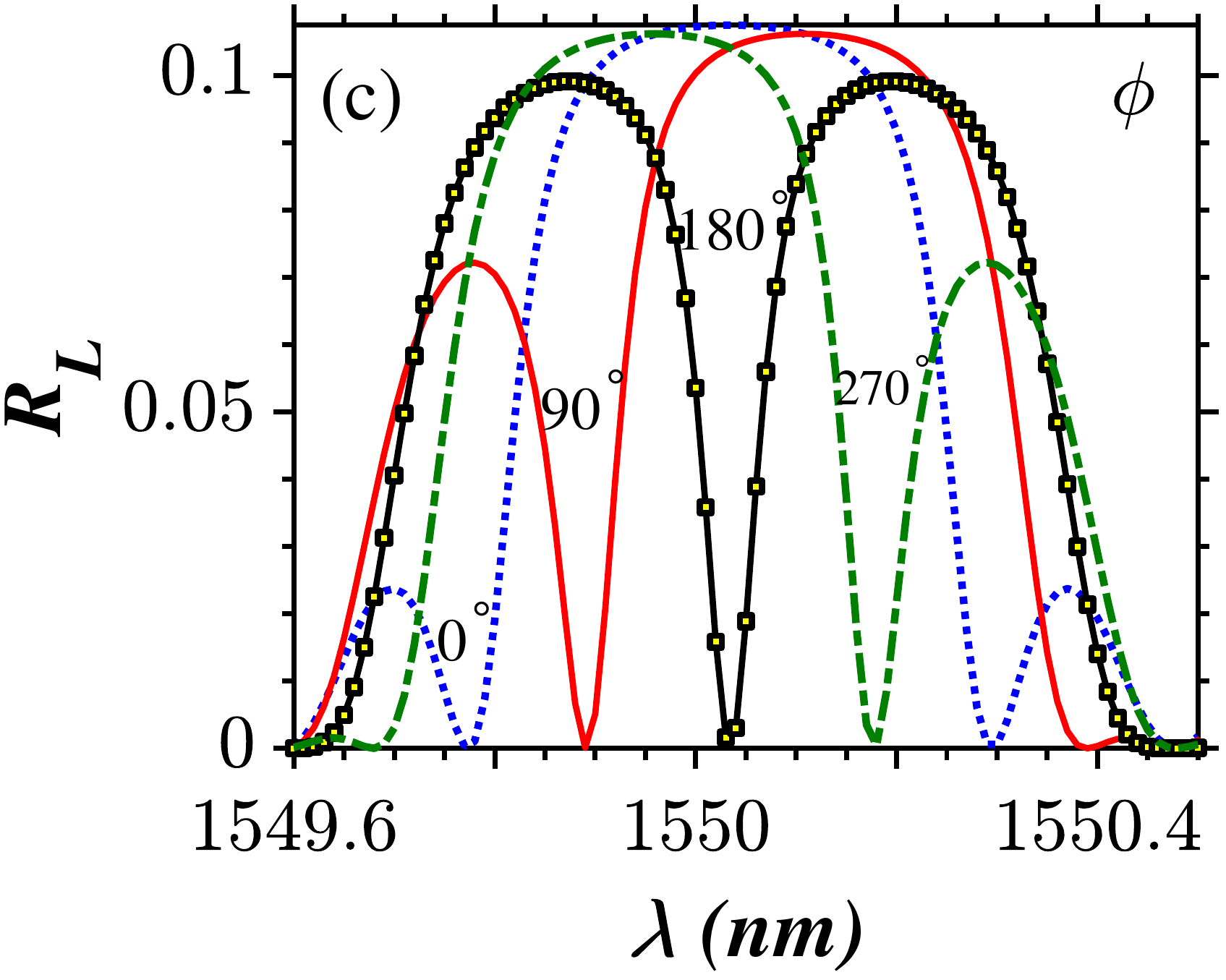}\includegraphics[width=0.5\linewidth]{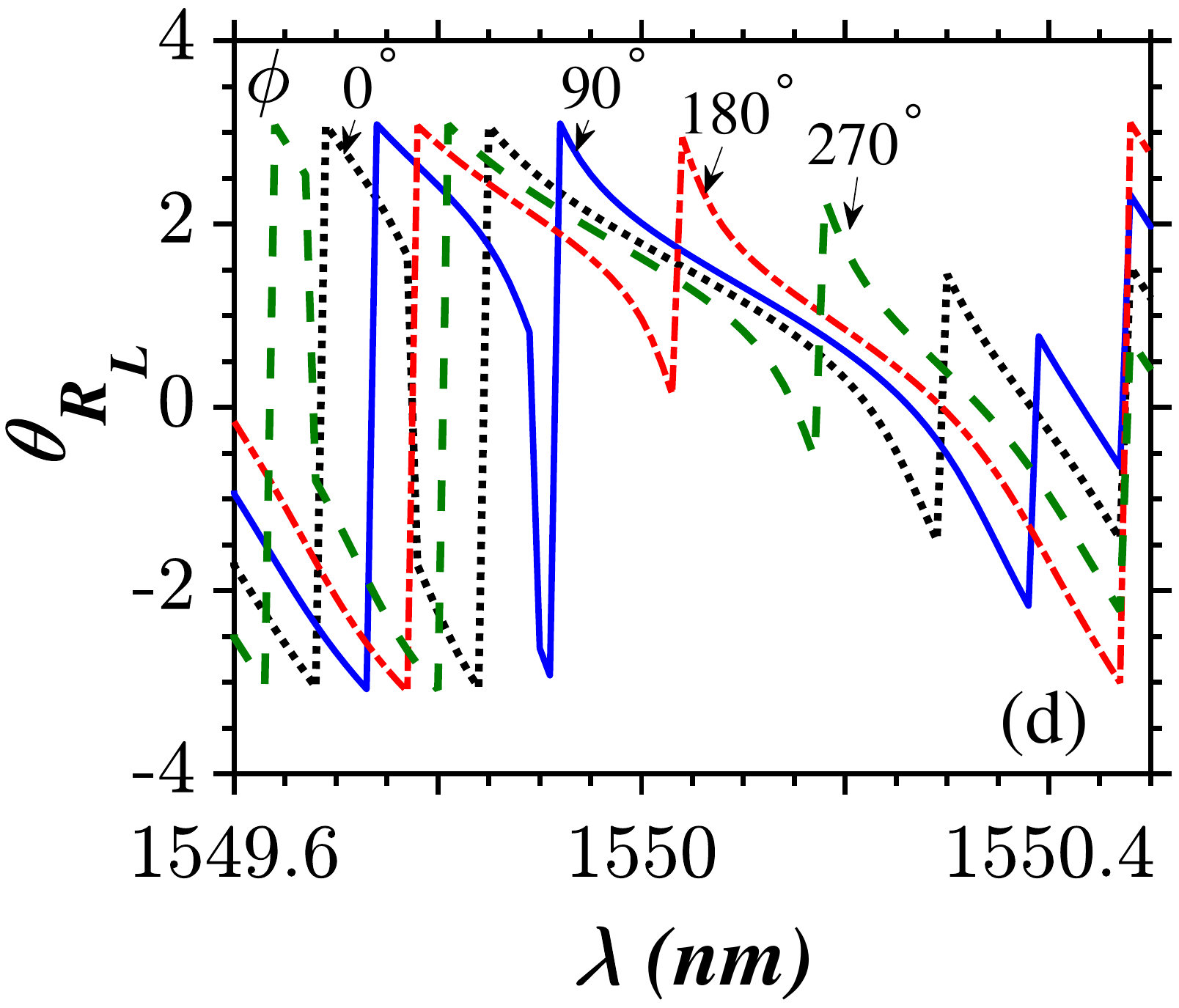}\\\includegraphics[width=0.5\linewidth]{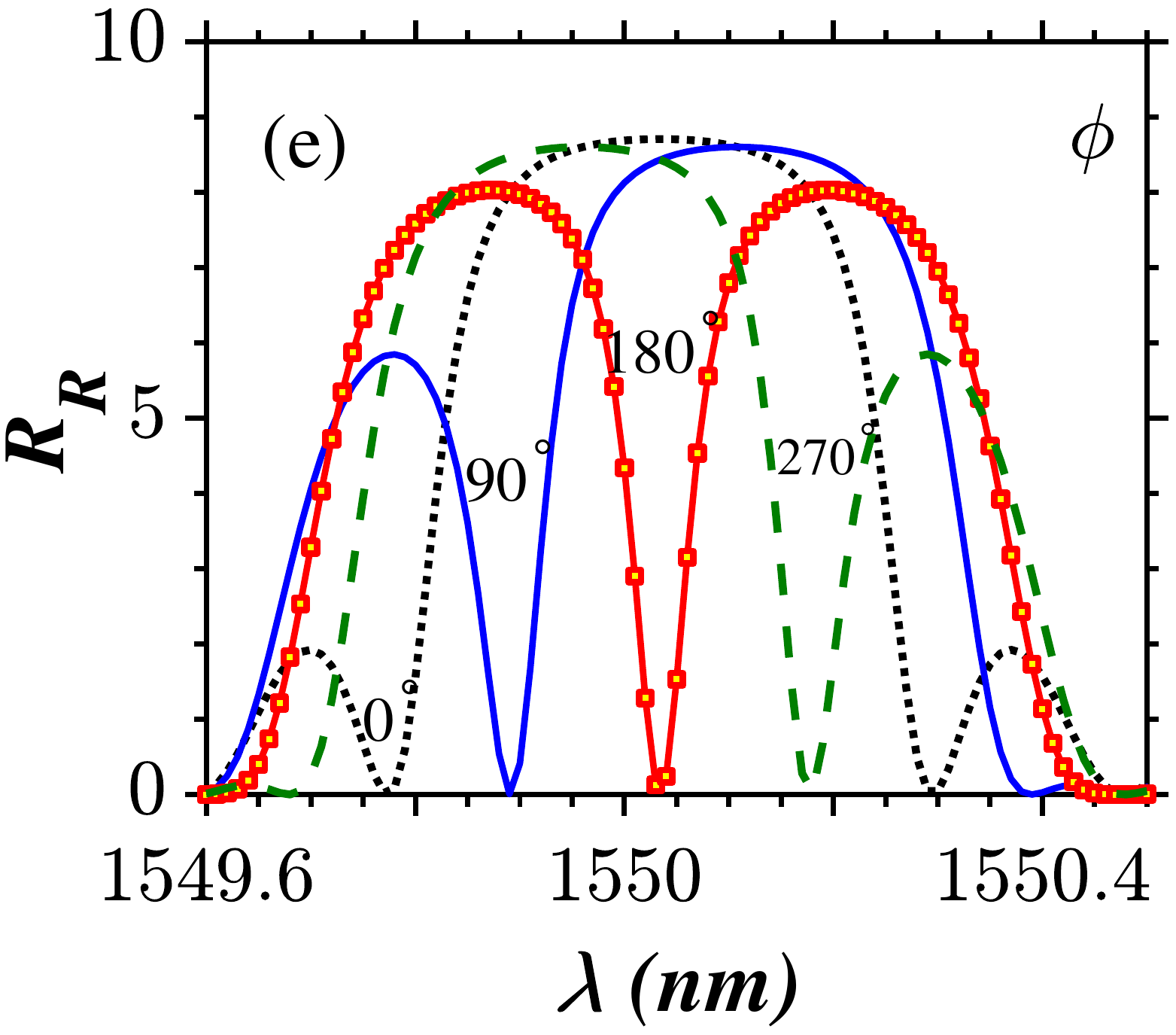}\includegraphics[width=0.5\linewidth]{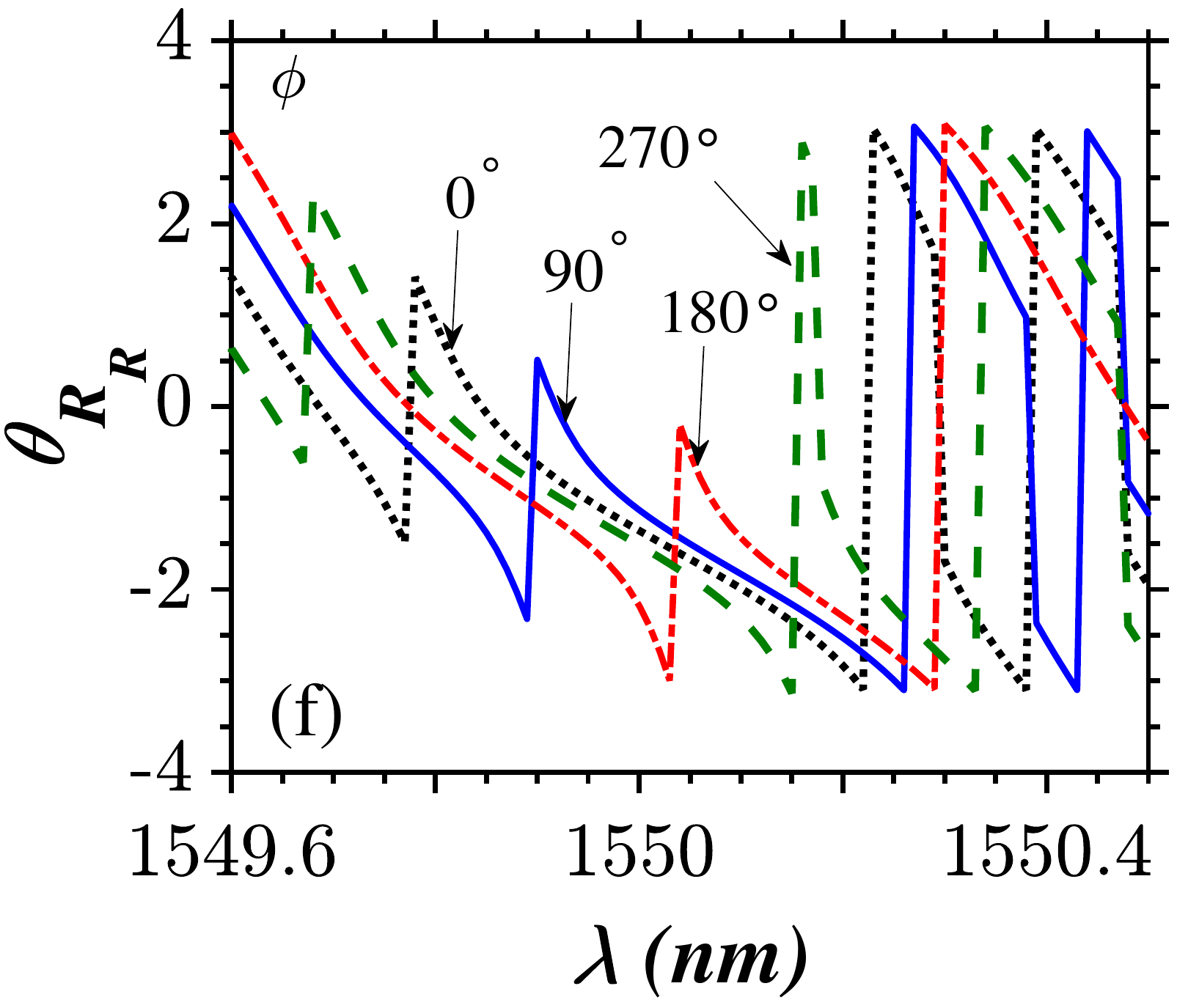}\\\includegraphics[width=0.5\linewidth]{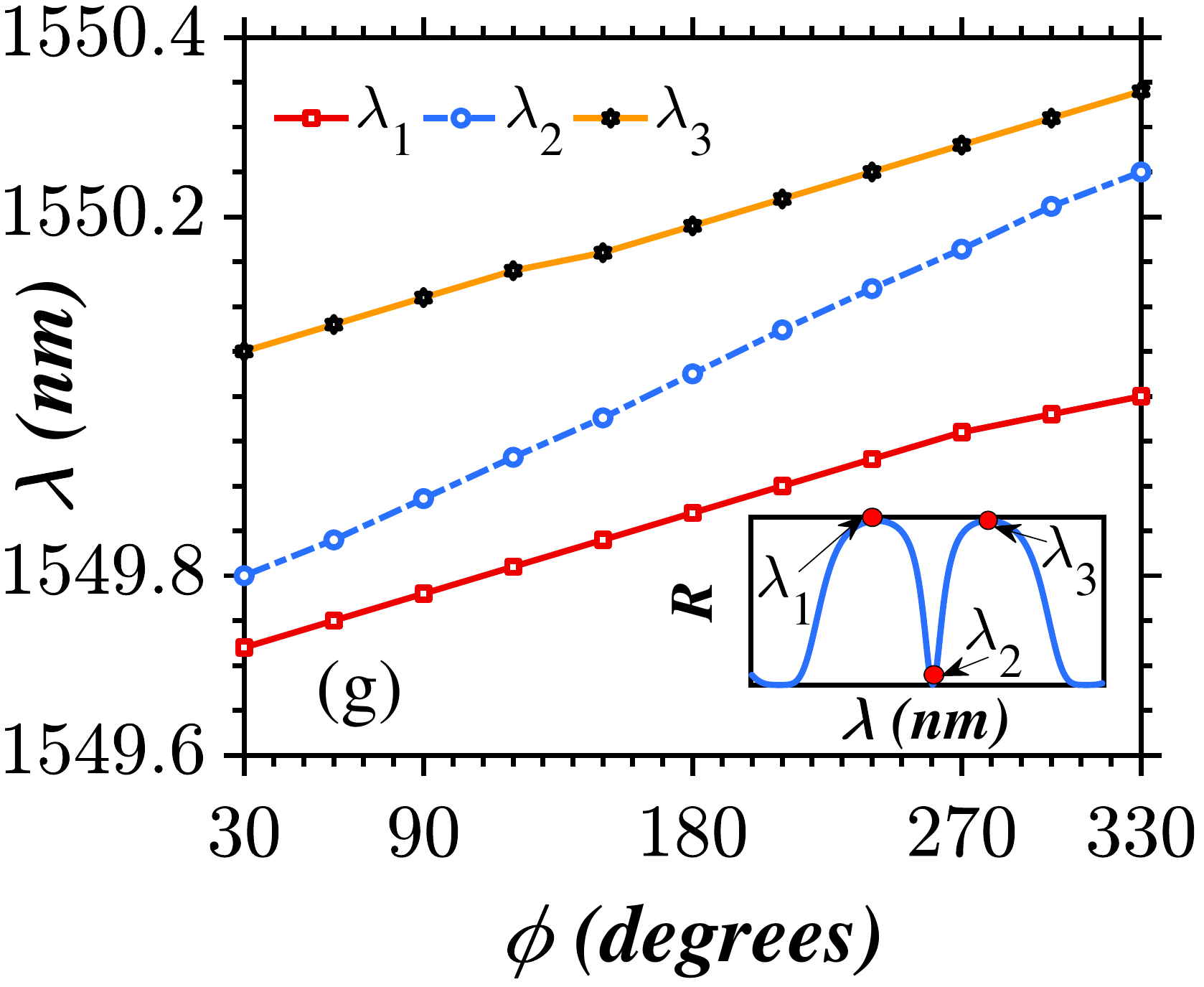}\includegraphics[width=0.5\linewidth]{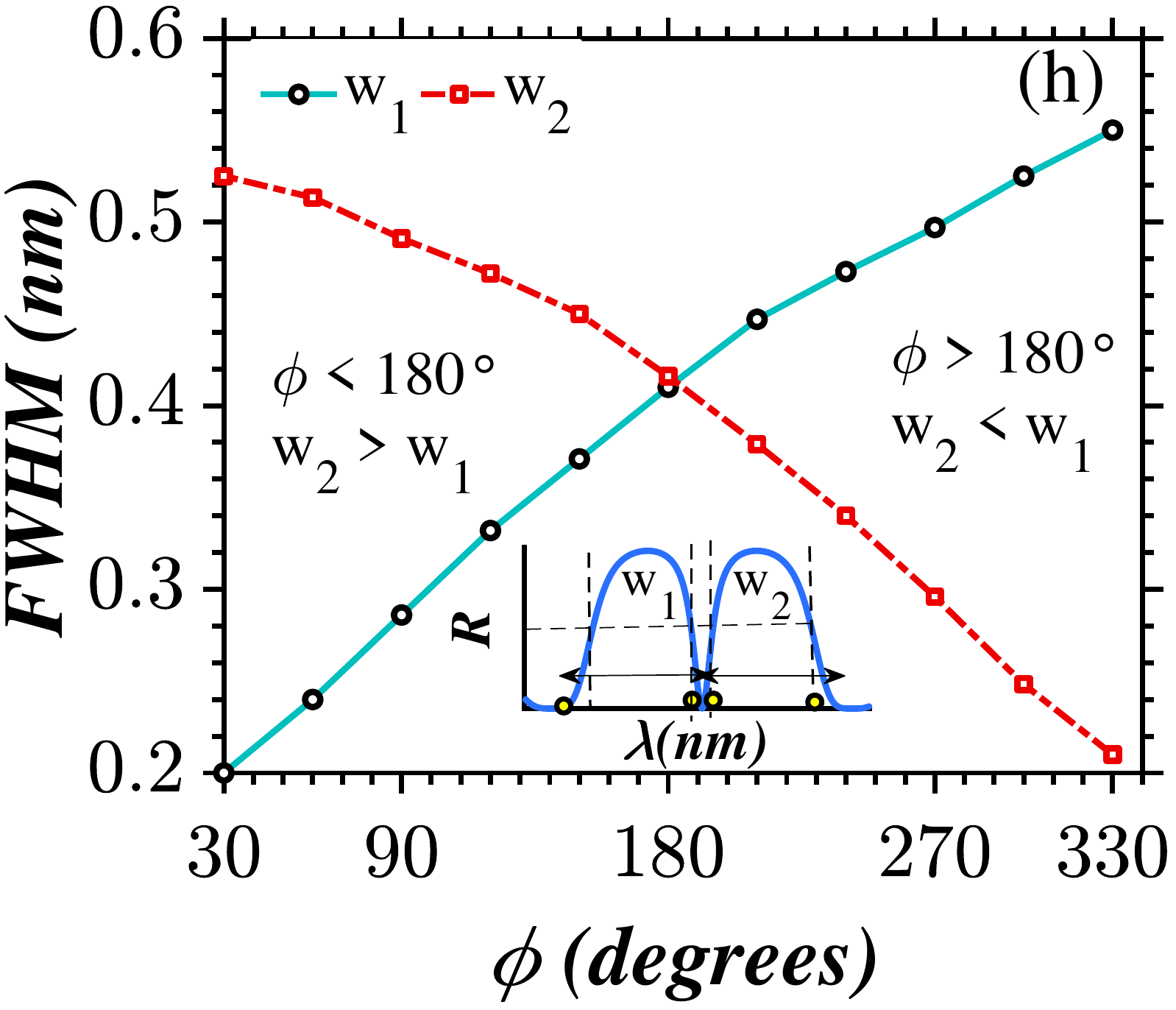}
	\caption{(a) -- (f) Illustrates the reflection and transmission spectra of a unbroken PPTFBG for different values of phase shift  $\phi = 0^\circ$, $90^\circ$, $180^\circ$, and $270^\circ$, respectively. The intensity plots are given in the left panel whereas the right panels correspond to phase plots. (g) Shows the variation in $\lambda_1$, $\lambda_2$ and  $\lambda_3$ with increase in phase ($\phi$) for $R_L$. The variations in the full width half maximum (FWHM) of the spectra  on either sides of $\lambda_2$ ($w_1$ and $w_2$) with respect to change in phase ($\phi$) is shown in (h).}
	\label{fig1}  
\end{figure}

\begin{figure}
	\centering
	\includegraphics[width=0.5\linewidth]{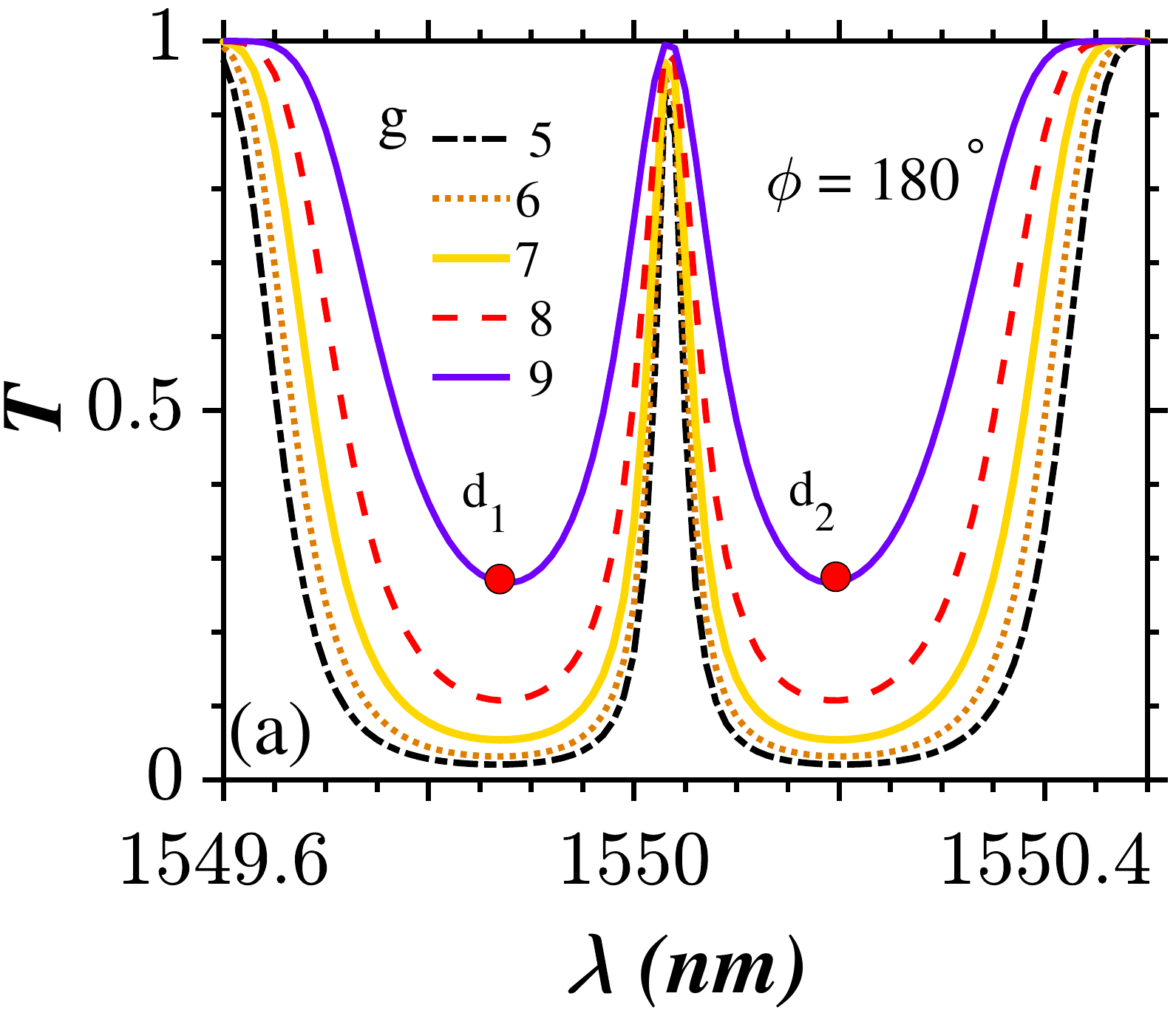}\includegraphics[width=0.5\linewidth]{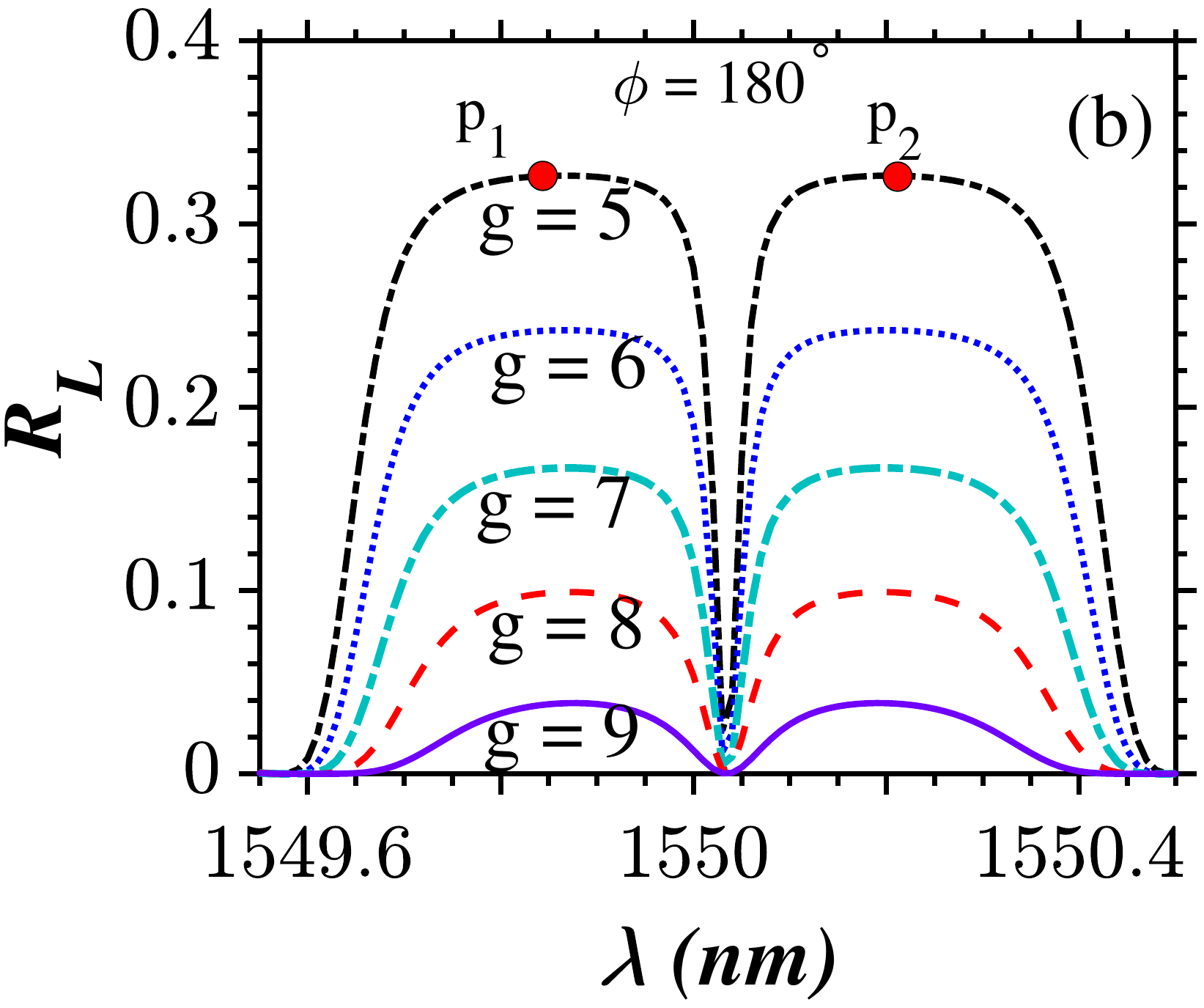}\\\includegraphics[width=0.5\linewidth]{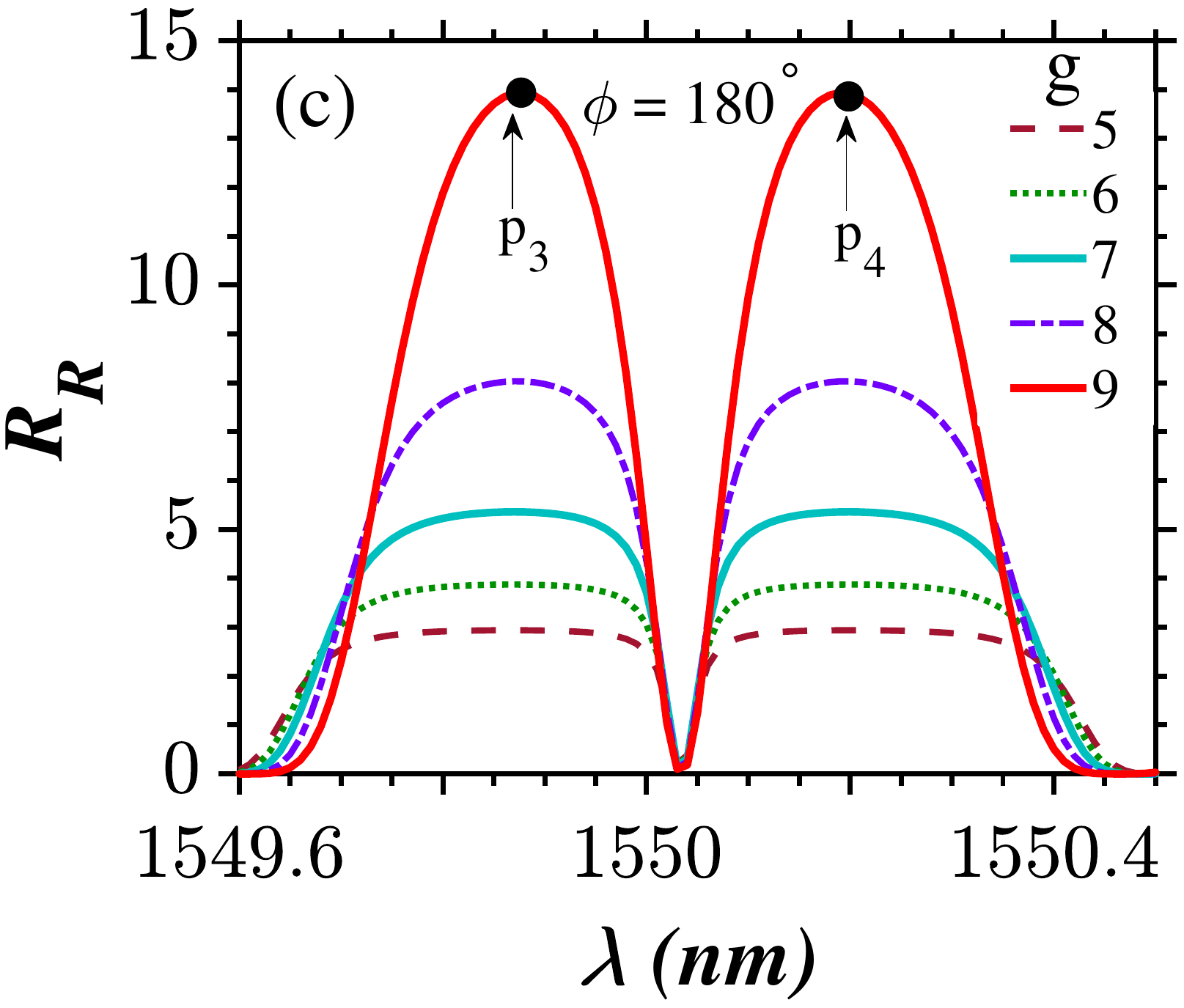}\includegraphics[width=0.5\linewidth]{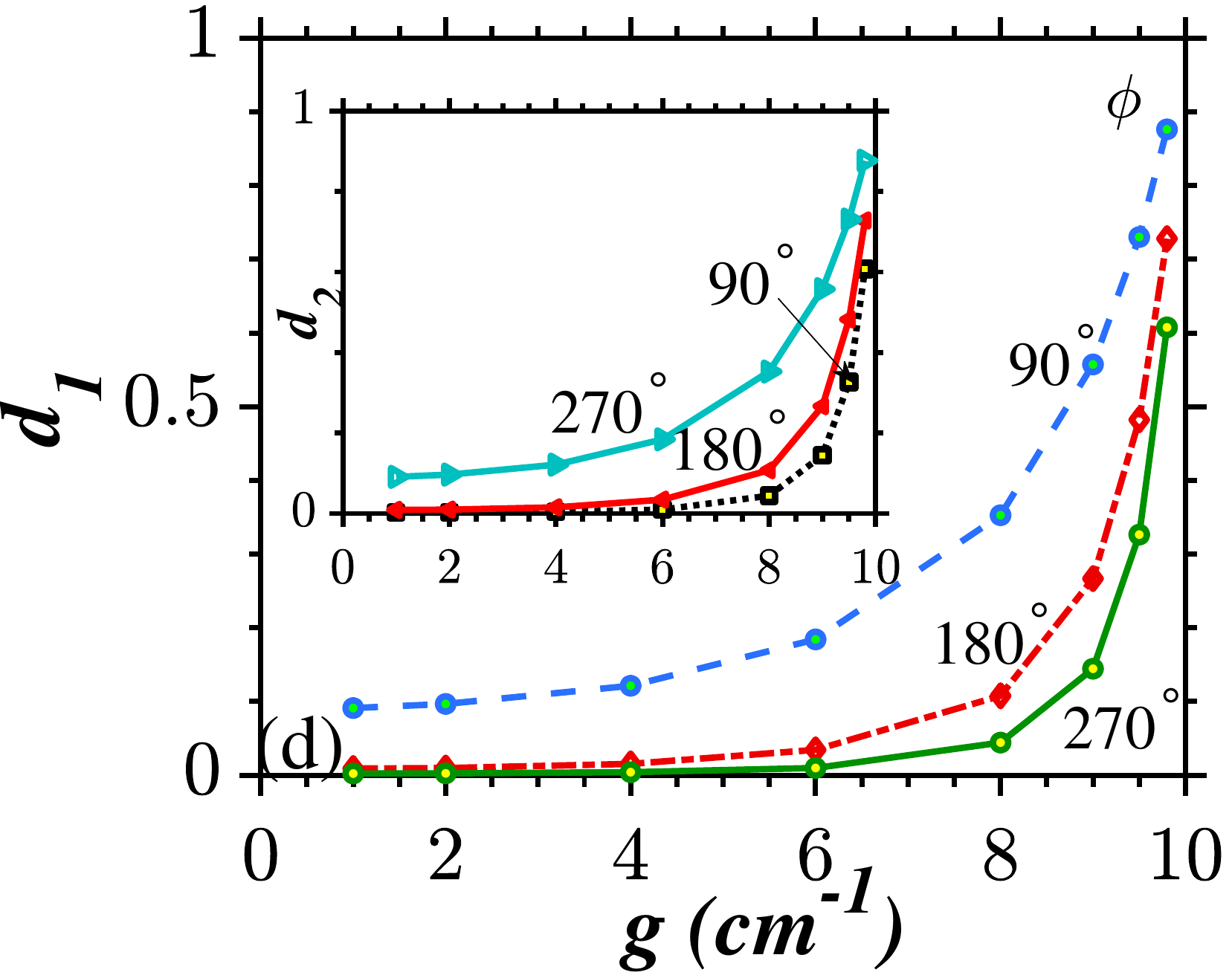}\\\includegraphics[width=0.5\linewidth]{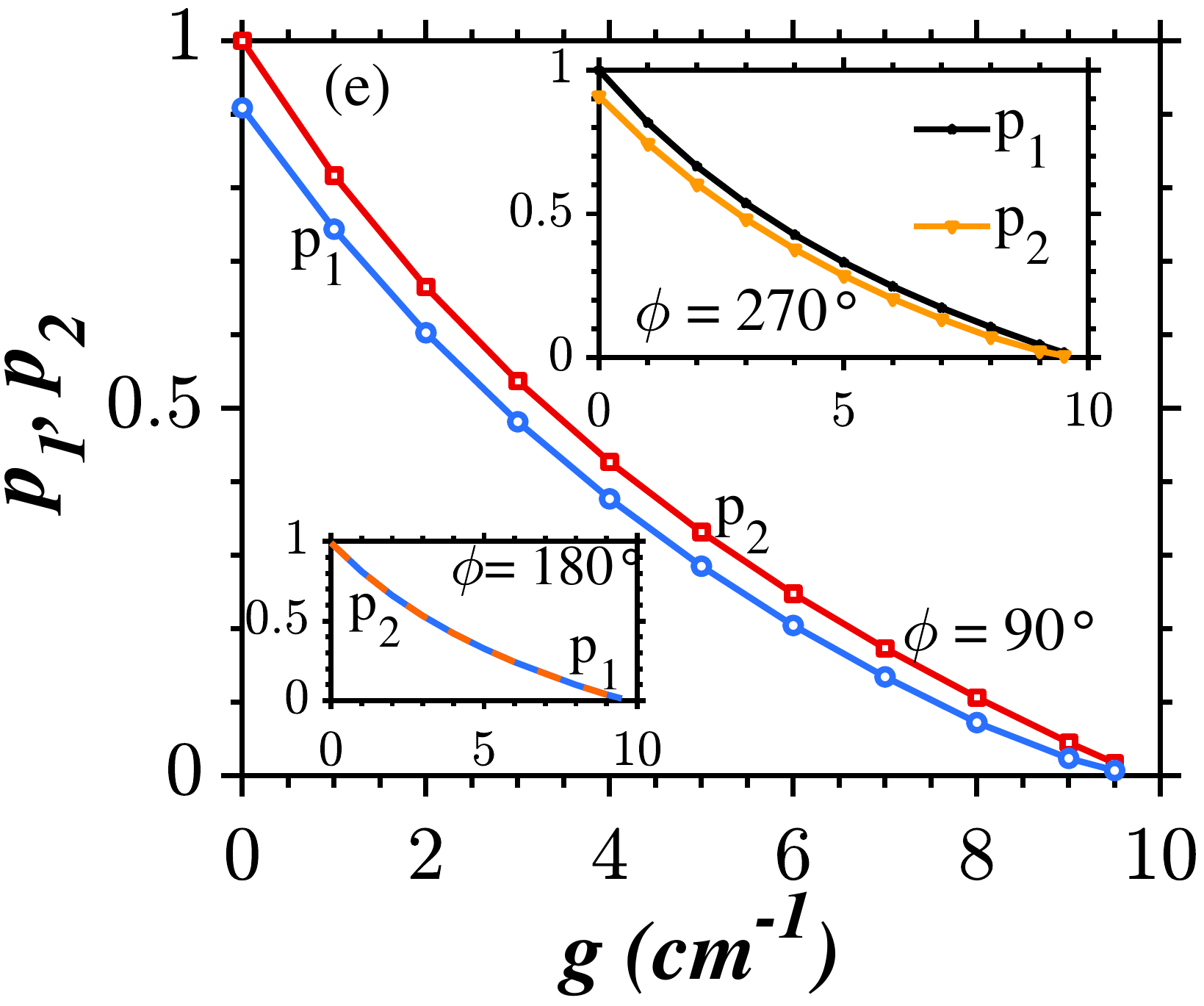}\includegraphics[width=0.5\linewidth]{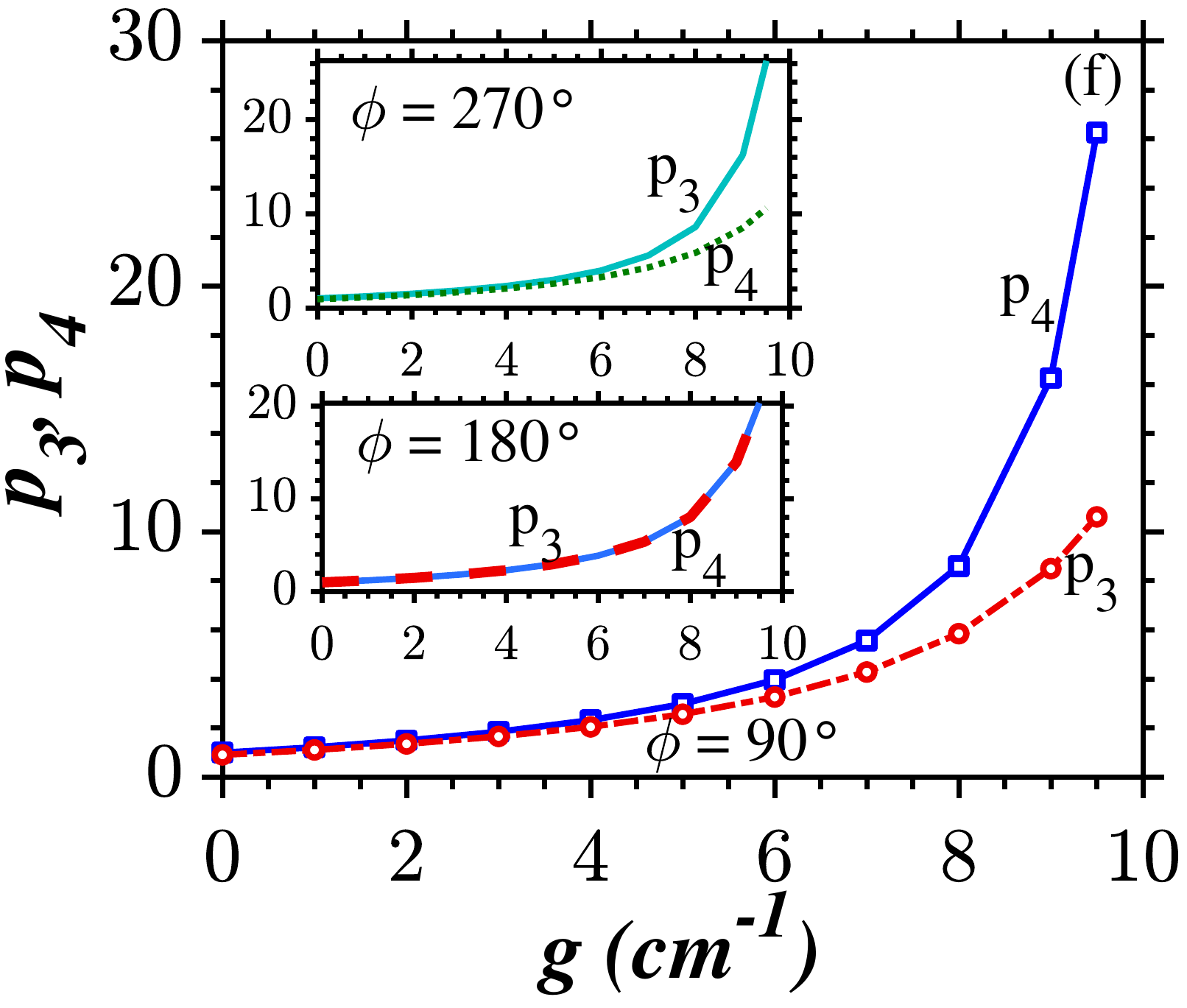}
	\caption{The variation in the transmission and reflection characteristics of a unbroken PPTFBG with respect to variation in gain and loss parameter ($g$) is plotted in (a) -- (c). The continuous variation of minimum of transmitted intensity ($d_1$ and $d_2$) against the  parameter $g$ is plotted in (d). (e) and (f) Show the continuous variation of maximum of reflected intensities ($p_1$, $p_2$, $p_3$ and $p_4$) against variation in gain-loss parameter ($g$).}
	\label{fig2}  
\end{figure}

In this section, the value of gain and loss parameter is varied in the range $0 < g < 10$ cm $^{-1}$ to maintain the unbroken $\mathcal{PT}$-symmetric condition $\kappa>g$. We also recall that a phase shifted FBG behaves like a transmission filter with a very narrow bandwidth. In this section, we consider such a phase-shifted $\mathcal{PT}$-symmetric FBG  with a truncated spectral response between 1549.6 to 1550.4 nm and thus the plots are scaled between these wavelengths as shown in Figs. \ref{fig1} and \ref{fig2}. 

\begin{figure}
	\includegraphics[width=1\linewidth]{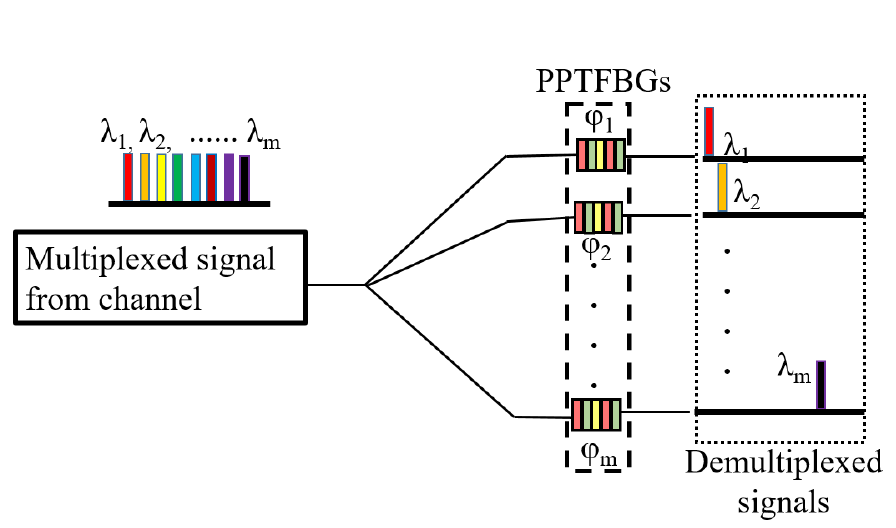}
	\caption{Schematic of all-optical demultiplexer with m-array of PPTFBGs having different phase-shift values ($\phi_1$, $\phi_2$,...$\phi_m$) for selecting m-individual wavelengths ($\lambda_1$, $\lambda_2$,.... $\lambda_m$) from a multiplexed input.}
	\label{fig3_1}
\end{figure}

\begin{figure}[t]
	\centering
	\includegraphics[width=0.5\linewidth]{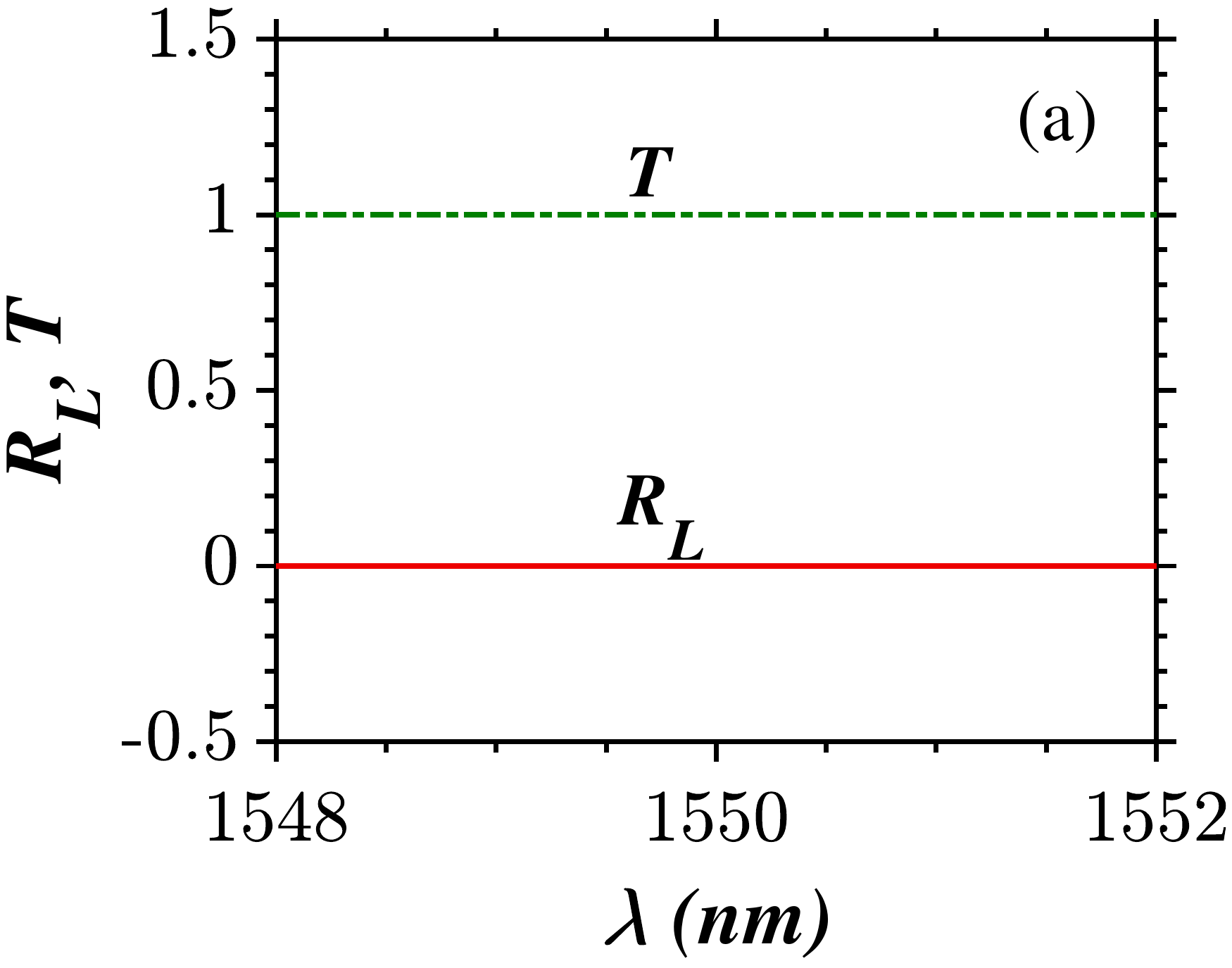}\includegraphics[width=0.5\linewidth]{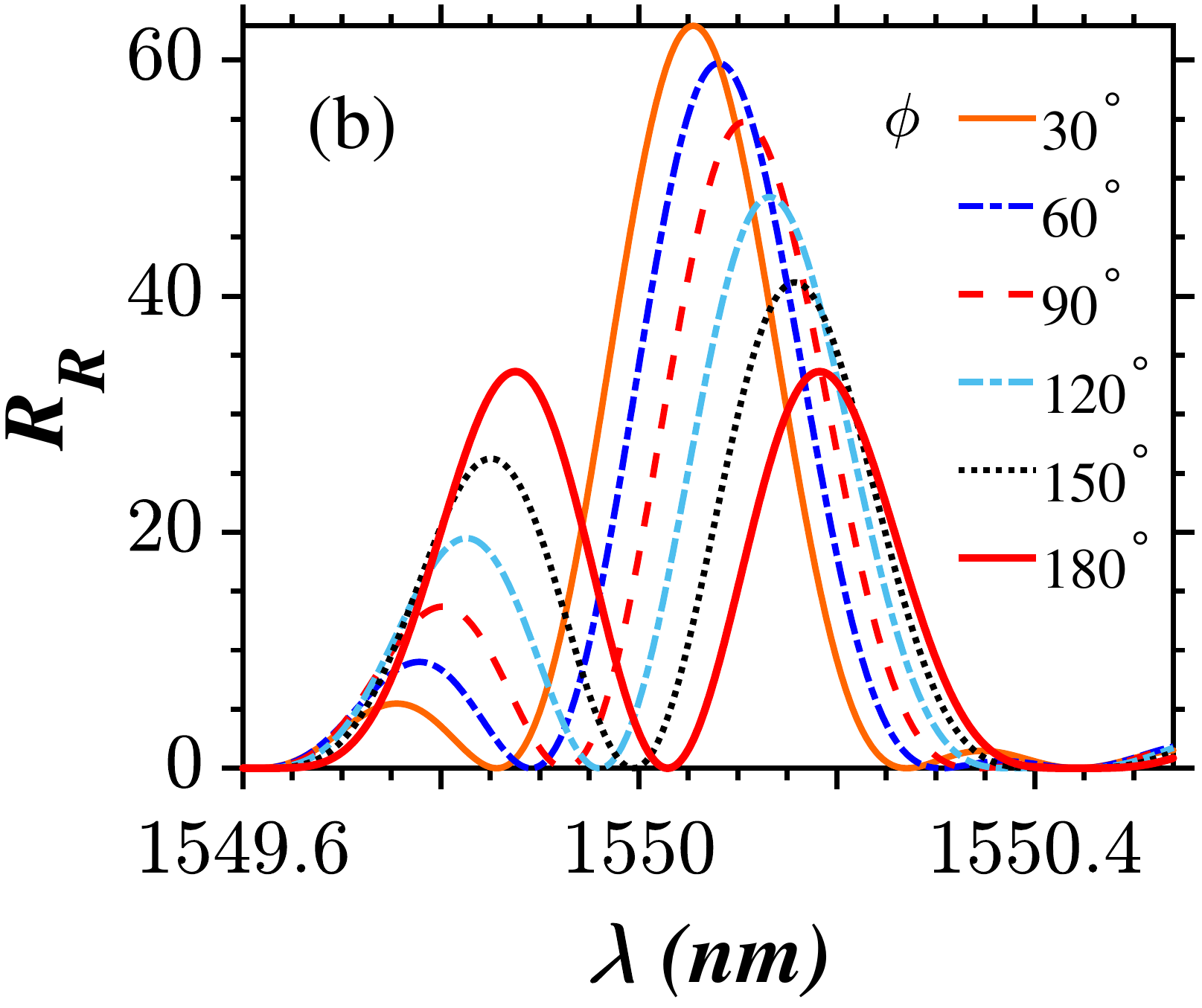}\\\includegraphics[width=0.5\linewidth]{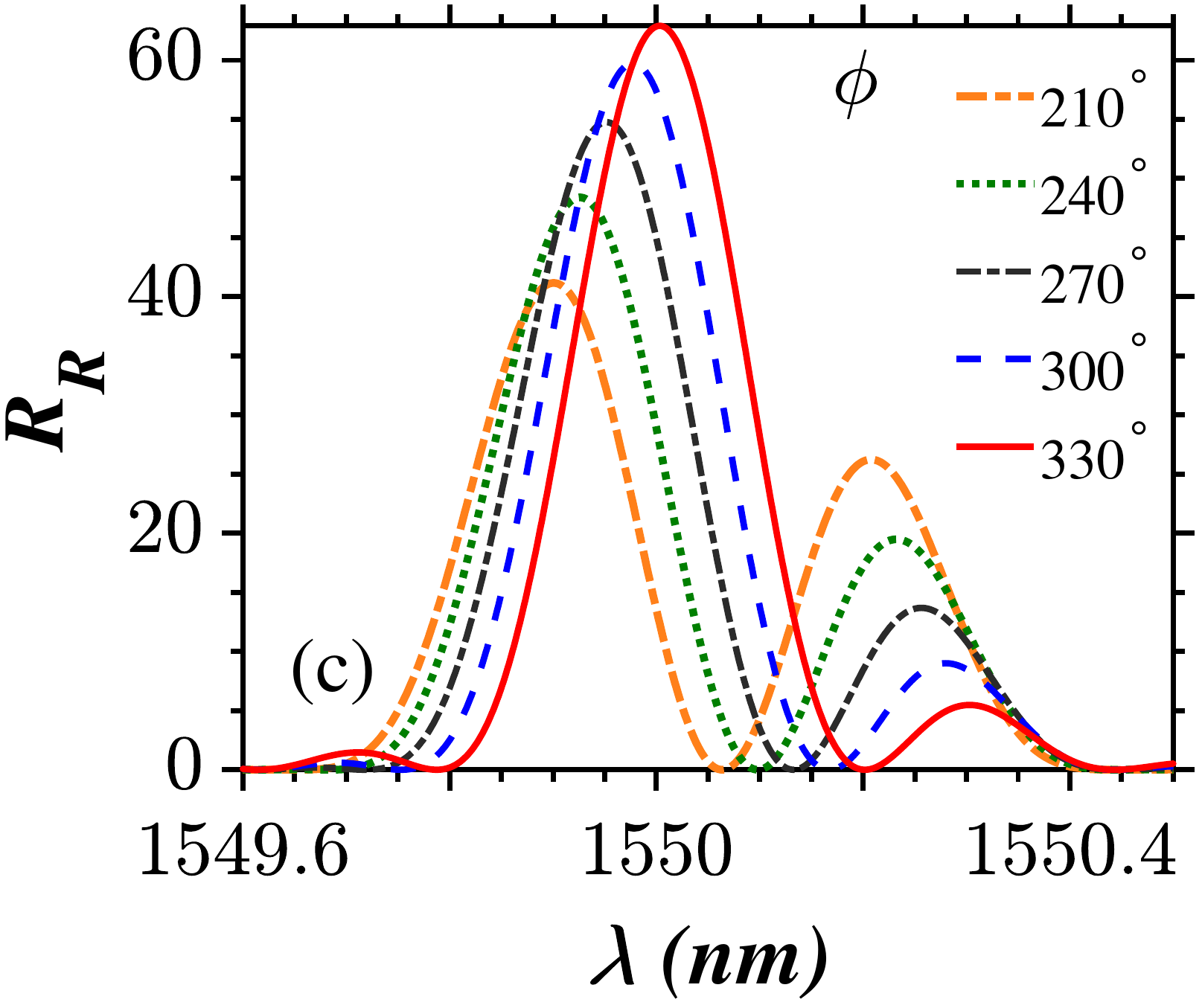}\includegraphics[width=0.5\linewidth]{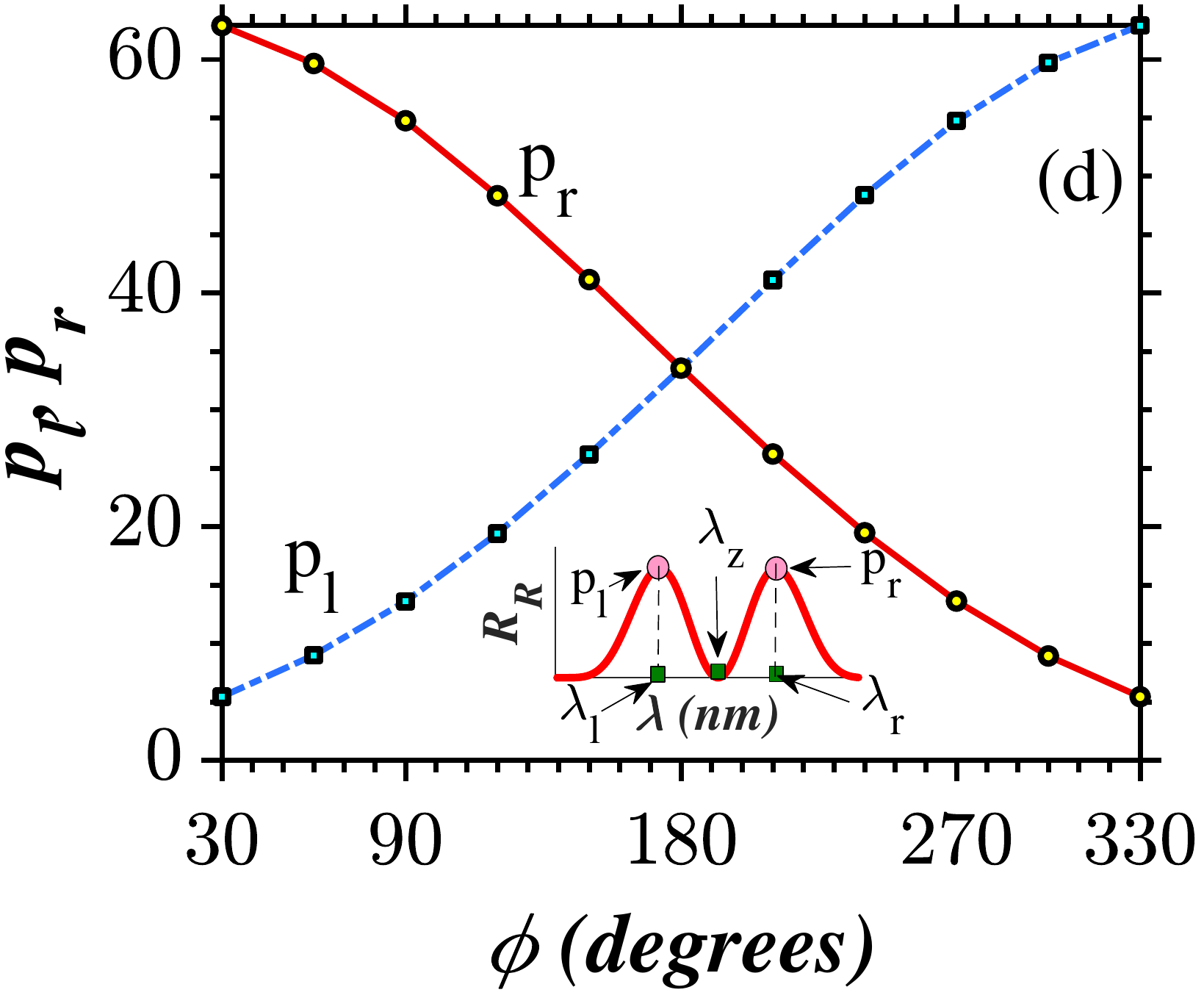}\\\includegraphics[width=0.5\linewidth]{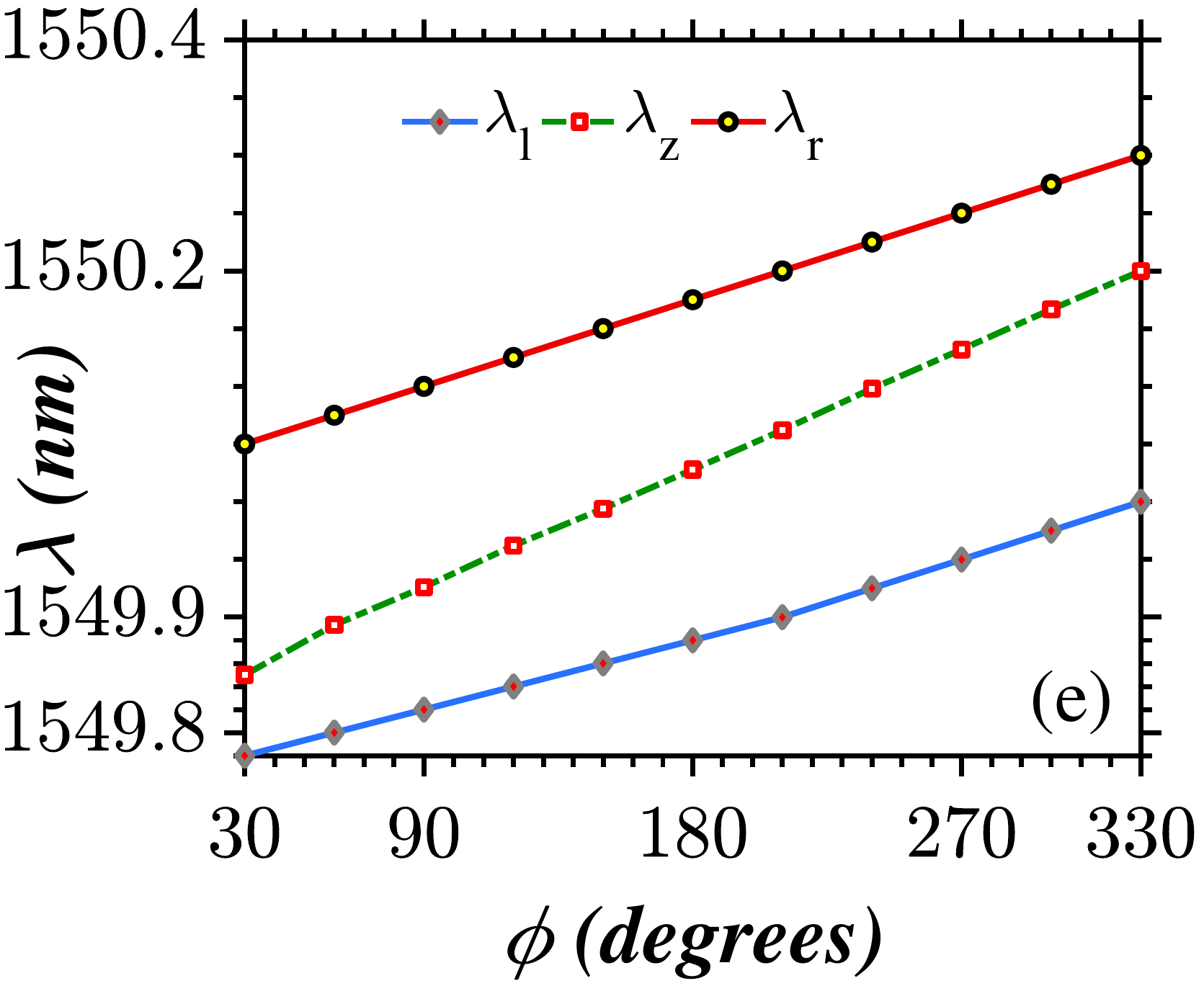}
	\caption{Phase independent unidirectional wave transport in a PPTFBG is shown in (a). (b) and (c) Depict the phase dependent  reflection spectra for the right light incidence at the exceptional point of a PPTFBG. The continuous variation of the reflection peaks ($p_l$ and $p_r$) against variation in phase ($\phi$) is plotted in (d). The wavelengths ($\lambda_l$, $\lambda_r$) corresponding to the peaks ($p_l$, $p_r$) in (d)  and the dip wavelength ($\lambda_z$) are plotted in (e). }
	\label{fig3}
\end{figure}

The magnitude and phase response of an unbroken $\mathcal{PT}$-symmetric FBG ($g = 8$ cm$^{-1}$) is very much similar to a conventional phase shifted FBG ($g = 0$) except that it is possible to alter the transmittivity and reflectivity by tuning the value of gain-loss. In the absence of phase shift ($\phi$ = 0), the device is predominantly reflective rather than transparent for the wavelengths between $1549.7$ - $1550.3$ nm as indicated by the dotted lines in Figs. \ref{fig1}(c) and \ref{fig1}(e). But with the introduction of phase shift in the middle, a narrow band of  wavelengths within the stop band transmits the incoming light as indicated by the dashed ($\phi = 90^\circ$), solid ($\phi = 180^\circ$), dash-dotted lines ($\phi = 270^\circ$) in Fig. \ref{fig1}(a).   At one particular wavelength ($\lambda_2$) (see inset of Fig. \ref{fig1}(g), which is same for $T$, and $R_R$ also), the transmittivity reaches maximum (unity) inside the stopband of the grating. At this wavelength,  we can observe a peak in the transmission spectra as seen in Fig. \ref{fig1}(a) and dip in the corresponding reflection spectra as shown in Figs. \ref{fig1} (c) and \ref{fig1}(e). The response of the grating is asymmetric for all the values of phase except for $\phi = 180^\circ$ as referred from these plots. For $\phi = 180^\circ$, the narrow transmission window occurs exactly in the middle of the spectra and on either side of the transmission peak, the spectrum is symmetrical. The wavelength at which maximum transmission within the stopband occurs ($\lambda_2$) is shifted towards longer wavelengths by increasing the value of phase. Moreover, the wavelengths at which peak reflection occurs on either side of $\lambda_2$ ($\lambda_1$, $\lambda_3$) are also shifted towards higher wavelengths as shown in Fig. \ref{fig1}(g). The magnitude plots obtained for phase value of $\phi = 270^\circ$ is exactly the mirror image of the spectra obtained for a phase of $\phi = 90^\circ$ as shown in Figs. \ref{fig1}(a),  \ref{fig1}(c) and  \ref{fig1}(e). 

To illustrate the spectral response further, we define  another useful parameter, namely \textit{Full width at half maximum} (FWHM). It indicates the difference between the two wavelengths corresponding to $R = 1/2$  $R_{max}$. 
For $\phi < 180^\circ$, the FWHM of the reflection spectra on the longer wavelength ($w_2$) side of $\lambda_2$ is broader than the one ($w_1$)  on the shorter wavelength side of $\lambda_2$ as shown in Fig. \ref{fig1}(h). At, $\phi = 180^\circ$, these FWHMs  are equal ($w_1 = w_2$). On the other hand, when $\phi > 180^\circ$, the FWHM of the spectra on the right side of $\lambda_2$ ($w_2$) is narrow compared to the one on the left ($w_1$) of $\lambda_2$. A well known application of phase shifted FBG is that it can be used as channel selector (demultiplexer). To implement such a scheme, a multiplexed signal emerging from a transport fiber is then passed into the array of FBGs (m-FBGs for m-channels) with each of the FBGs having a particular value of phase shift ($\phi$) for selecting a particular channel. This is facilitated by the inherent property of phase shifted FBGs to allow a shift in $\lambda_2$ and the wavelength of the peaks ($\lambda_1$ and $\lambda_3$) on either side of $\lambda_2$. An unbroken PPTFBG provides an additional degree of freedom to tune the intensity of these peaks as illustrated in Fig. \ref{fig2}. From Figs. \ref{fig1}(b), \ref{fig1}(d), and \ref{fig1}(f), we find that the phase ($\theta_T, \theta_{R_L}$, and $\theta_{R_R}$) responses of the system rely on the the gain-loss parameter ($g$) and the magnitude of the phase shift ($\phi$) in the middle. Physically, this would mean that the phase of the reflected and transmitted signals is strongly influenced by the changes in the energy of the wave packet as a consequence of variation in gain-loss potential \cite{lin2011unidirectional} and the value of phase shift.

Like any other $\mathcal{PT}$-symmetric FBG configuration, our system also demonstrates an increase (decrease) in the reflectivity for right (left) light incidence with an increase in the value of gain-loss parameter ($g$) as illustrated in Figs. \ref{fig2}(e) and \ref{fig2}(f). For $\phi < 90^\circ$, the magnitude of the reflection spectra on the longer wavelength side of $\lambda_2$ is larger than the peak of reflection on the shorter wavelength side and this is true for both left ($p_2$ $>$ $p_1$) and right ($p_4$ $>$ $p_3$) light incidences. When $\phi = 180^\circ$, the magnitudes of these peaks are equal ($p_1 = p_2$ and $p_3 = p_4$). However, when $\phi > 180^\circ$, the opposite effect occurs ($p_1$ $>$ $p_2$ and $p_3$ $>$ $p_4$). The increase in the value of gain-loss also contributes to the increase in the magnitude of the dip ($d_1$ and $d_2$) with increase in gain-loss in the transmission spectra as portrayed by Fig. \ref{fig2}(d). Thus, it is possible to construct all-optical demultiplexers with tunable intensity and spectral width with the proposed $\mathcal{PT}$-symmetric system as shown in Fig. \ref{fig3_1}.

\subsection{Unidirectional wave transport at the exceptional point}
\label{Sec:4}  

Mathematically, if a $\mathcal{PT}$-symmetric device satisfies the condition $\kappa = g$, then it is known to be working at the exceptional point. It is reported in the literature that different $\mathcal{PT}$-symmetric FBGs, namely uniform \cite{lin2011unidirectional}, apodized \cite{lupu2016tailoring}, chirped and apodized FBGs \cite{raja2020tailoring} exhibit unidirectional wave transport at the exact $\mathcal{PT}$-symmetric phase. From the numerical study presented here, we also confirm that this phenomenon persists even in the presence of phase shift in the middle of the grating. The incident light travels inside the device and emanates at the other end of the grating for the left light incidence as if there is no grating present in the propagating path to reflect the signal ($T = 1$ and $R = 0$) as shown in Fig. \ref{fig3}(a). However, the right incident light shows changes with respect to variations in the value of phase-shift ($\phi$) as depicted in the remaining plots from Figs. \ref{fig3}(b) -- \ref{fig3}(e). The wavelength corresponding to zero reflectivity within the stop band at the exceptional point is designated as $\lambda_z$. On either side of this wavelength, the reflected spectra exhibit peaks ($p_l$ and $p_r$) which are very similar to the dynamics obtained in the unbroken $\mathcal{PT}$-symmetric regime, except that the magnitude of these peaks are larger as the system operates at a higher value of gain-loss ($g = 10$ cm$^{-1}$). For example, $R_{R_{max}}$ measures a magnitude of $54.72$ at the exceptional point, whereas in the unbroken regime ($g = 9.5$ cm $^{-1}$) it is measured to be $26.25$, for a phase value of $\phi = 90^\circ$. Unlike the reflection dynamics for the right light incidence reported in other PTFBGs structures at the exceptional point \cite{lupu2016tailoring, huang2014type, lin2011unidirectional}, these structures are not completely reflective to all wavelengths within the stopband, thanks to the presence of phase shift in the middle of the grating. To illustrate this point, for a phase of $\phi = 180^\circ$ and $\lambda = 1550.04$ nm, $R_R$ is measured to be zero. We designate the wavelength corresponding to the zero reflectivity dip within the stopband as $\lambda_z$, at the exact $\mathcal{PT}$-symmetric phase. The wavelengths corresponding to the peaks $p_l$ and $p_r$ are given by $\lambda_l$ and $\lambda_r$. All these wavelengths show shift towards longer wavelengths of the spectra with increase in the value of $\phi$ as shown in Fig. \ref{fig3}(e).  

\subsection{Broken $\mathcal{PT}$-symmetric regime}
\label{Sec:5}

\begin{figure}[t]
 	\centering
 	\includegraphics[width=0.5\linewidth]{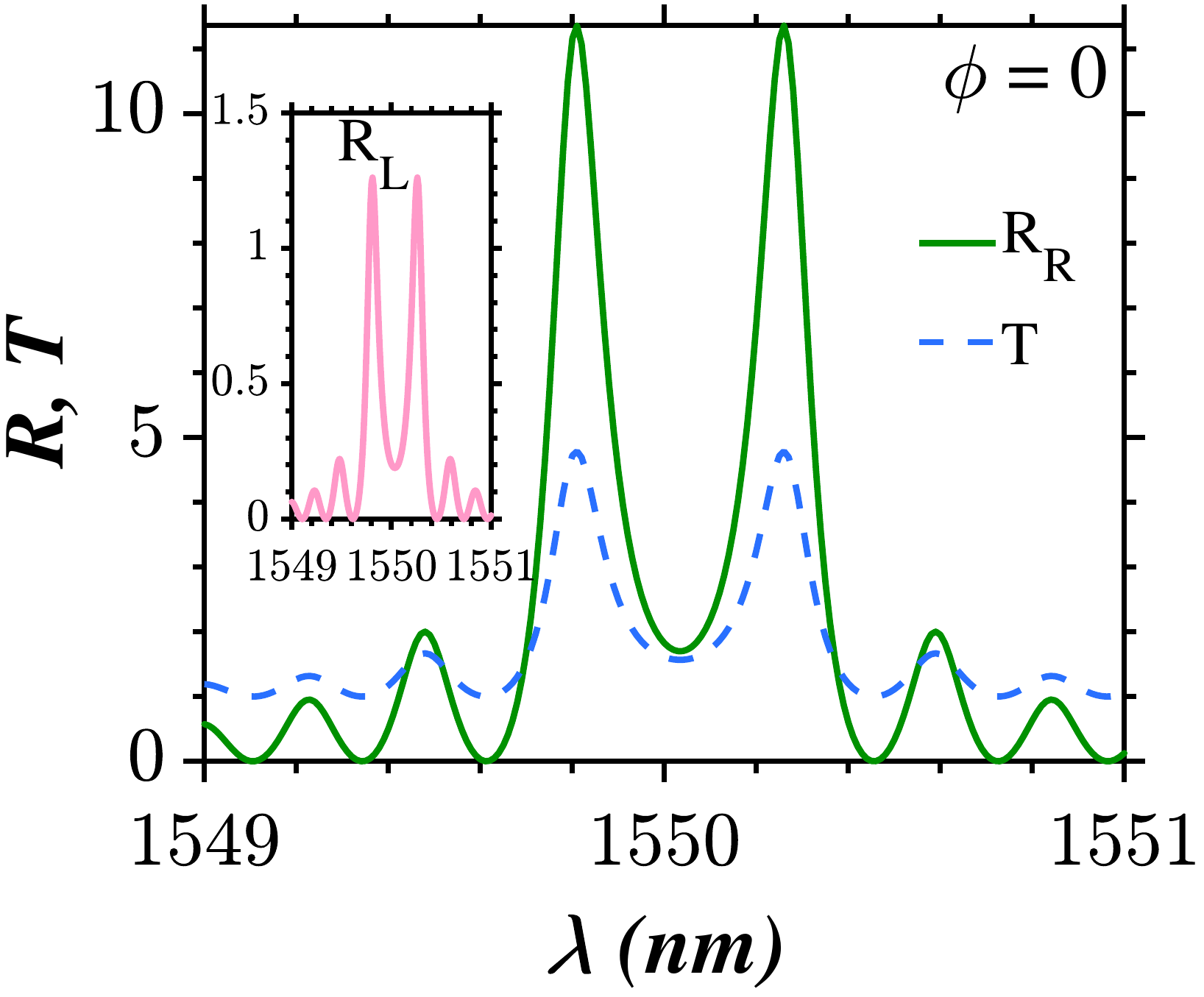}
 	\caption{Spectral characteristics of a broken $\mathcal{PT}$-symmetric FBG in the absence of phase shift ($\phi = 0^\circ$)}
 	\label{fig4} 
 \end{figure}

 As well known, a broken $\mathcal{PT}$-symmetric FBG needs to obey the mathematical condition $g>\kappa$ and hence the gain-loss parameter value is kept at $g = 20$ $cm^{-1}$ throughout this section and the tuning is achieved by varying the phase ($\phi$).  Before investigating the phase shifted system, it is important to look at the spectral response of the system in the absence of phase shift ($\phi = 0^\circ$). Instead of amplification at the center wavelength, the maximum amplification occurs on both longer and shorter wavelength sides of the Bragg wavelength and the amplification at $\lambda_b$ is suppressed but not to zero. Also, amplification at the side lobes are visible 
 in Fig. \ref{fig4}.
 \subsubsection{\textbf{Single mode lasing behavior}}
 \begin{figure}
 	
 	\centering
 	\includegraphics[width=0.5\linewidth]{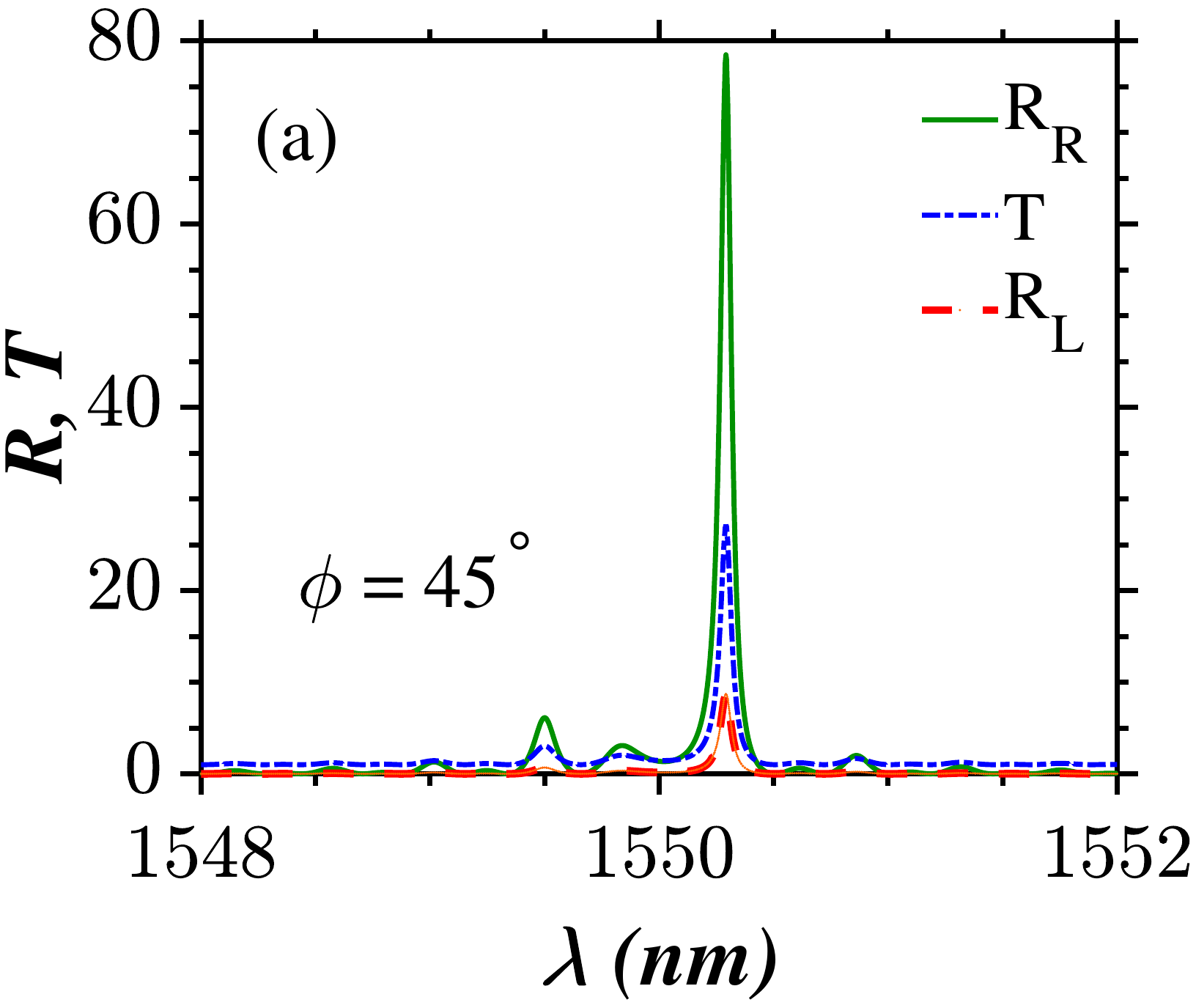}\includegraphics[width=0.5\linewidth]{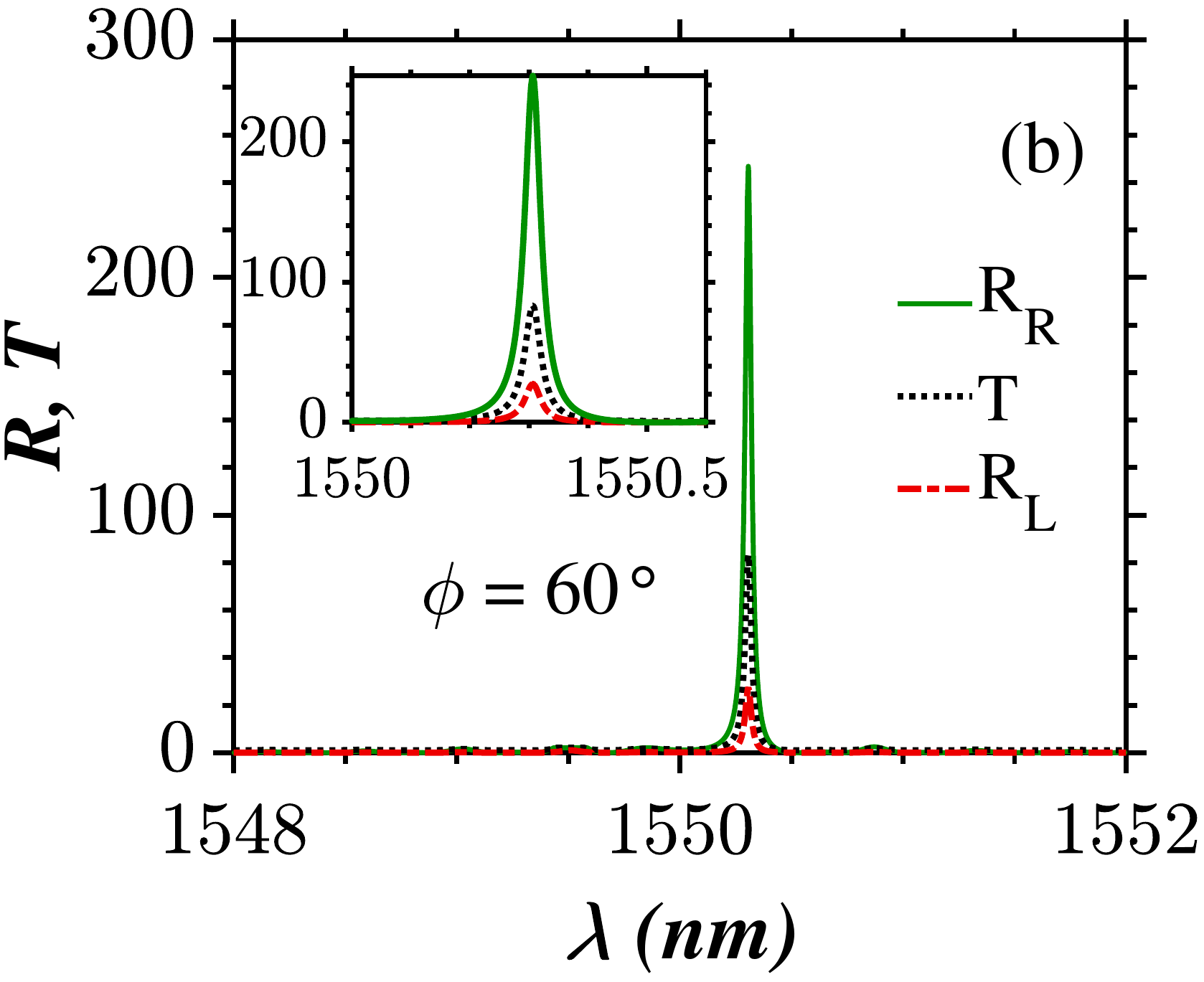}\\\includegraphics[width=0.5\linewidth]{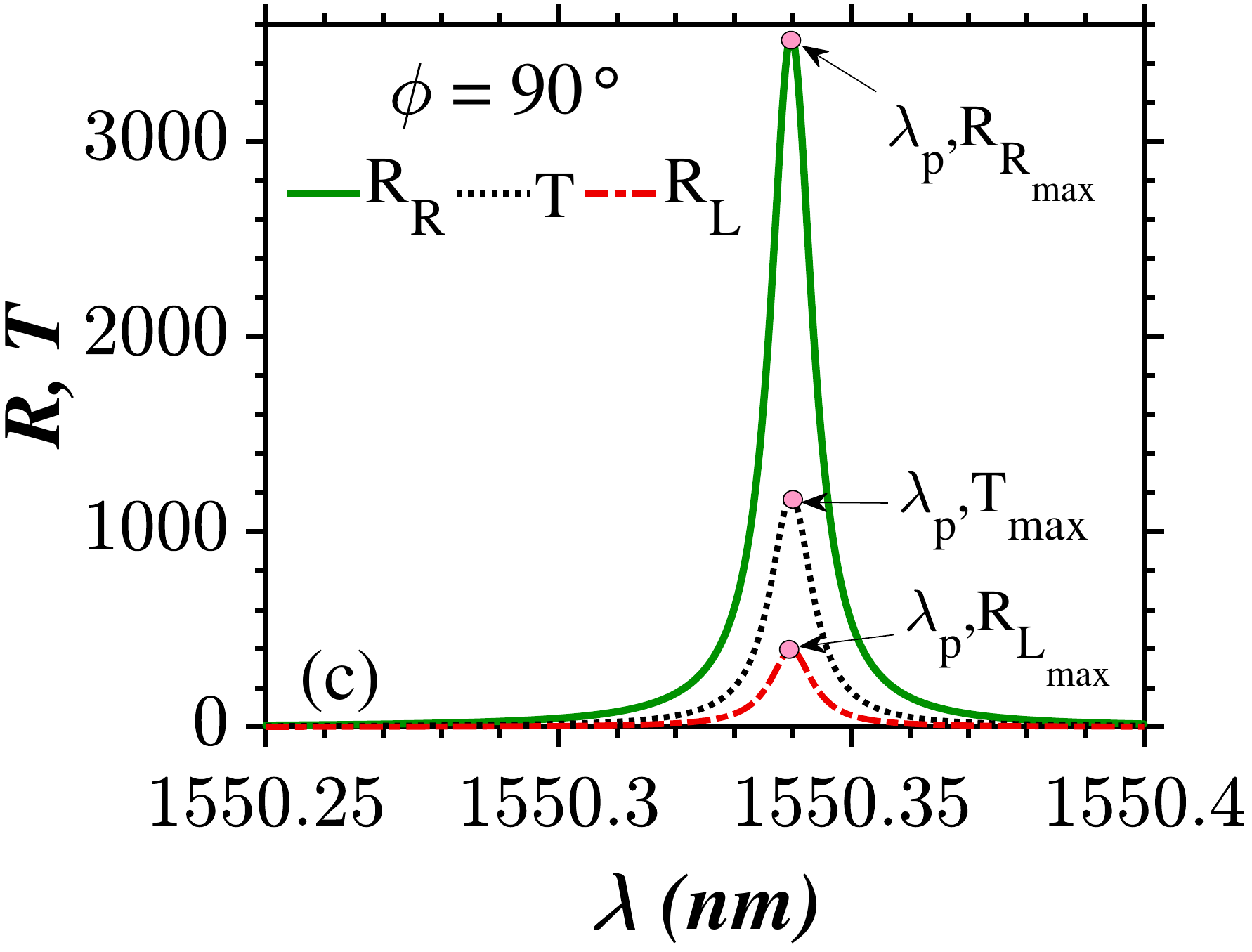}\includegraphics[width=0.5\linewidth]{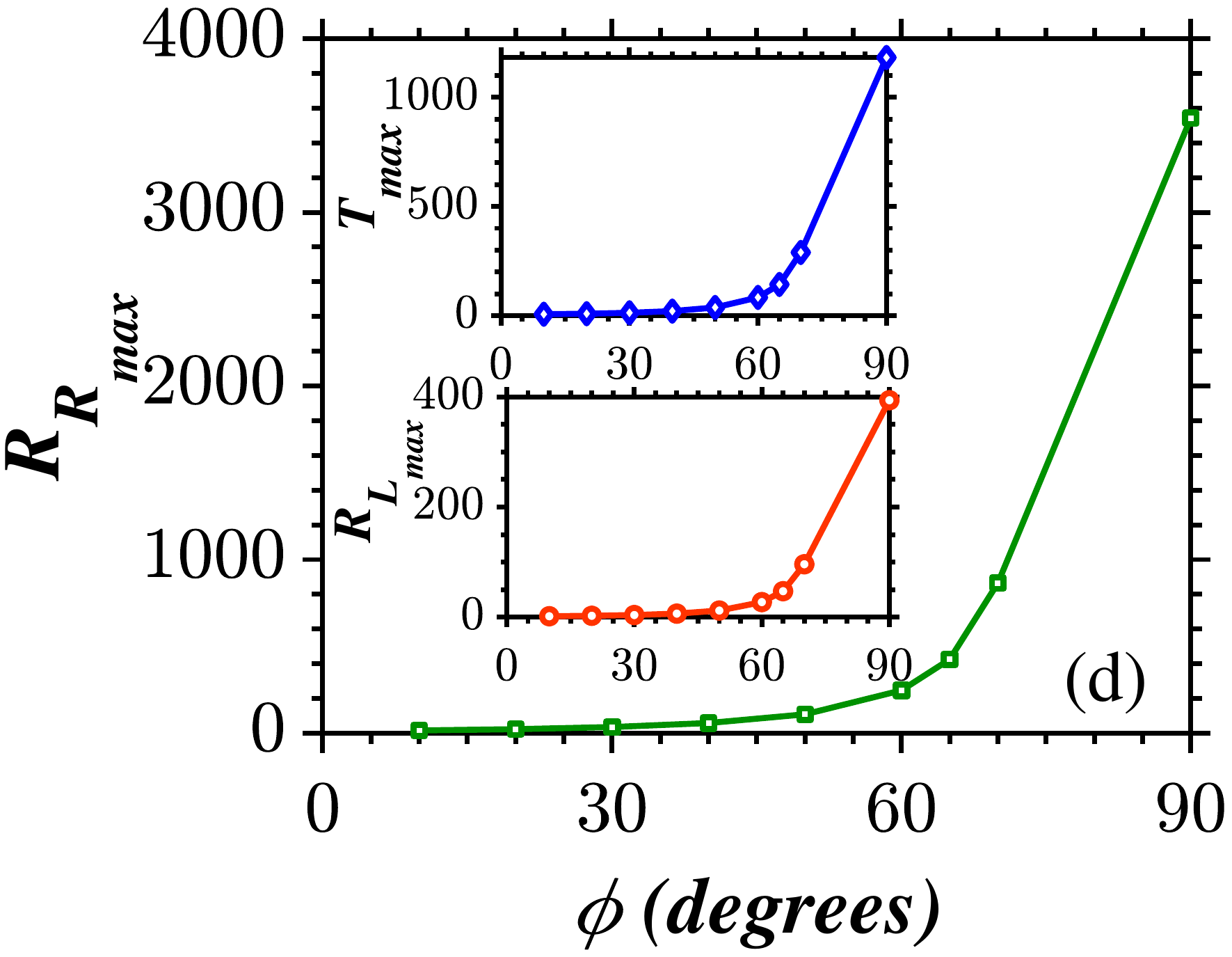}\\\includegraphics[width=0.5\linewidth]{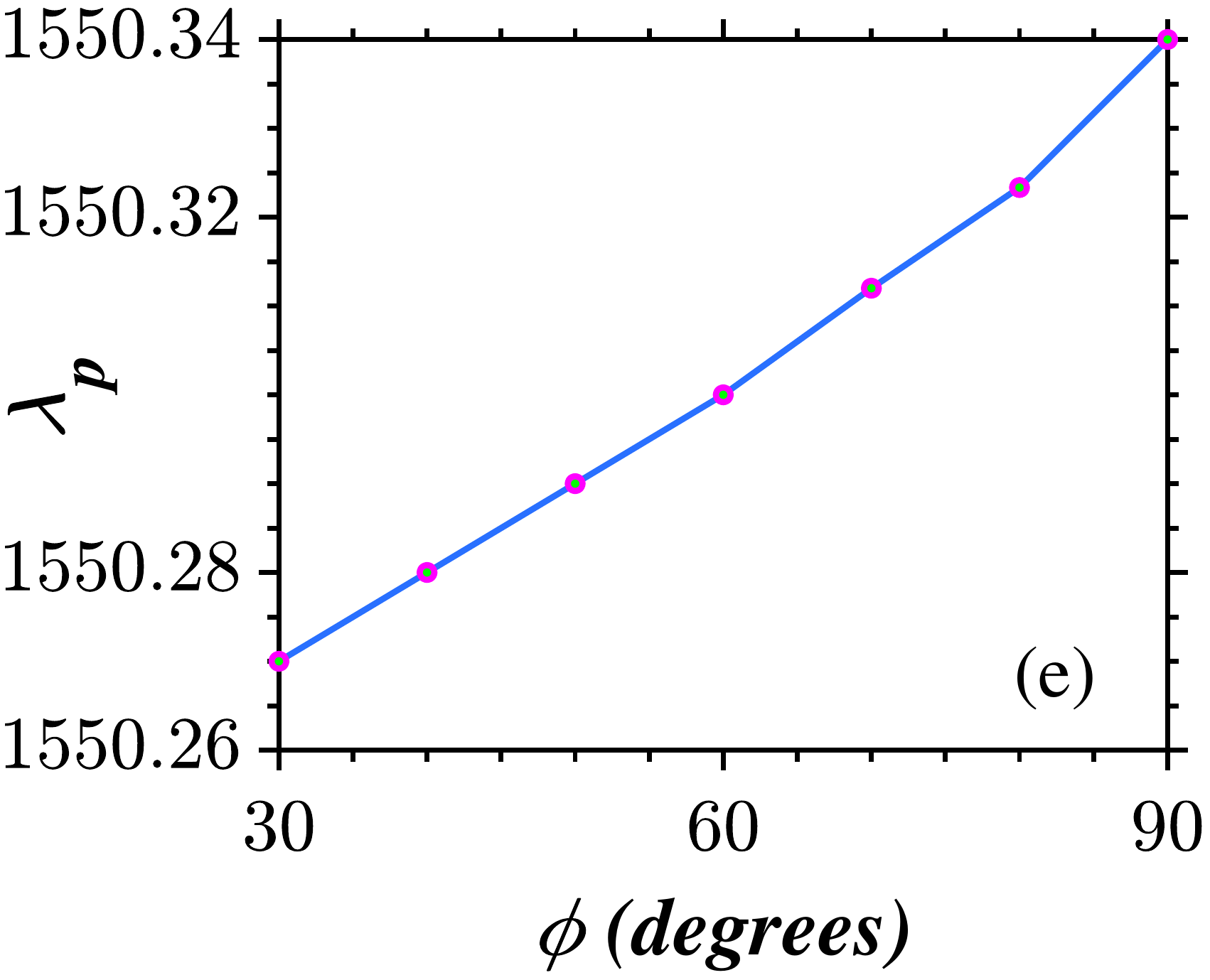}\includegraphics[width=0.5\linewidth]{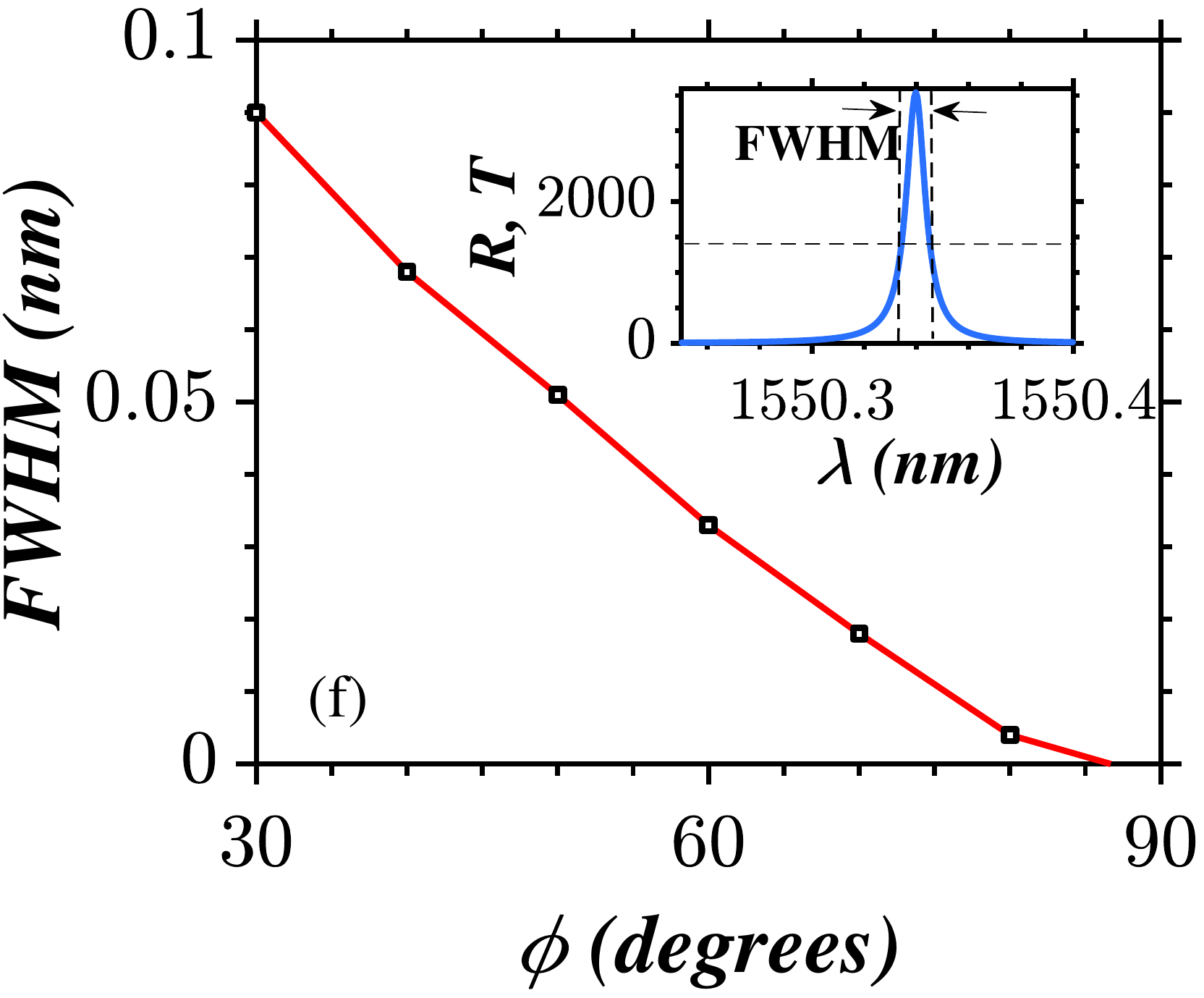}
 	\caption{Phase controlled single mode lasing behavior of a broken PTFBG against the variation in phase shift at $g = 20$ for different phase values $\phi = 45^\circ$, $60^\circ$ and $90^\circ$ is shown in  (a) -- (c), respectively. The continuous variation of maximum of transmitted ($T_{max}$) and reflected intensities ($R_{max}$) with respect to change in the values of $\phi$ is shown in (d) and the wavelengths ($\lambda_2$) corresponding to the peaks are shown in (e). The change in the FWHM of the transmitted and reflected spectra against the variation in $\phi$ is plotted in (f).}
 		\label{fig5}
 \end{figure}
When $\phi$ is increased to $45^\circ$, the reflections in the side lobes which were existing in the absence of phase shift ($\phi = 0^\circ$) are inhibited to a large extent as shown in Fig. \ref{fig5}(a). But some weak reflections still persist in the spectra corresponding to the right incidence but it is comparably very less to the maximum intensity ($R_{R_{max}}$). Even these weak reflections are  suppressed further when $\phi$ is increased ($\phi = 60^\circ$) and thus we obtain a pure single mode lasing behavior  as shown in Fig. \ref{fig5}(b). The maximum transmittivity (reflectivity) of single mode lasing spectra is observed for $\phi = 90^\circ$ as shown in Figs. \ref{fig5}(c) and \ref{fig5}(d). It is important to mention that the side lobes are significantly reduced by tuning the value of phase shift without a need for an apodization technique. Tuning the value of phase also has a drastic effect on the wavelength ($\lambda_p$) corresponding to the peaks of the lasing spectra. From Fig. \ref{fig5}(e), we confirm that $\lambda_p$ is shifted to longer wavelengths of the spectra with an increase in the value of $\phi$.  The reflectivity and transmittivity get increased proportionally with an increase in $\phi$, whereas the FWHM shows an inverse relationship with the increase in $\phi$ as shown in Fig. \ref{fig5}(f). This means that higher the peak reflectivity ($R_{max}$) and transmittivity ($T_{max}$), narrower the FWHM of the lasing spectra and this is indeed the most desired characteristic feature of a typical lasing spectrum.
 \subsubsection{\textbf{Dual mode lasing behavior}}
 \begin{figure}
 	\centering
 	\includegraphics[width=0.5\linewidth]{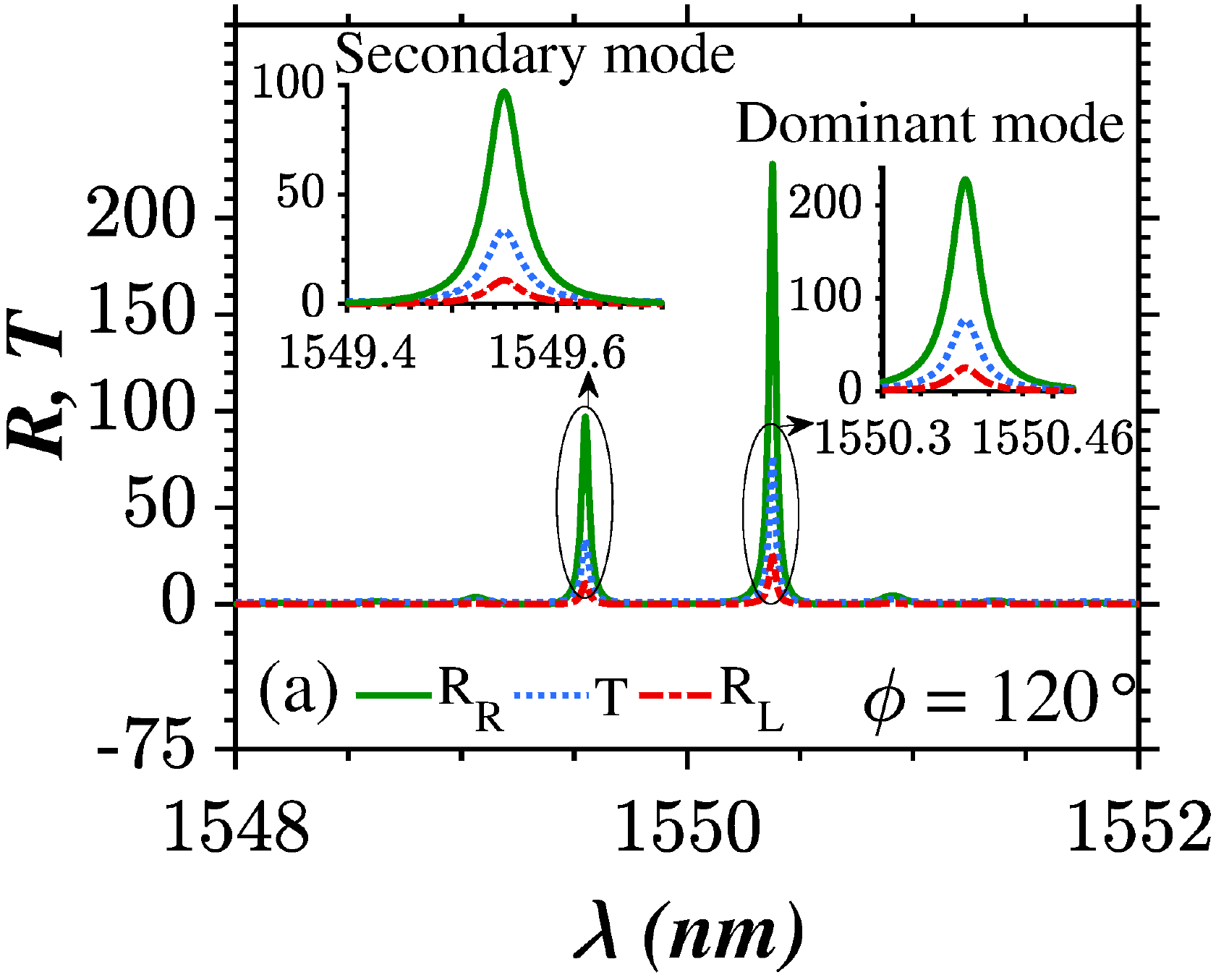}\includegraphics[width=0.5\linewidth]{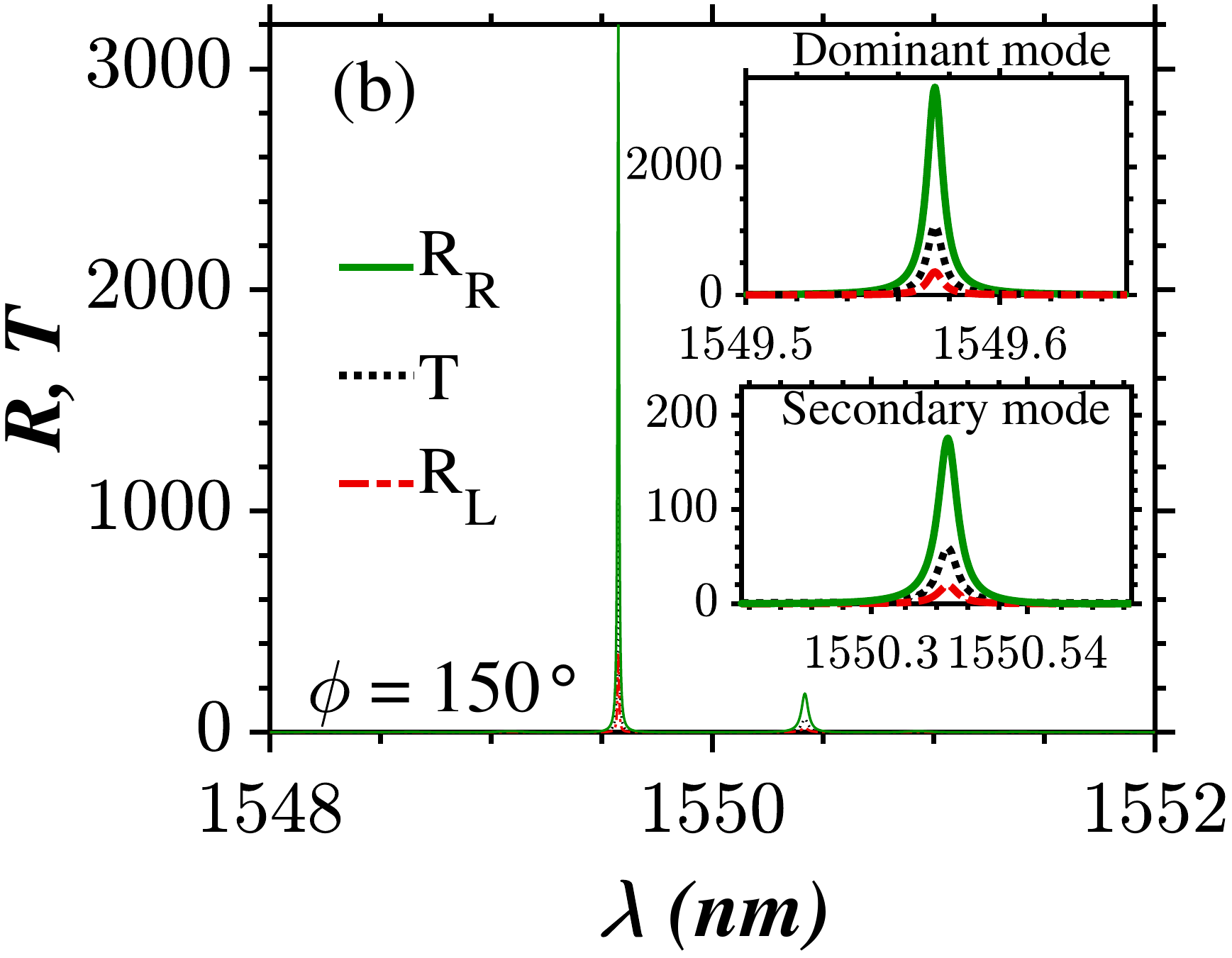}\\\includegraphics[width=0.5\linewidth]{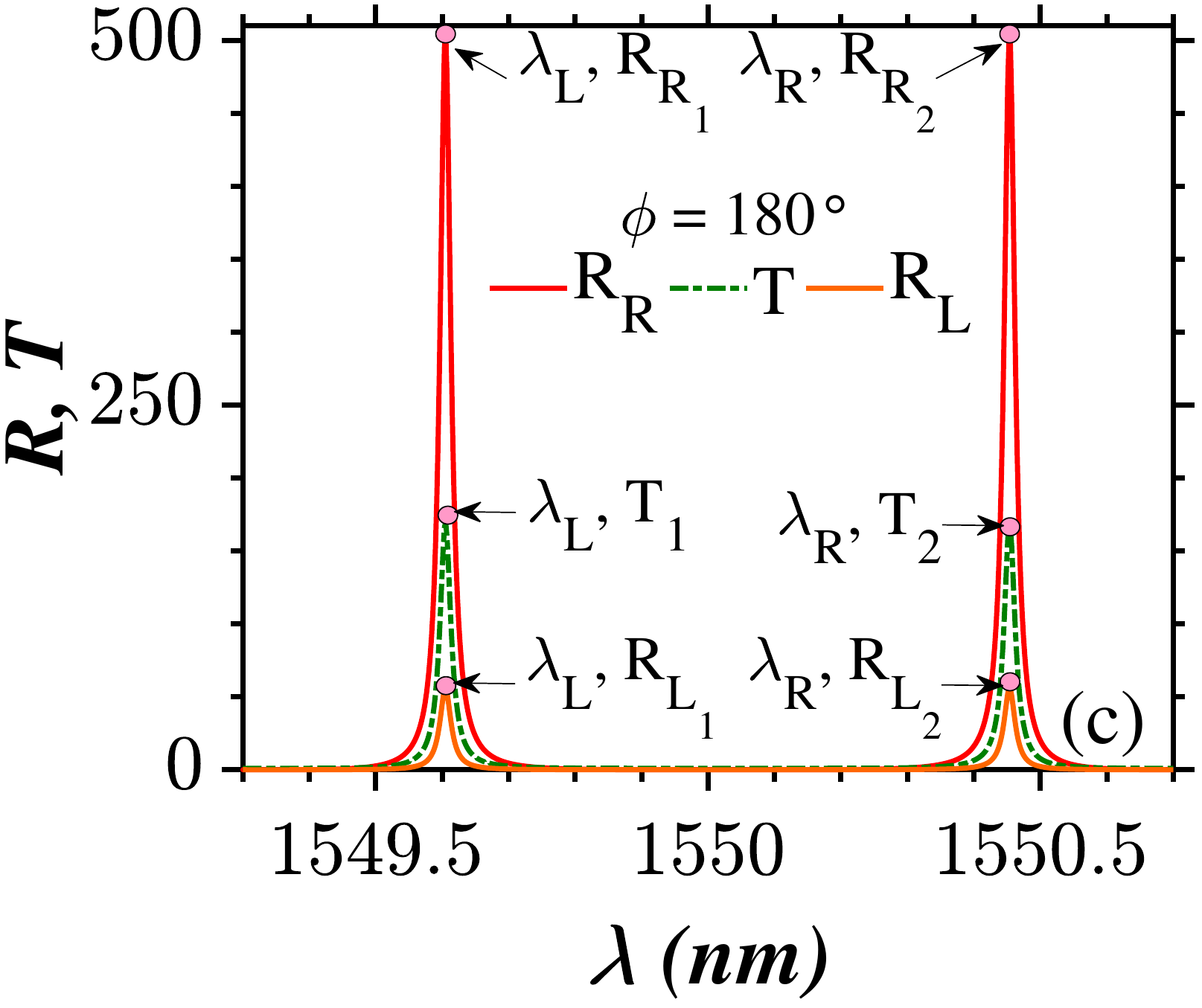}\includegraphics[width=0.5\linewidth]{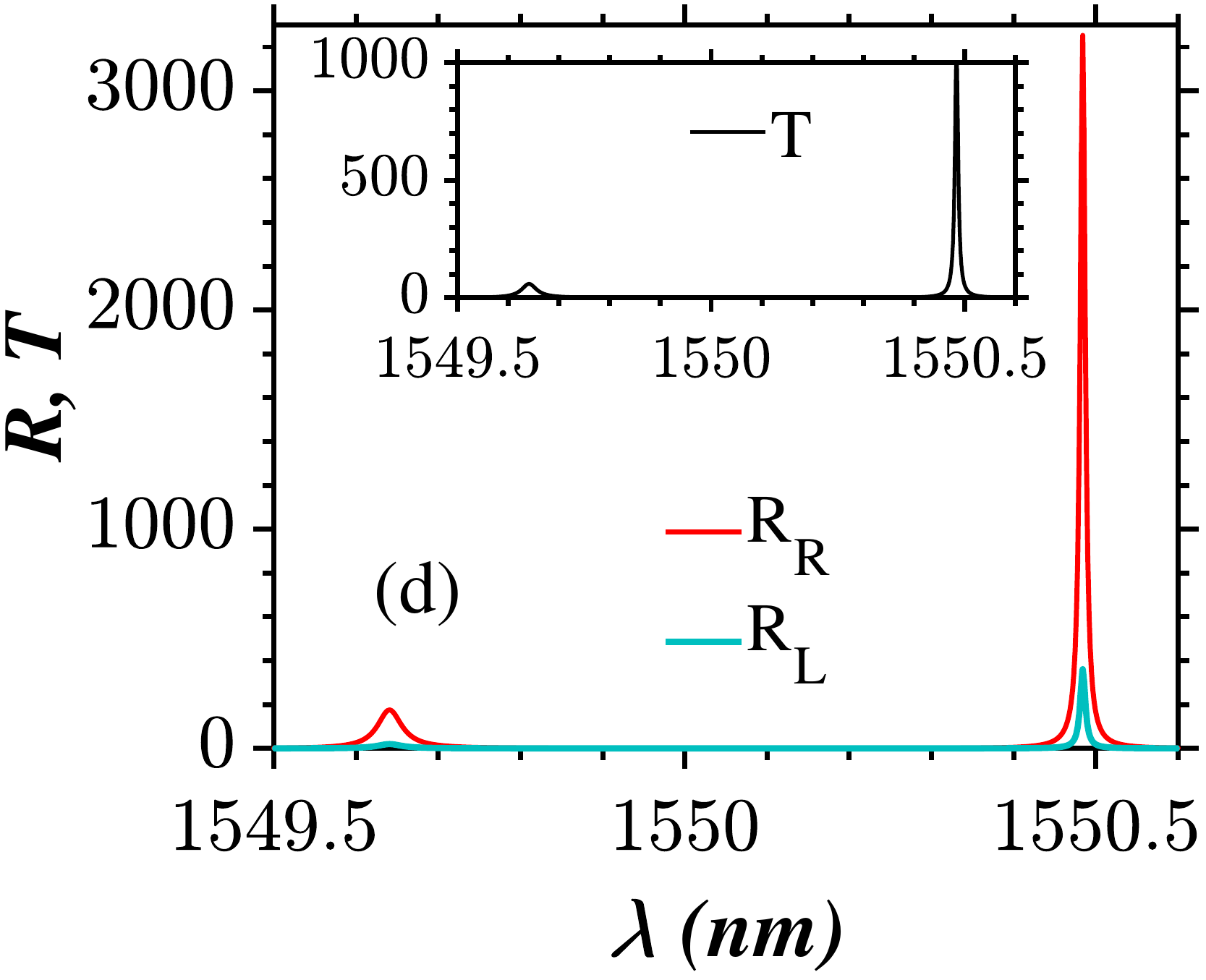}\\\includegraphics[width=0.5\linewidth]{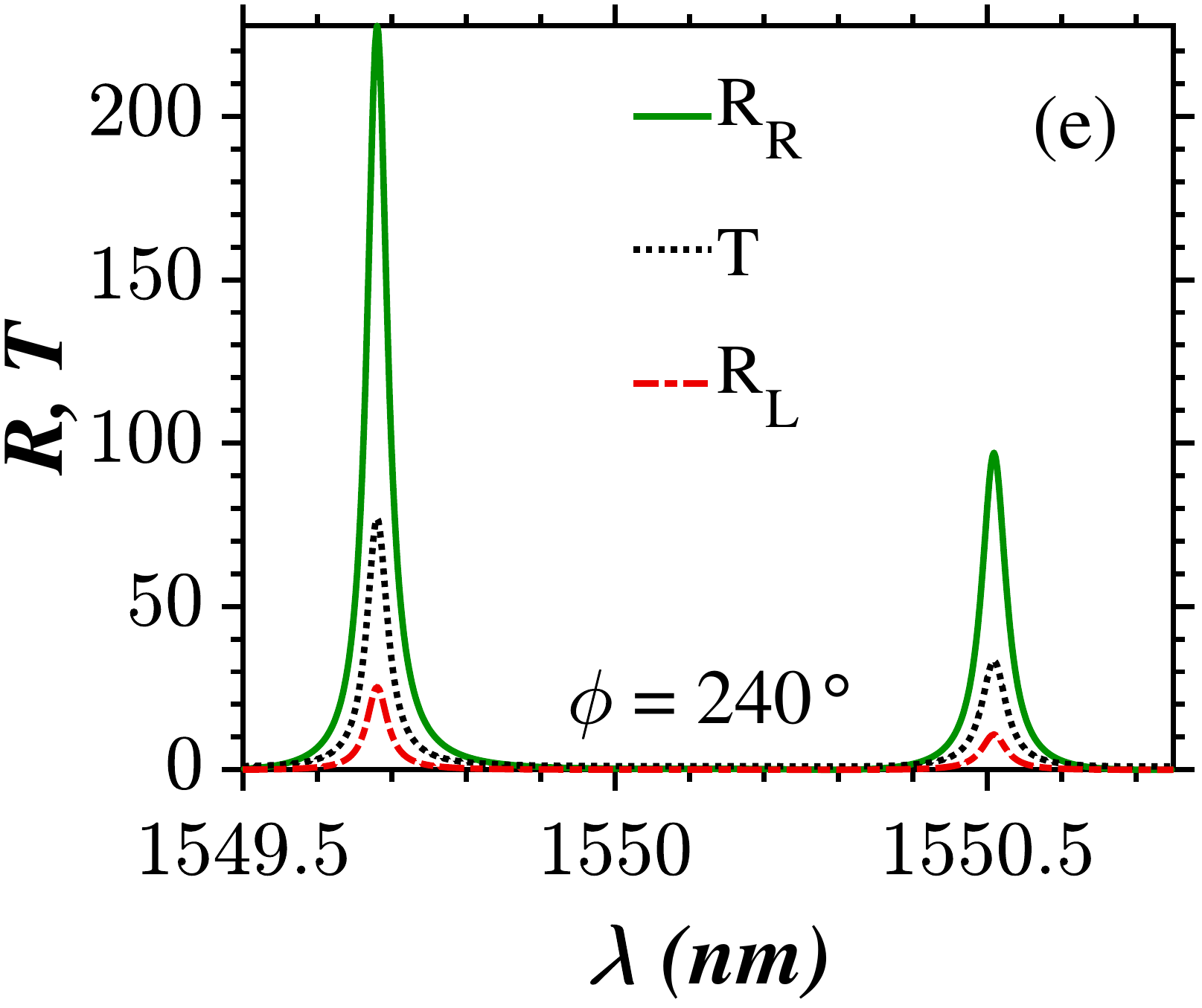}\includegraphics[width=0.5\linewidth]{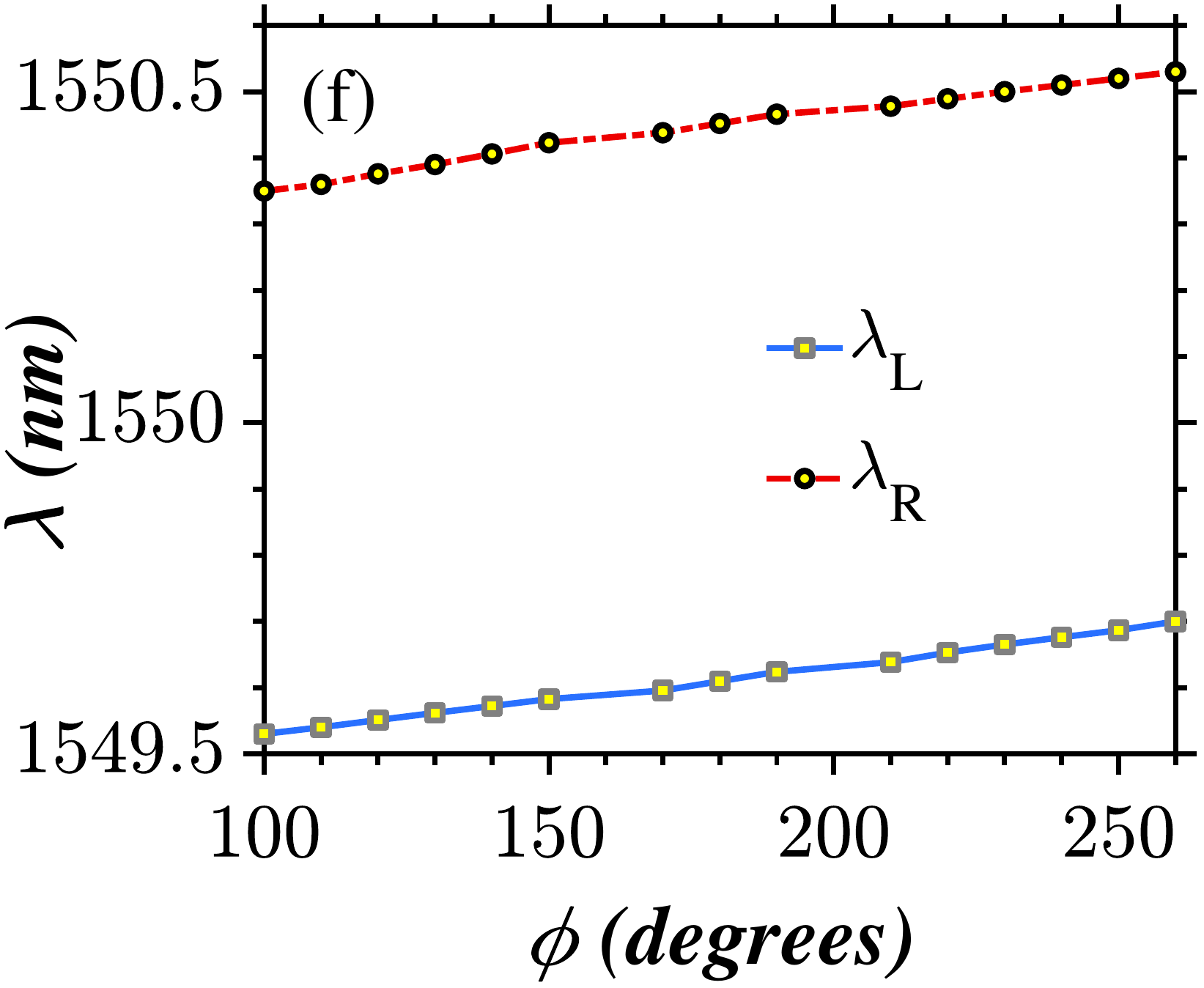}
 	\caption{Dual mode lasing behavior exhibited by a broken $\mathcal{PT}$-symmetric FBG at $g = 20$ against the variation in phase shift with $\phi = 120^\circ$, $150^\circ$, $180^\circ$, $210^\circ$, $240^\circ$ is shown in (a) -- (e), respectively. The variation in the wavelengths corresponding to the reflectivity peaks ($\lambda_l$ and $\lambda_r$) of the spectra is plotted in (f).}
 	\label{fig6}
 \end{figure}
\begin{figure}
	
	\centering
	\includegraphics[width=0.5\linewidth]{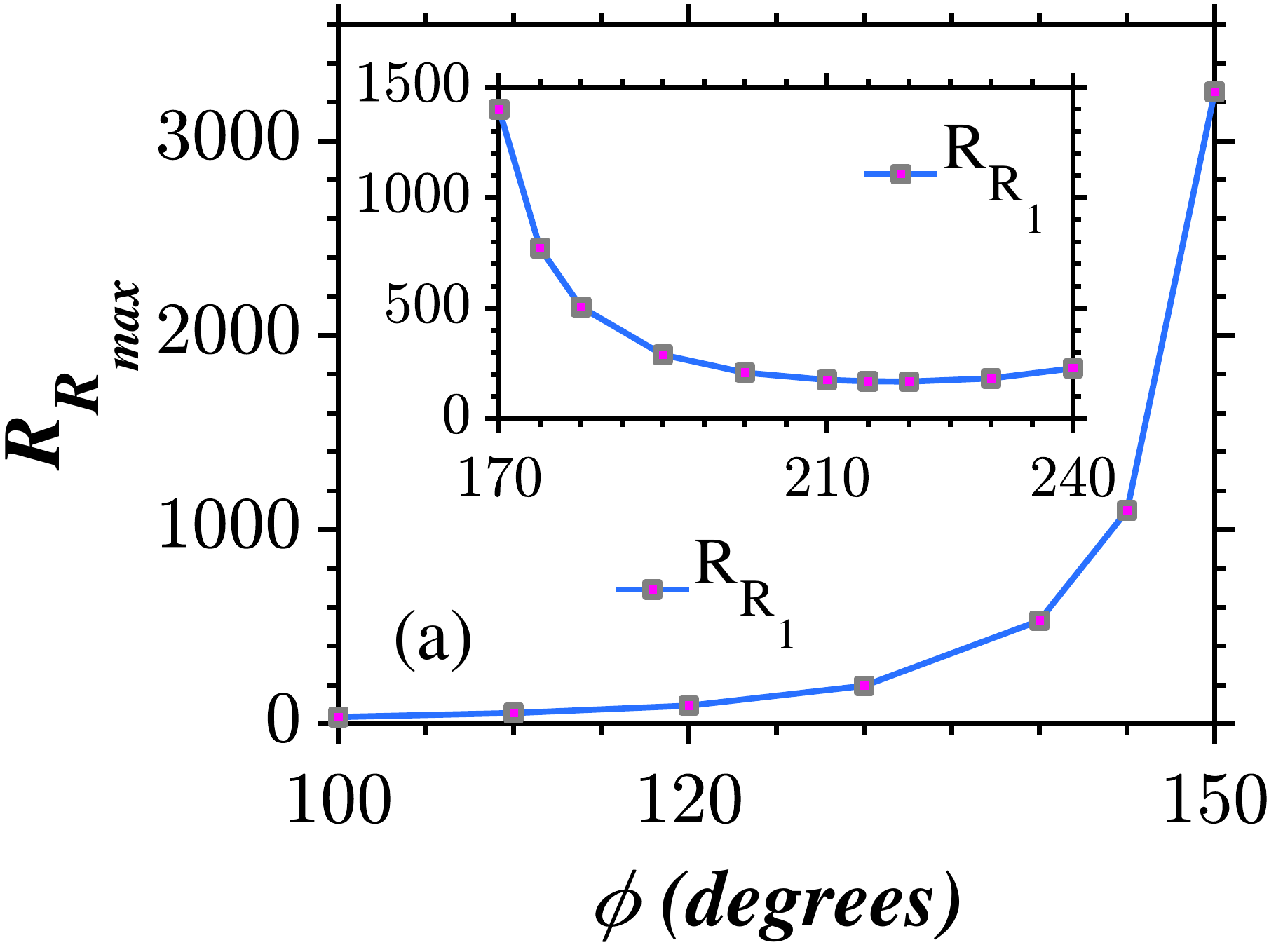}\includegraphics[width=0.5\linewidth]{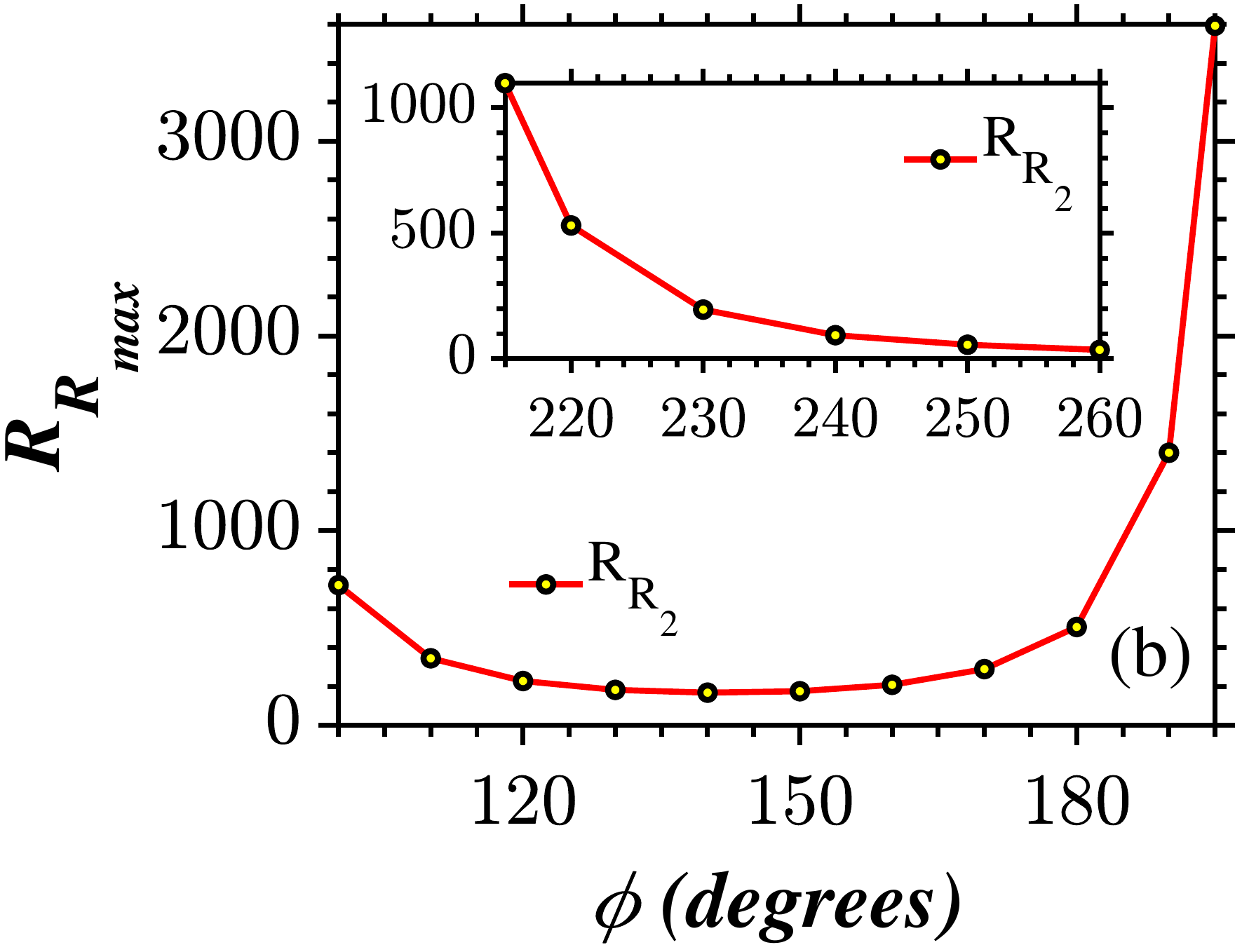}\\\includegraphics[width=0.5\linewidth]{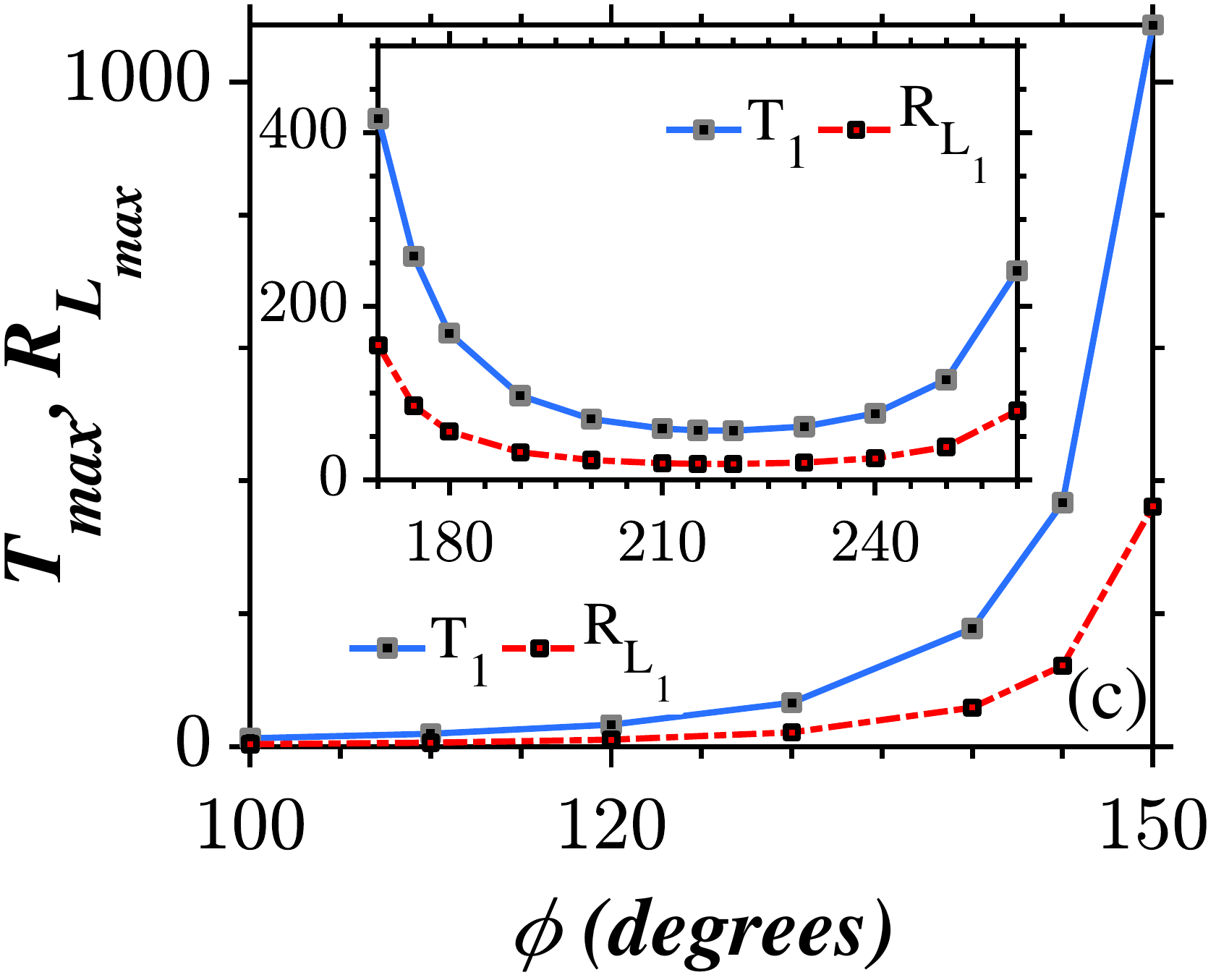}\includegraphics[width=0.5\linewidth]{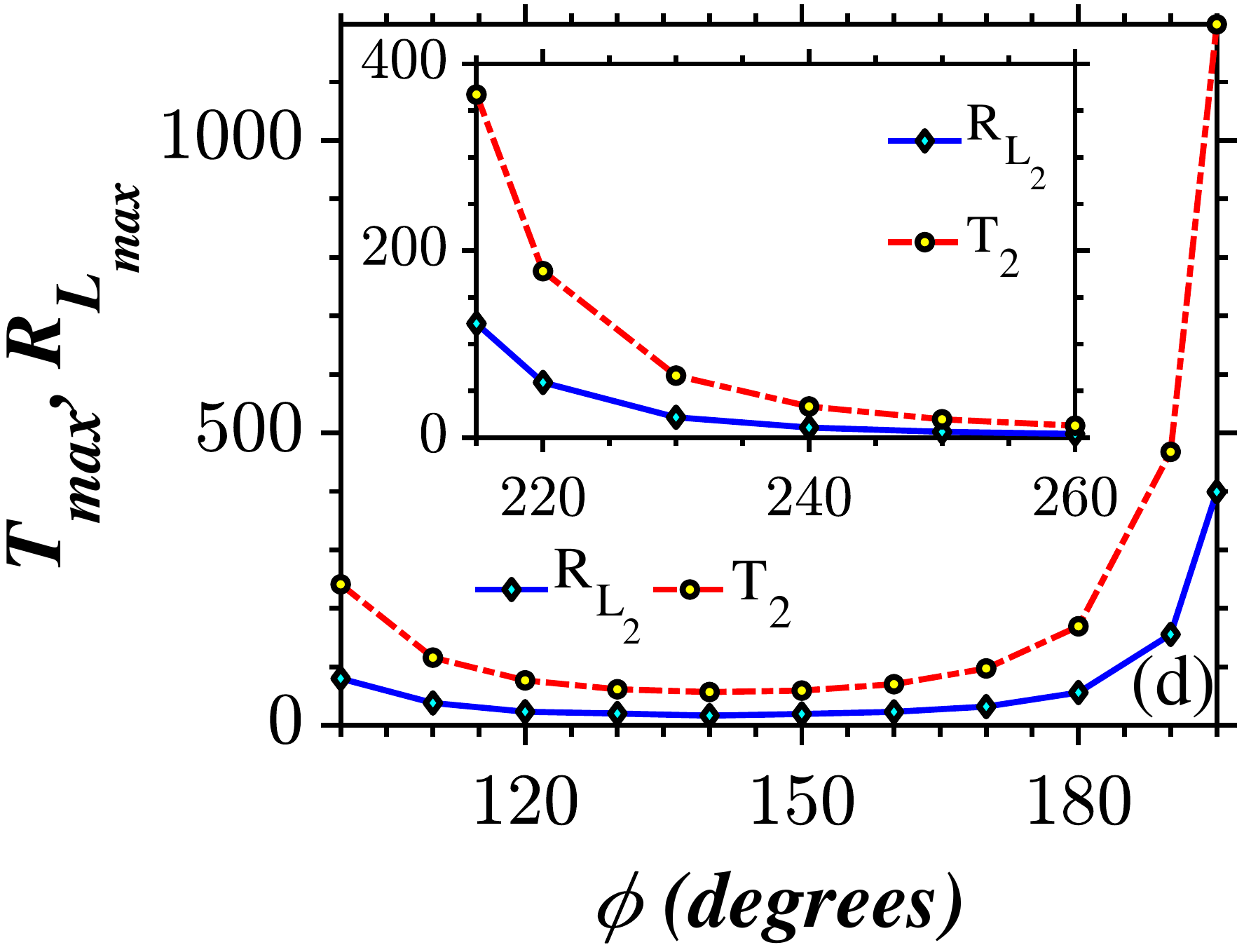}\\\includegraphics[width=0.5\linewidth]{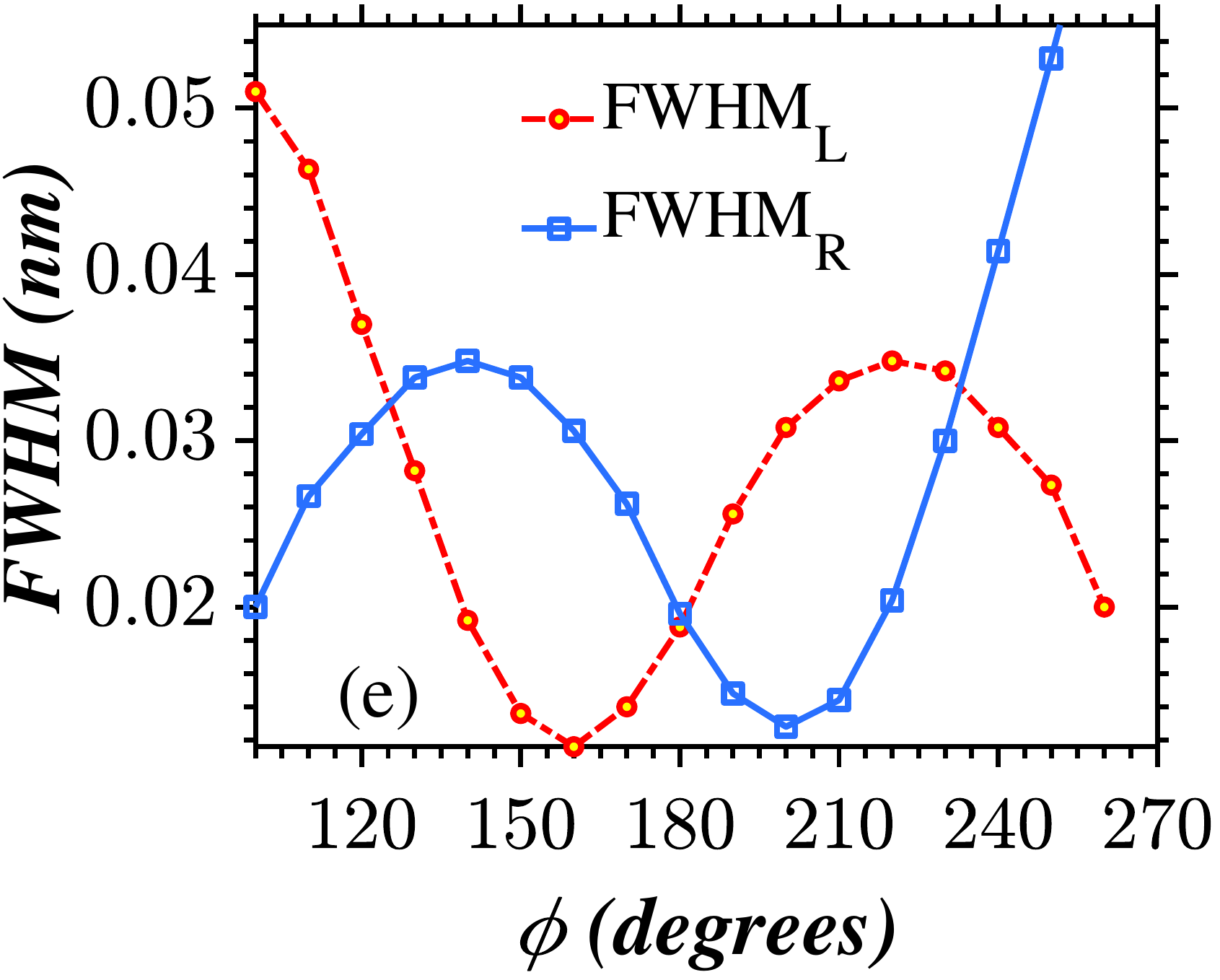}
	\caption{ The continuous variation of transmitted ($T_{max}$) and reflected intensity peaks ($R_{max}$) with respect to change in the values of $\phi$ is shown in (a) -- (c). (d) and (e) Show the decrease in the peak reflectivity and transmittivity when length of the system ($L$) is increased. The changes in the FWHM of the transmitted and reflected spectra against the variation of $\phi$ are plotted in  (e) and the notations FWHM$_L$ and FWHM$_R$ denote the full width half maximum of the modes on the left and right side of the $1550$ nm wavelength, respectively. }
	\label{fig7}
\end{figure}

\begin{figure}
	\centering
	\includegraphics[width=0.5\linewidth]{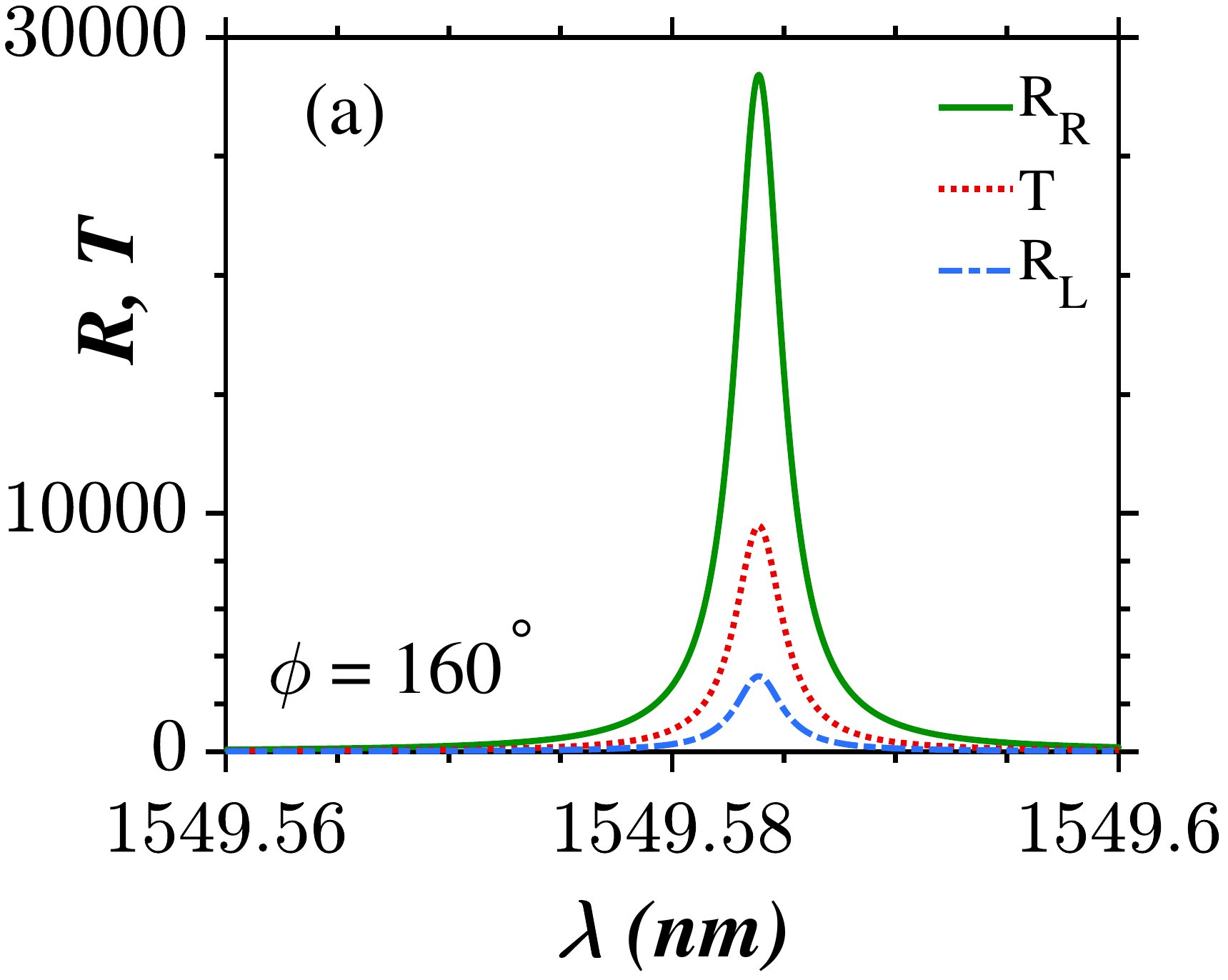}\includegraphics[width=0.5\linewidth]{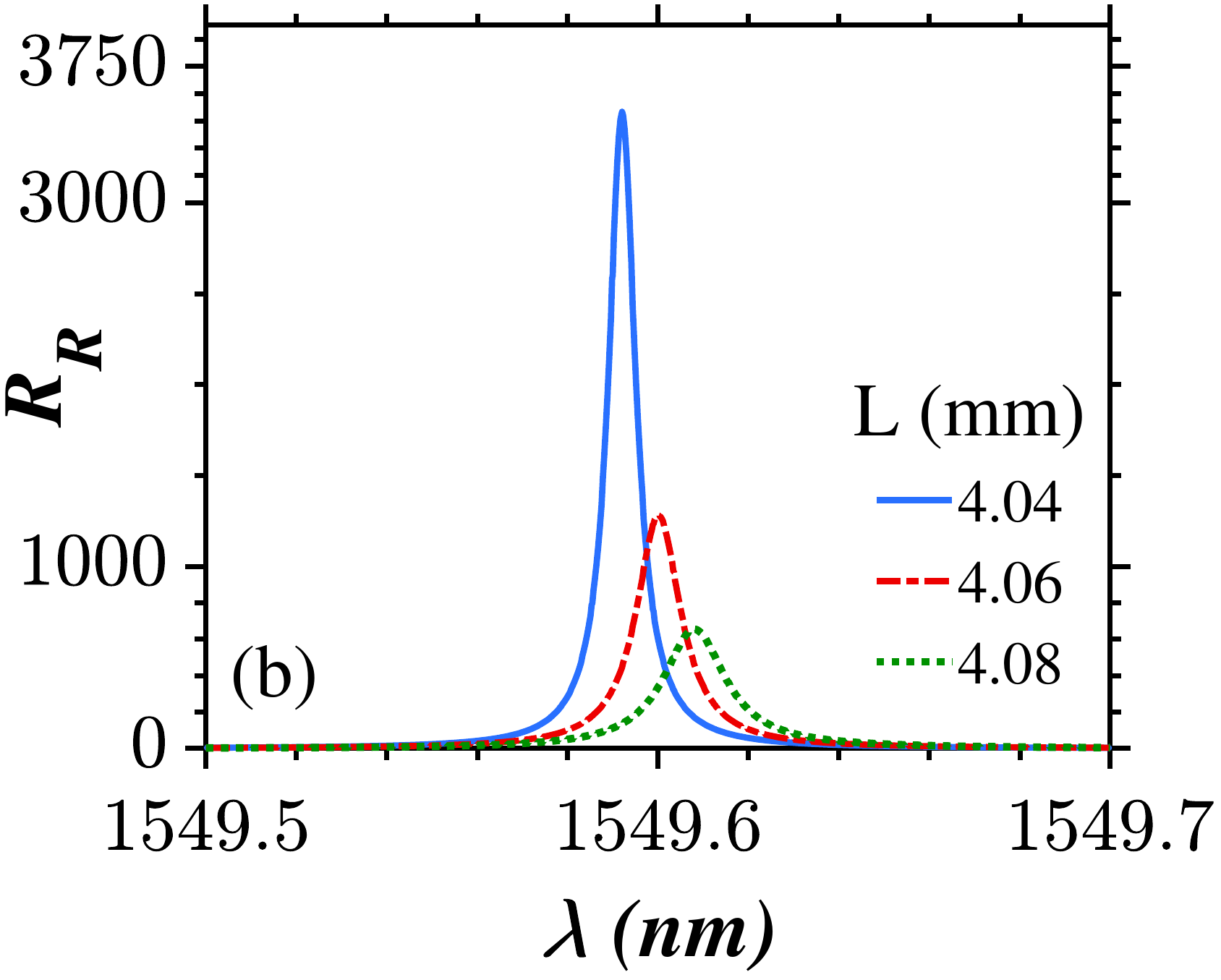}\\
	\caption{ Reflection and transmission spectra of a broken $\mathcal{PT}$-symmetric FBG at $g = 20$ cm$^{-1}$ and $\phi = 160^\circ$ are shown in (a). (b) Illustrates the decrease in the peak reflectivity for the right light incidence and shift in the corresponding peak wavelengths when length of the system ($L$) is increased.}
	\label{fig7l}
\end{figure}

The peaks of reflectivity and transmittivity  in the spectra are found to occur at two wavelengths for values of phase falling in the range $90^\circ < \phi < 270^\circ$. To differentiate them, they are designated here as dominant and secondary modes  as shown in Figs. \ref{fig6}(a) -- \ref{fig6}(c). The former has larger amplification at the peak ($R_{max}$ and $T_{max}$) compared to the latter one. The reflectivity and transmittivity at the peaks between these two modes are always asymmetrical except for certain values of $\phi$, say $\phi= 180^\circ$. For a phase shift of $\phi = 120^\circ$, the dominant (secondary) mode occurs on the left (right) of the Bragg wavelength (1550 nm) as illustrated in Fig. \ref{fig6}(a). On the other hand, when $\phi = 150^\circ$ the dominant and the secondary modes are seen on the shorter and longer wavelength sides of the Bragg wavelength, respectively, as shown in Fig. \ref{fig6}(b). However, both these modes are equally amplified at the phase of $\phi = 180^\circ$ as depicted in Fig. \ref{fig6}(c). Post the occurrence of the symmetric amplification at $180^\circ$, the asymmetric nature of amplification between these two modes begins to appear.  As portrayed in Fig. \ref{fig7}(d), the appearance of dominant and secondary modes at a phase of $\phi = 210^\circ$, looks alike the plots obtained for a phase of $\phi = 150^\circ$ in terms of intensity with the center wavelength interchanged. It is confirmed from Fig. \ref{fig6}(e) that the spectrum shown by the system for a phase of $\phi = 240^\circ$ is exactly the mirror image of the plot at $\phi = 120^\circ$ about the Bragg wavelength ($1550$ nm). Finally, we look into the impact of phase on the wavelength at which peak reflectivity occurs in Fig. \ref{fig6}(f). The reflectivity and transmittivity peaks on the shorter wavelength side ($\lambda_L$) of $1550$ nm are shifted towards the Bragg wavelength, whereas the corresponding peaks on the longer wavelength side ($\lambda_R$) of $1550$ nm are shifted further away from the Bragg wavelength. From the continuous variation of the intensity plots against the variation in the value of phase ($\phi$) depicted in Figs. \ref{fig7}(a) -- \ref{fig7}(d), we infer that the PPTFBG system exhibits both increasing and decreasing lasing behavior in both reflection and transmission spectra depending on the value of phase shift ($\phi$).  As stated in the previous section, the FWHM is narrower for those values of $\phi$ for which reflectivity and transmittivity of the spectra is larger and vice-versa and this behavior is plotted in Fig. \ref{fig7}(e). For some values of $\phi$, the device shows lasing spectra with very huge reflectivity of the order of $10^4$ in its light propagation characteristics ($R$ and $T$) when operated in the broken $\mathcal{PT}$-symmetric regime. Such a behavior is plotted in Fig. \ref{fig7l}(a) at a phase of $\phi = 160^\circ$. These kinds of uncontrollable amplification in the lasing spectra  are found to decrease at higher length of the grating as shown in Fig. \ref{fig7l}(b). 
  \subsubsection{\textbf{Single mode lasing behavior}}
 \begin{figure}
  	\centering
  	\includegraphics[width=0.5\linewidth]{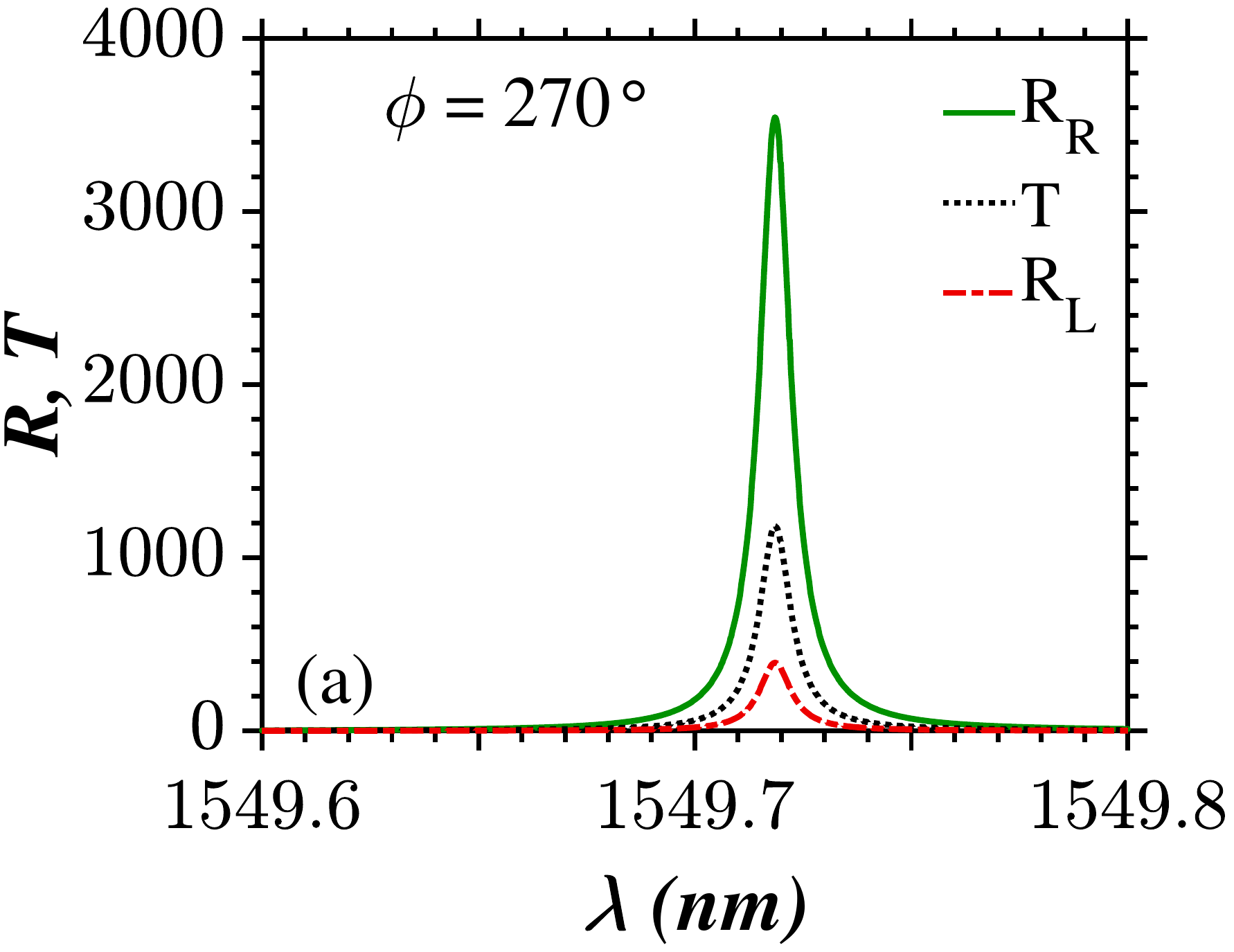}\includegraphics[width=0.5\linewidth]{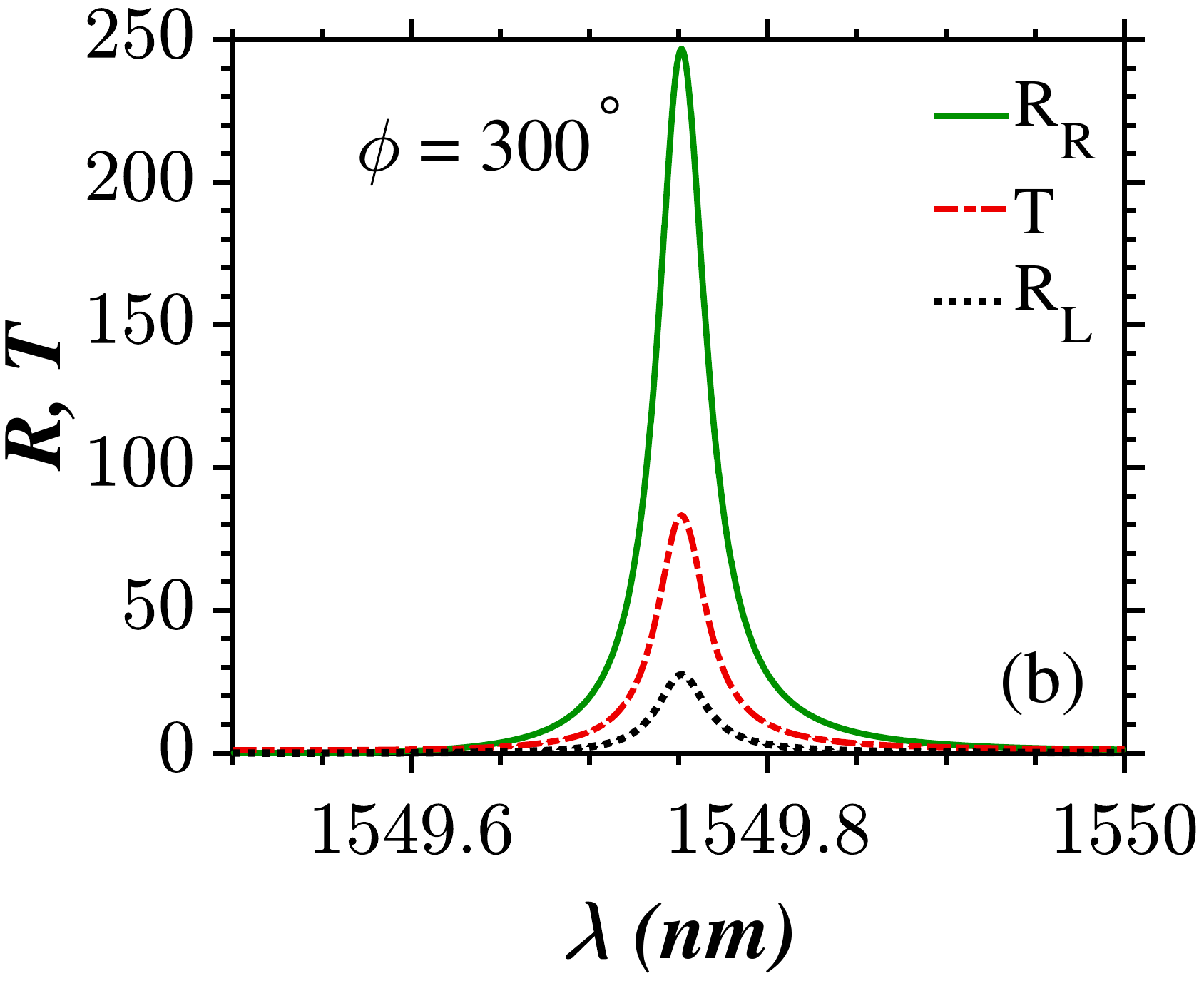}\\\includegraphics[width=0.5\linewidth]{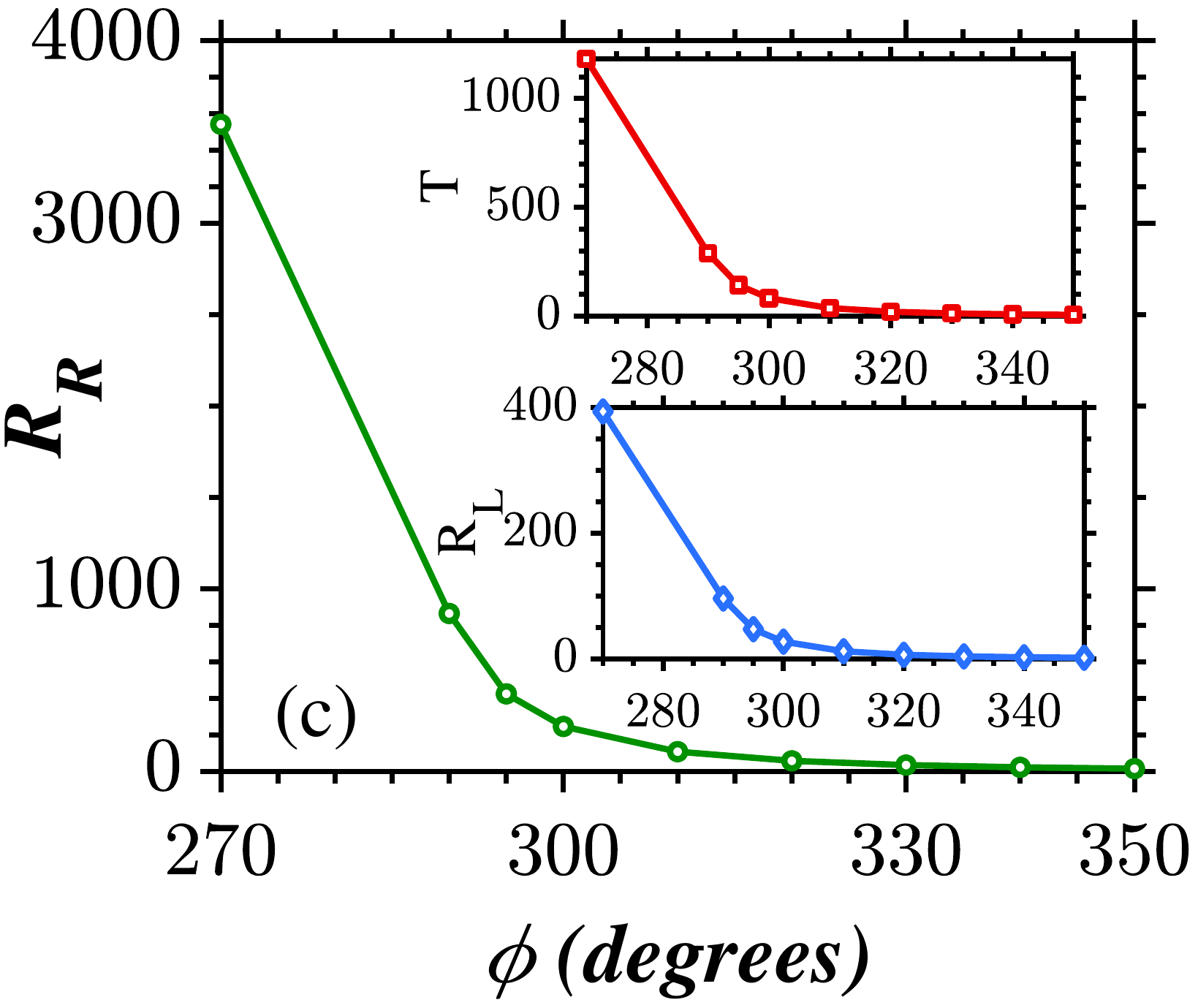}\includegraphics[width=0.5\linewidth]{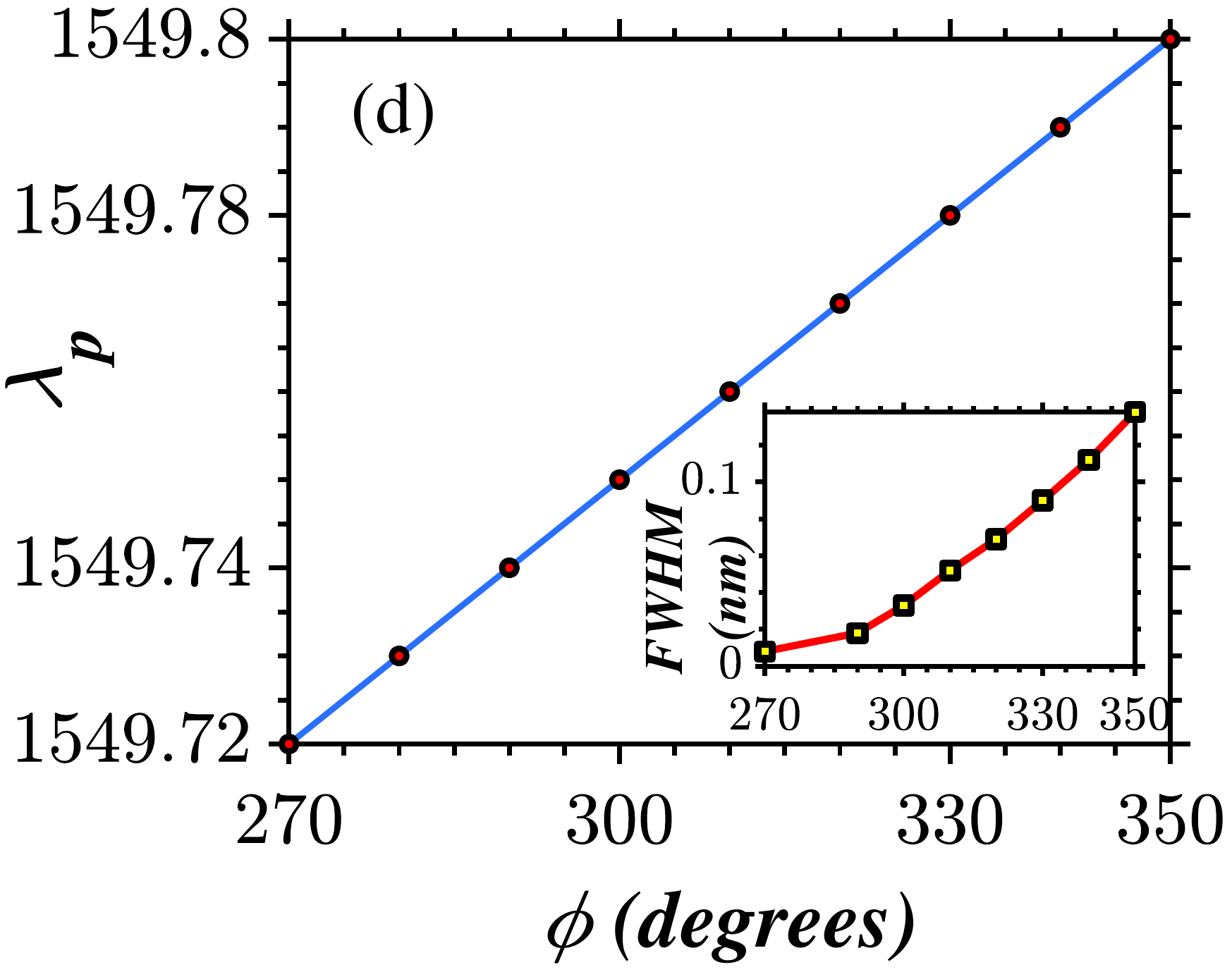}\\
  	\caption{Phase controlled single mode lasing behavior of a broken $\mathcal{PT}$-symmetric FBGat $g = 20$ against the variation in phase shift values $\phi = 270^\circ$ and $300^\circ$ is shown in (a) and (b), respectively. The continuous variation of maximum of transmittivity ($T_{max}$) and reflectivity ($R_{max}$) with respect to change in the values of $\phi$ is shown in (c). The wavelength ($\lambda_2$) corresponding to these peaks is shown in (d) and the FWHM of the spectra is plotted in the inset.}
  	\label{fig8}
  \end{figure}
 If the phase is continuously tuned further, the secondary modes are not observed in the lasing spectra and the reflectivity and transmittivity peaks are found to appear only in one distinct mode as shown in Fig. \ref{fig8}(a). When $\phi = 270^\circ$, the lasing spectra is centered at 1549.72 nm with the FWHM of 0.008 nm is observed in the transmission as well as reflected spectra. The spectra also resemble the spectra plotted at $\phi = 90^\circ$ in terms of intensity and FWHM except for a difference that it occurs on the shorter wavelength side of the Bragg wavelength in the spectra. Any increase in the value of $\phi$ leads to a decrease in the reflectivity and transmittivity and increase in the FWHM as shown in Figs. \ref{fig8}(b) and \ref{fig8}(c). The wavelength over which the lasing spectrum is centered is shifted towards the Bragg wavelength as depicted in Fig. \ref{fig8}(c). Thus we can conclude that by carefully tuning the value of the phase shift in the middle of the grating, it is possible to control grating characteristics such as intensity, FWHM and the wavelength over which the lasing spectra is centered.
 
  \section{Spectral characteristics of a PPTFBG with multiple phase shift regions}
 It is worthwhile to mention that FBGs find their main application as channel selection filters in the light wave communication systems. Investigations on improving the channels selection characteristics of FBGs are mainly targeted at increasing the width of the stopband band in the middle. However, the ranges of the stop band in practically realizable FBGs are limited. The concept of cascading multiple PTFBGs with different grating periods can give rise to a good solution to  improve the transmission characteristics significantly. But it is limited by the difficulties in fabricating a compact system \cite{zengerle1995phase}. It should be remembered that it possible to tailor the stop band of a $\mathcal{PT}$-symmetric FBG by introducing the concepts of chirping and apodization \cite{raja2020linear}. Further, by introducing multiple phase shifts in the middle, the spacing between number of channels can be minimized without inflicting any additional penalties on the system.
 \subsection{Unbroken $\mathcal{PT}$-symmetric regime}
 \begin{figure}
 	\centering
 	\includegraphics[width=0.5\linewidth]{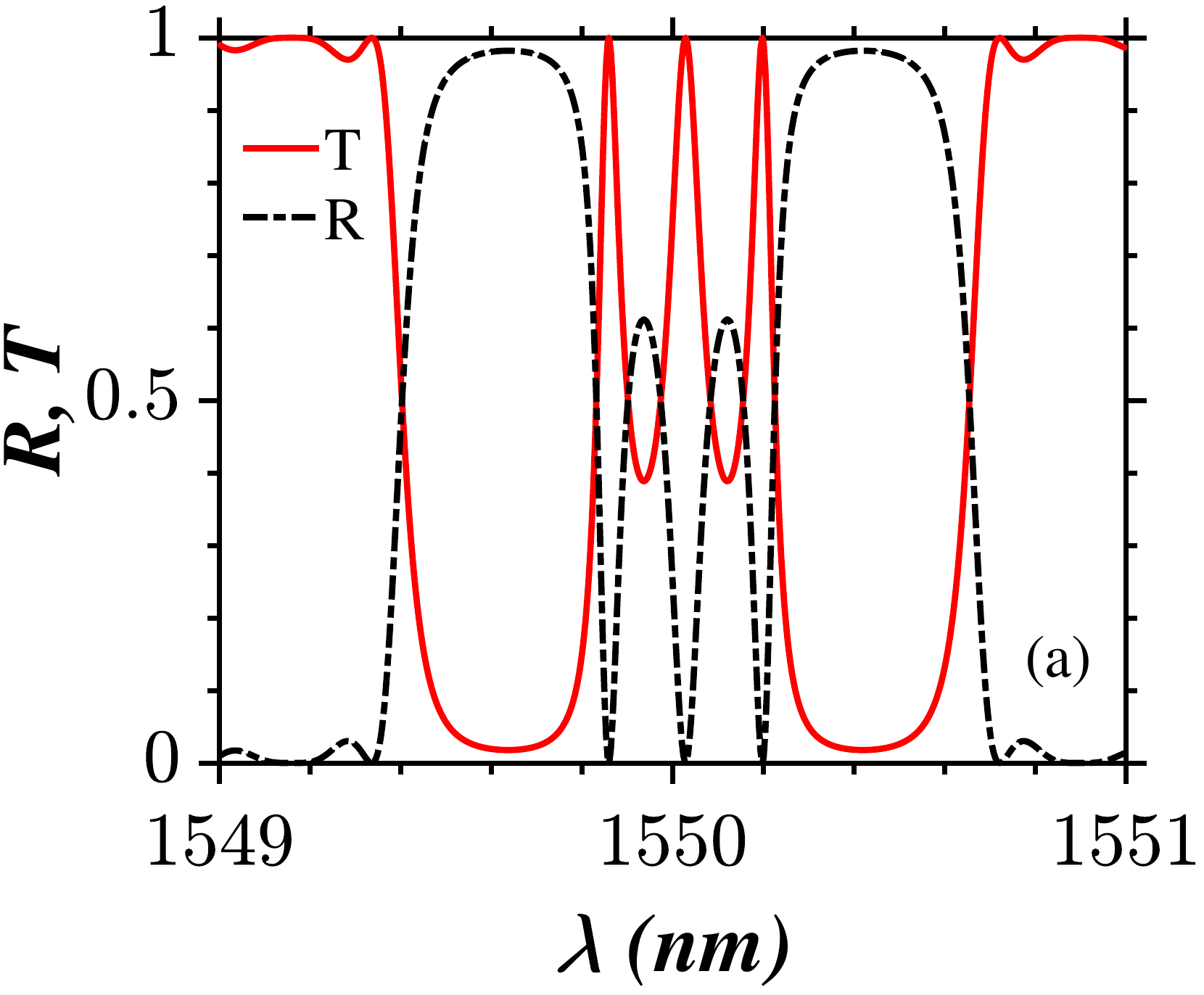}\includegraphics[width=0.5\linewidth]{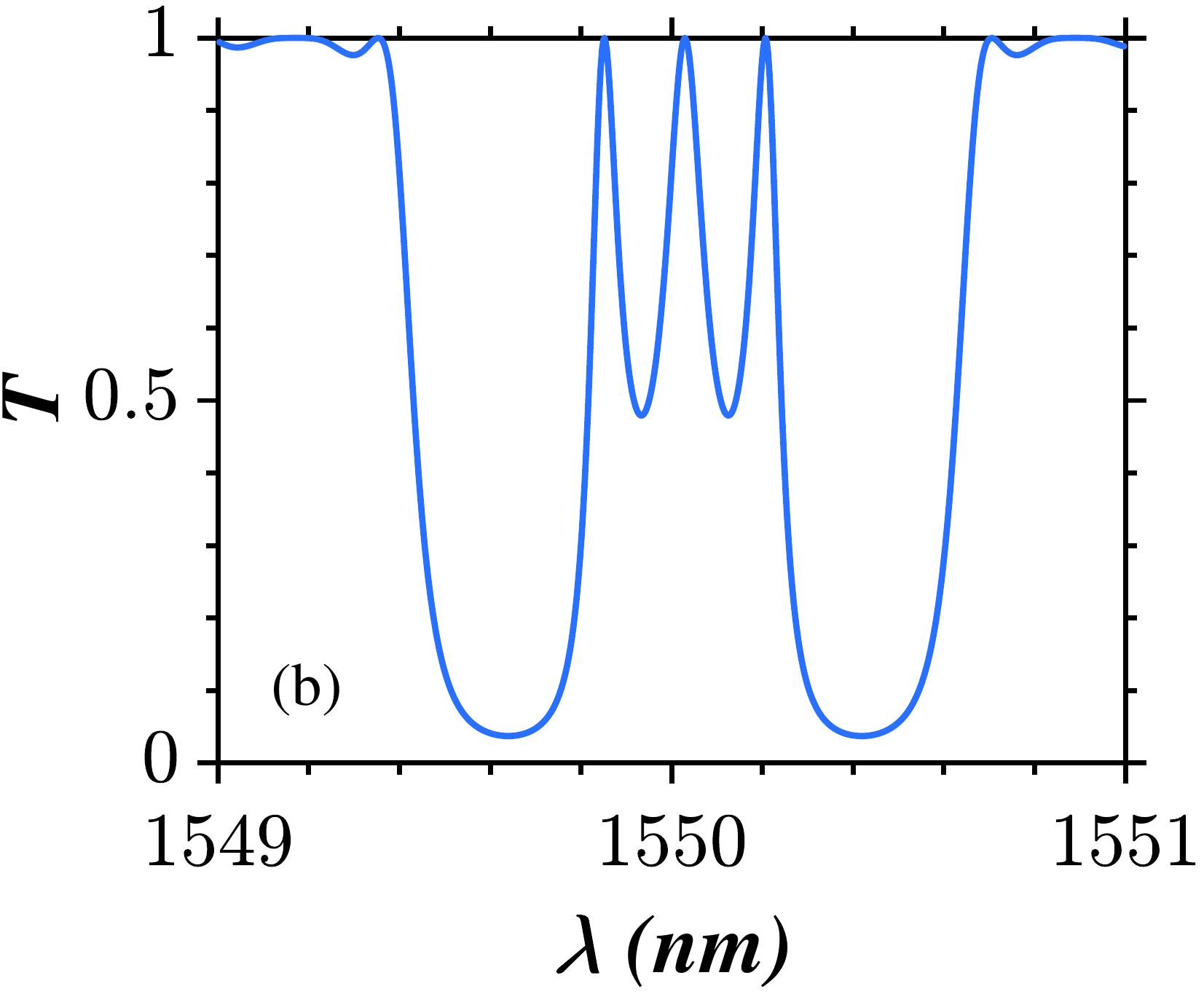}\\\includegraphics[width=0.5\linewidth]{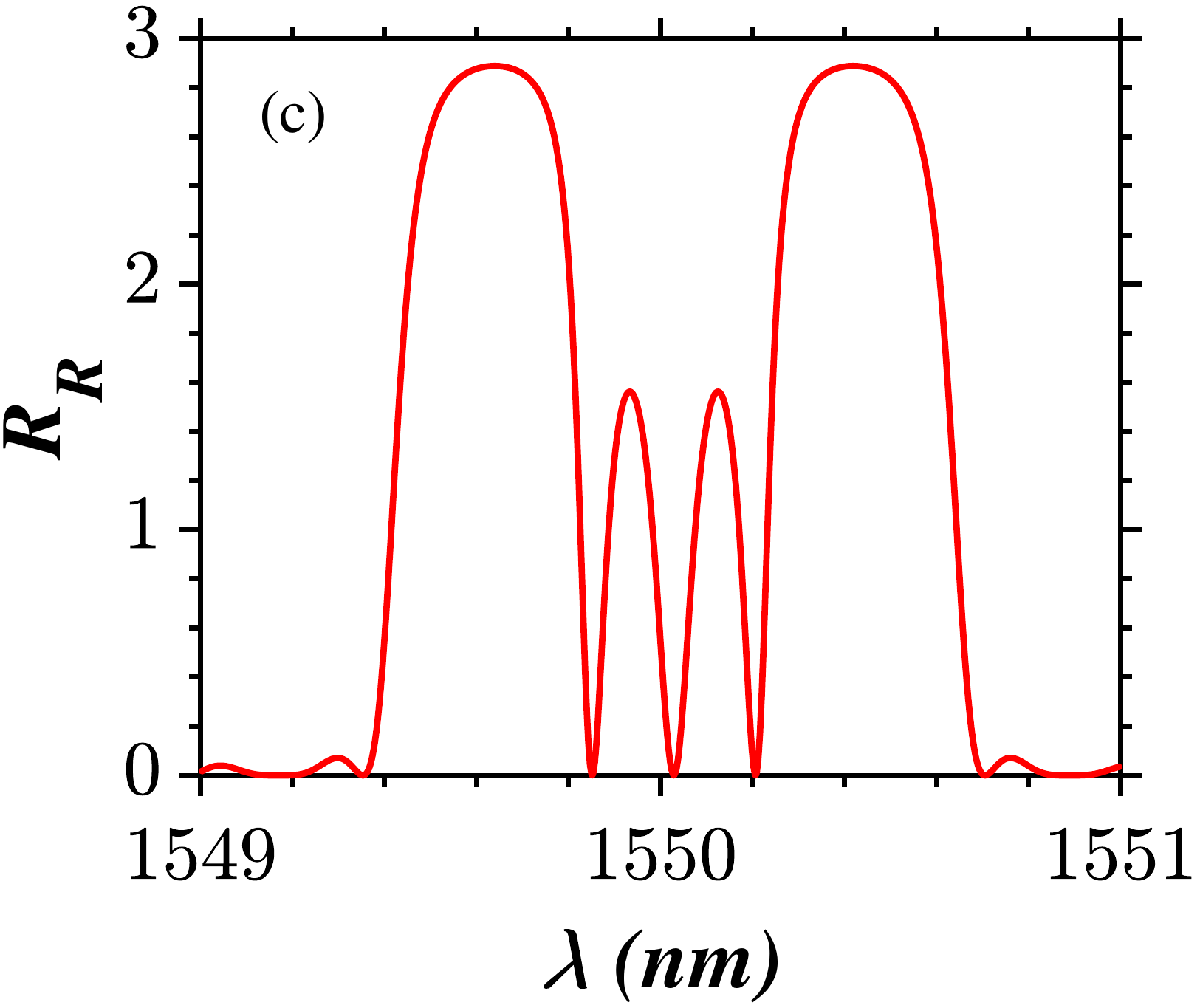}\includegraphics[width=0.5\linewidth]{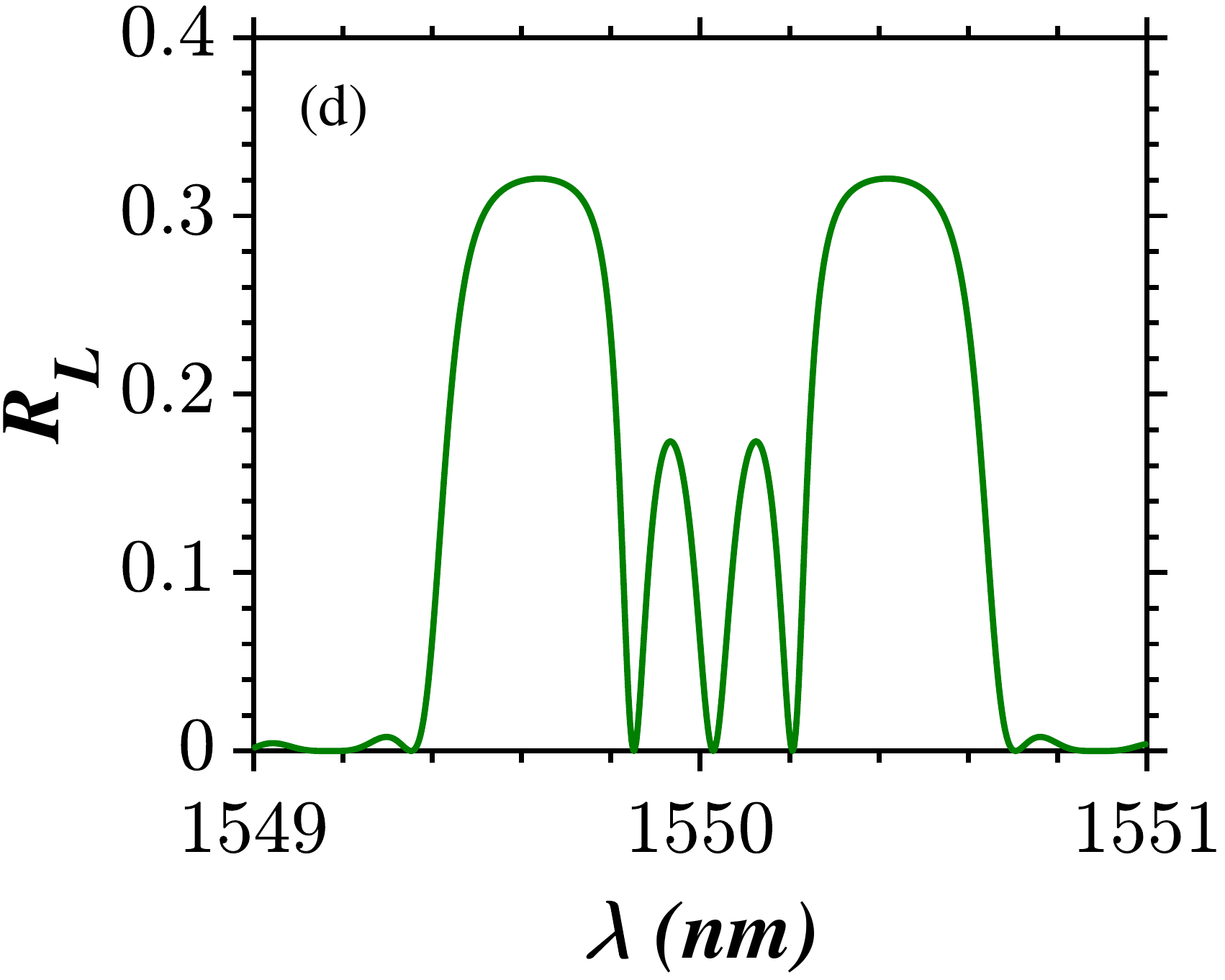}
 	\caption{(b-d) Reflection and transmission spectra of a unbroken PPTFBG ($g = 5$ cm $^{-1}$) with a phase shift ($\phi = 180^\circ$) located at $-L/4$, $0$, $L/4$. Plot (a) is simulated in the absence of gain-loss ($g=0$).}
 	\label{figm1}  
 \end{figure}

\begin{figure}
 	\centering
 	\includegraphics[width=0.5\linewidth]{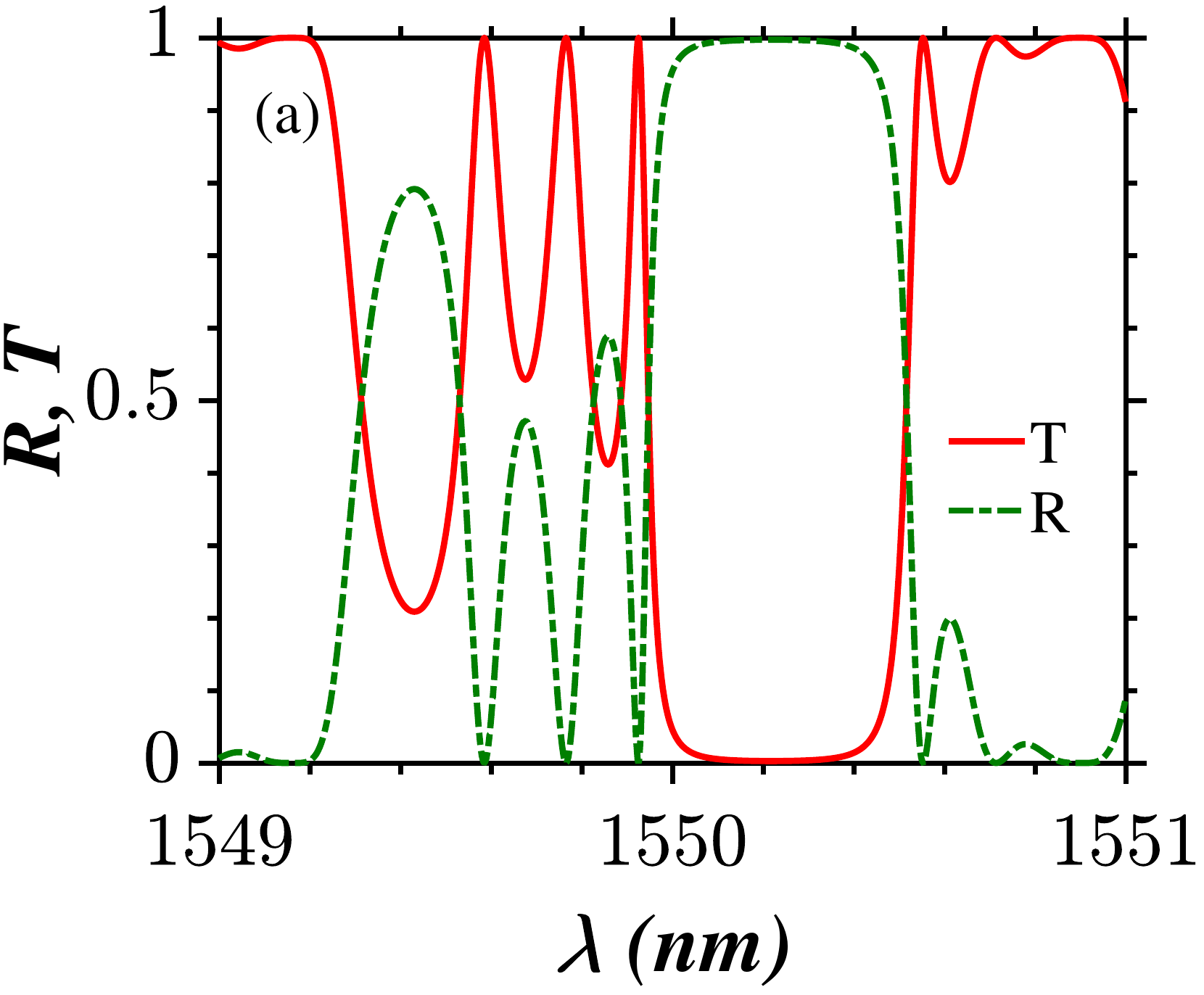}\includegraphics[width=0.5\linewidth]{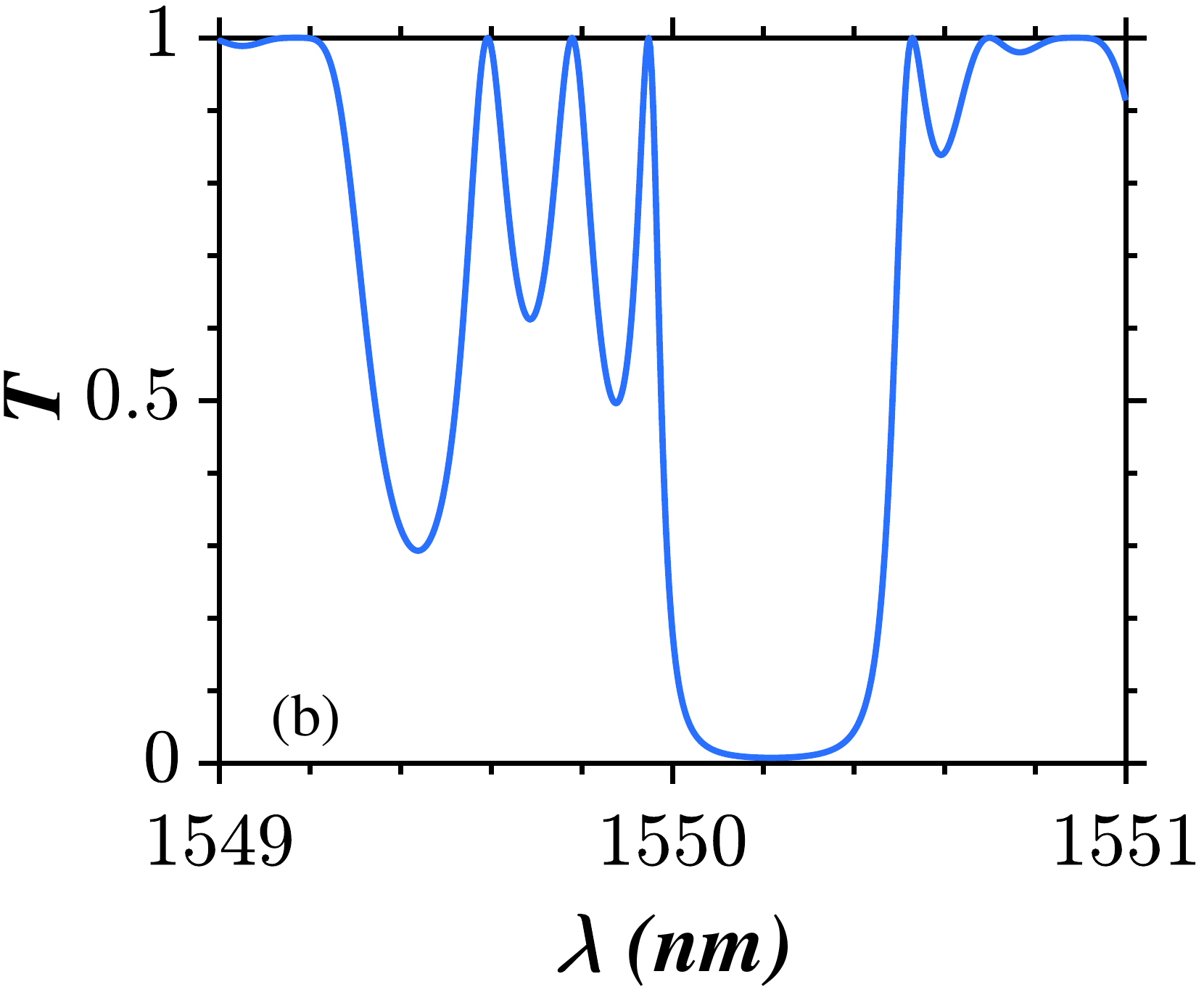}\\\includegraphics[width=0.5\linewidth]{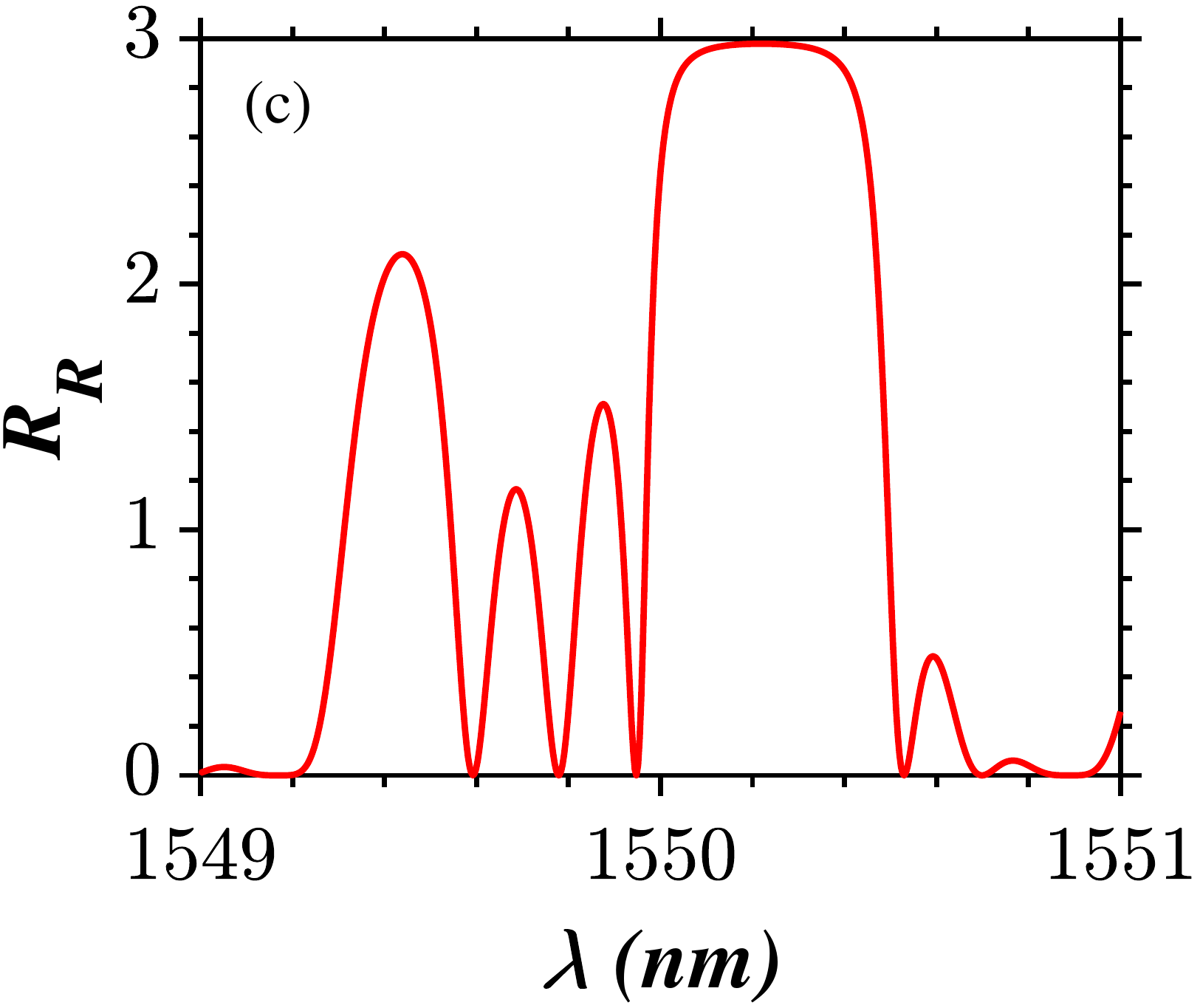}\includegraphics[width=0.5\linewidth]{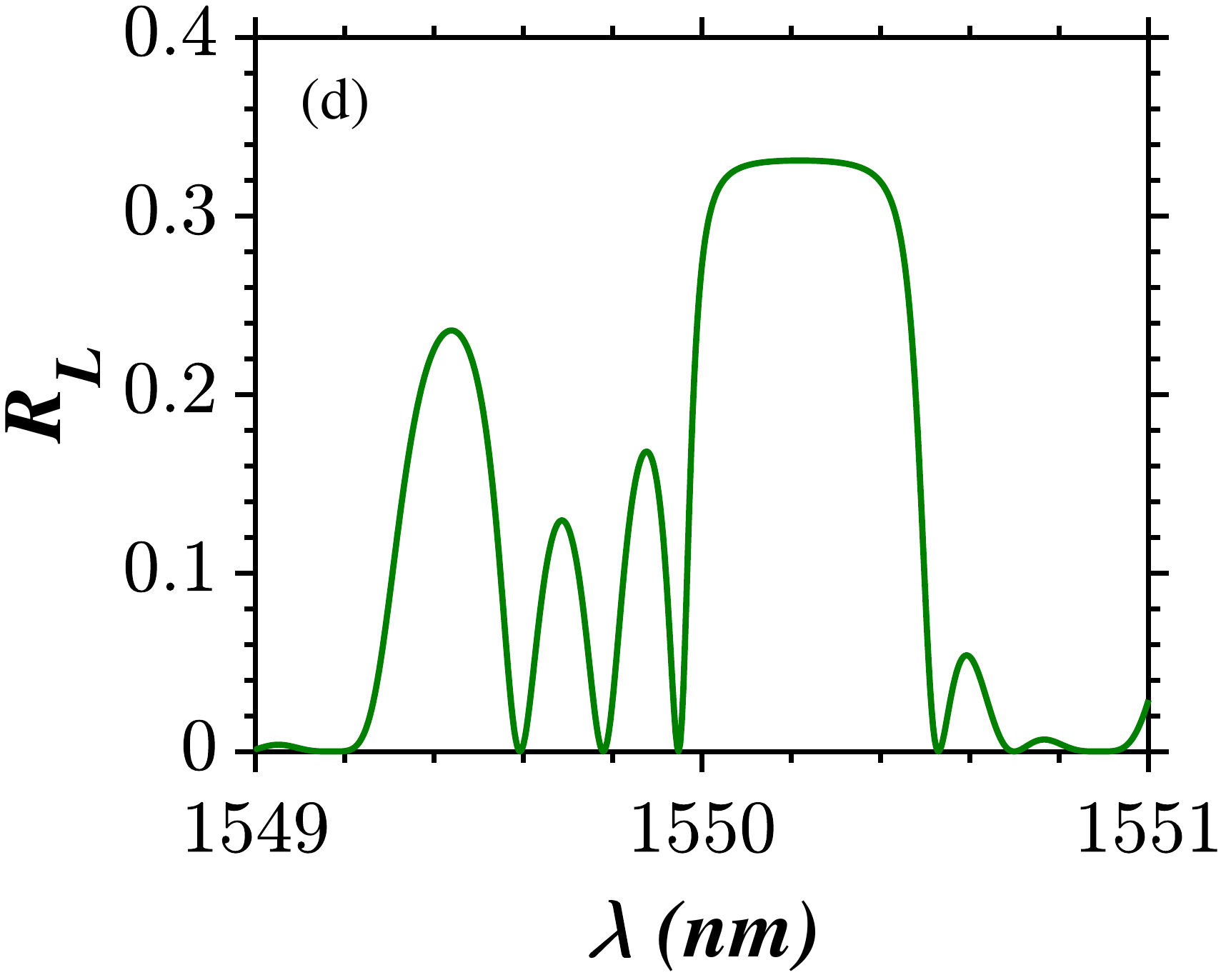}
 	\caption{(b-d) Reflection and transmission spectra of a unbroken PPTFBG ($g = 5$ cm $^{-1}$) with a phase shift ($\phi = 90^\circ$) located at $-L/4$, $0$, $L/4$. Plot (a) is simulated in the absence of gain-loss ($g=0$).}
 	\label{figm2}  
 \end{figure}

 Figure \ref{figm1} depicts the spectra of unbroken PPTFBG ($g = 5$ cm$^{-1}$) for an identical phase shift of $\phi = 180^\circ$ located at three locations ($z_1, z_2$ and $z_3$). Even though the locations of the phase shifts can be varied according to the requirement and thereby the position of peaks within the stop band can be controlled, we fix the location of multiple phase shift in our investigations as stated earlier in the section \ref{Sec:2}. The distance between each phase shift region is one and the same by virtue of positioning the phase shifts at $-L/4$, $0$, $L/4$ along $z$. Finally, the reflection and transmission spectra in the presence of multiple phase shift can be found using Eqs. (\ref{Eq: mul1})  -- (\ref{Eq: mul3}). The number of transmission peaks and the dips in the reflectivity within the stop band is dictated by the number of phase shift regions along the length of PPTFBG. For instance, one can find three transmission windows in Fig. \ref{figm1}(a) and Fig. \ref{figm2}(a) which are plotted in the absence of $\mathcal{PT}$-symmetry at $\phi = 180^\circ$ and $\phi = 90^\circ$, respectively.  Our primary aim is to tailor the spectra by varying the magnitude of phase shift rather than varying the location of the phase shift provided that the magnitude of phase shift at all the locations is same. Also, the inclusion of $\mathcal{PT}$-symmetry paves the way to control the magnitude of the peaks which is not reported so far in the literature. It has already been proven that the usage of 2 or 3 phase-shift regions inside a conventional FBG structure is optimal for practical light wave communication systems \cite{zengerle1995phase}. From our investigations, we proved that the above thumb role holds true in the presence of gain-loss also. One of the major improvements in the spectral characteristics of the $\mathcal{PT}$-symmetric systems is that these systems open a new door towards tailoring of the spectra via an additional degree of freedom in the form of gain and loss. In Figs. \ref{figm1}(b) -- \ref{figm1}(d), we observe that the peaks in the middle of the stop band are symmetric in nature as the magnitude of the phase shift is $\phi = 180^\circ$. If the same system is simulated at $\phi = 90^\circ$, these multiple peaks are asymmetric and also located to the left side of the Bragg wavelength ($1550$ nm). Previously, we have concluded that the plots simulated at $\phi = 270^\circ$, will resemble the mirror image of the plots drawn at $\phi = 90^\circ$. This conclusion holds true even in the presence of multiple phase-shifts. Like any other PTFBG system, multiple phase shifted PTFBG also shows the directional direction reflection characteristics (reflectivity is different for the two light incident directions) and this is portrayed in Figs. \ref{figm1}(c), \ref{figm1}(d), \ref{figm2}(c) and \ref{figm2}(d). For the left incidence, the reflectivity gets decreased and it increases for the light launching condition from the other side ($R_R$). Thus the system enables controlling the magnitude of the peaks in two different approaches: First, by tuning the magnitude of the phase shift and alternatively by varying the value of $g$.
 \subsection{Exceptional point dynamics}
\begin{figure}[h!]
 	\centering
 	\includegraphics[width=0.5\linewidth]{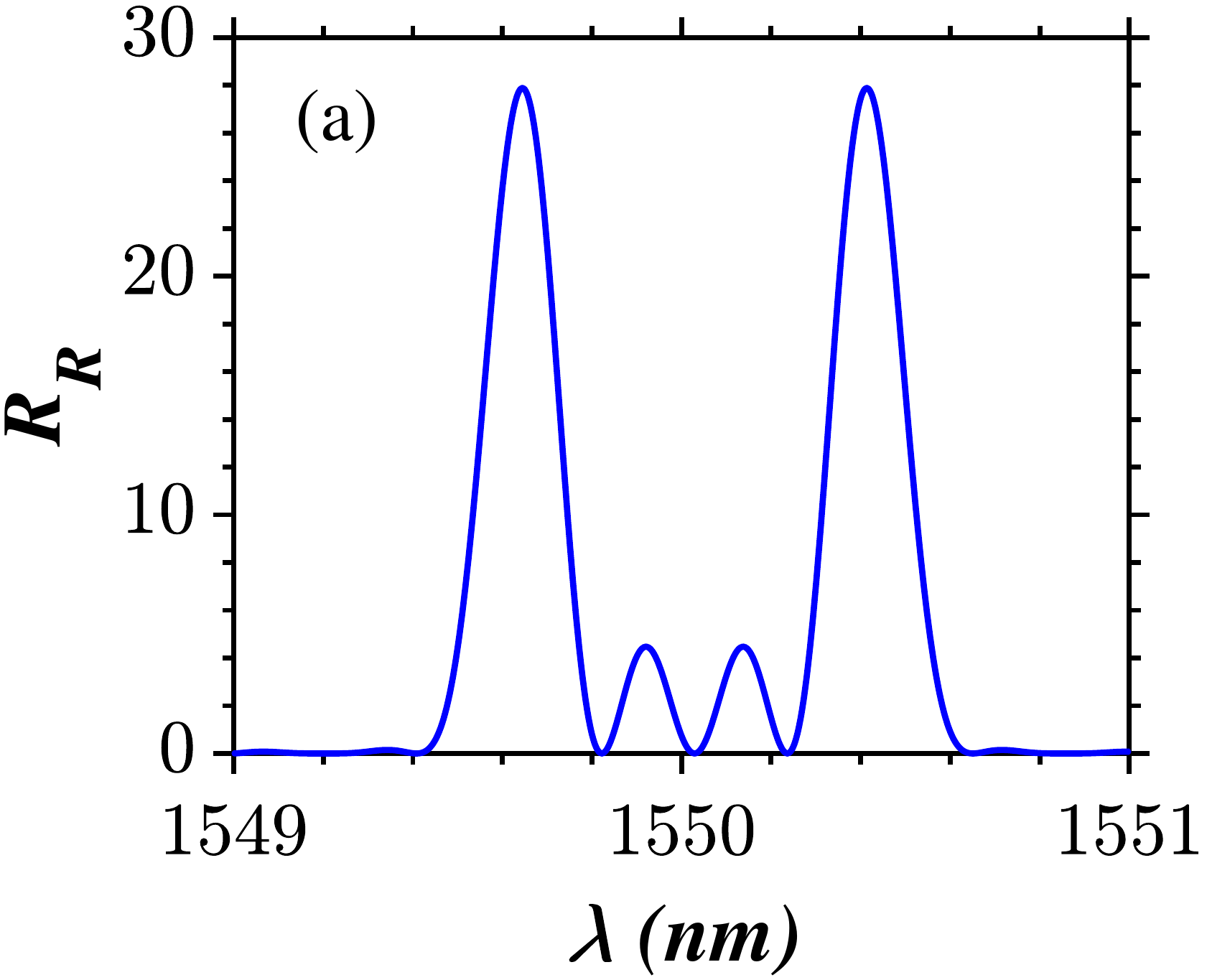}\includegraphics[width=0.5\linewidth]{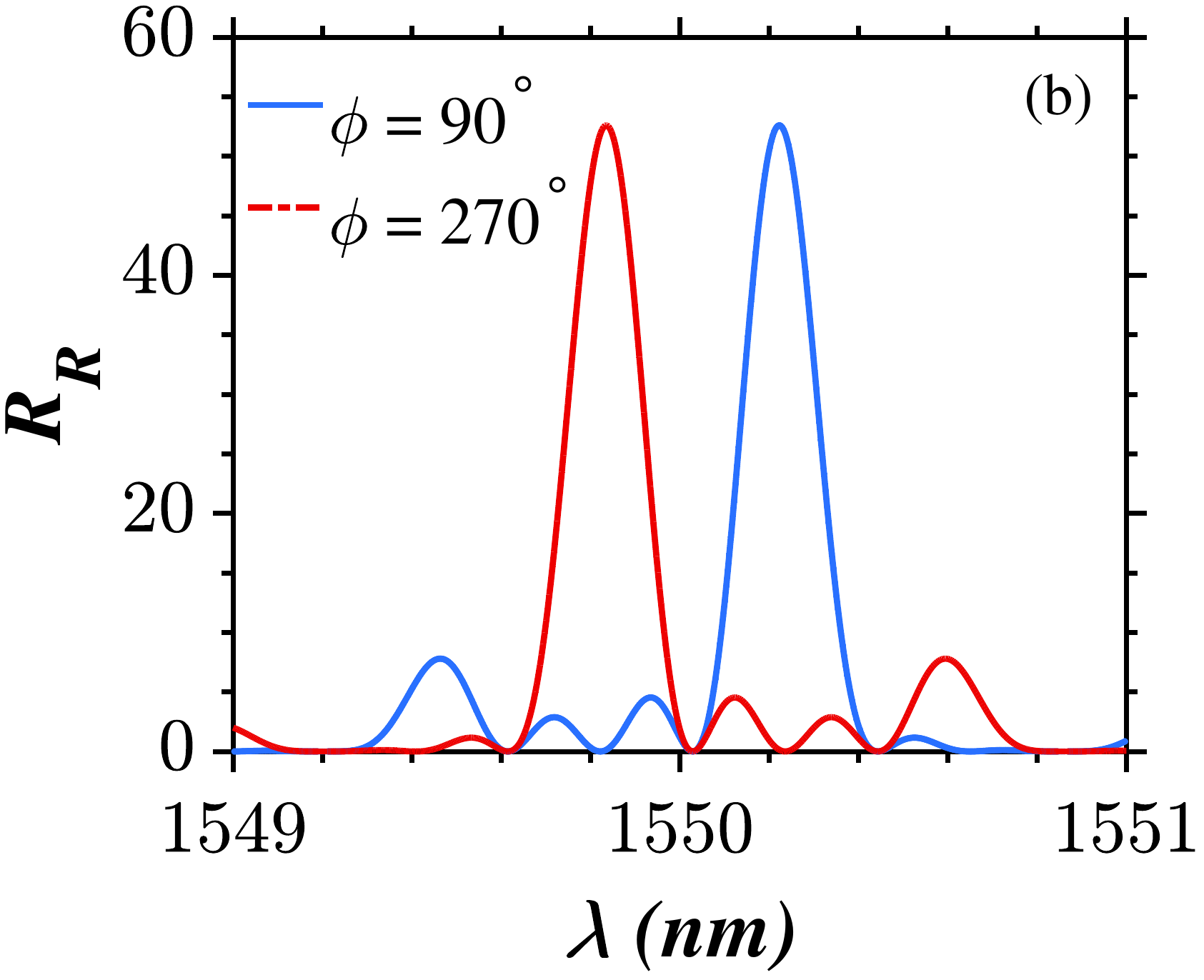}
 	\caption{Reflection (right) spectra of a PPTFBG at the exceptional point ($g = 10$ cm $^{-1}$) with phase shifts (equal) located at $-L/4$, $0$, $L/4$. Plots (a) is simulated at $\phi = 180^\circ$ and (b) is simulated at $\phi = 90^\circ$ and $270^\circ$.}
 	\label{figm3}  
 \end{figure}

 At the exceptional point, the phenomenon of unidirectional wave transport remains the same as before in the present case of multiple phase shifts also. This once again proves that this phenomenon purely relies on the equality between the coupling ($\kappa$) and gain-loss coefficient ($g$) and is independent of any variation in other control parameters including multiple phase shifts. On the other hand, the reflection spectra for the right incidence is influenced by the presence of multiple phase shift, \textit{i.e.}, their location and magnitude brings in notable changes in the spectra. We identify that there are three dips in the middle of the stop band in between two peaks on either sides for $\phi$ = 180$^\circ$ degrees as shown in Fig. \ref{figm3}(a). The same reflectivity dips are seen on the lower (higher) wavelength side of Bragg wavelength when $\phi= 270^\circ$ ($90^\circ$). This once again confirms that the magnitude of phase shift is a highly influential parameter in imposing the variations in the spectra.
 \subsection{Broken $\mathcal{PT}$-symmetric regime}
\begin{figure}
 	\centering
 	\includegraphics[width=0.5\linewidth]{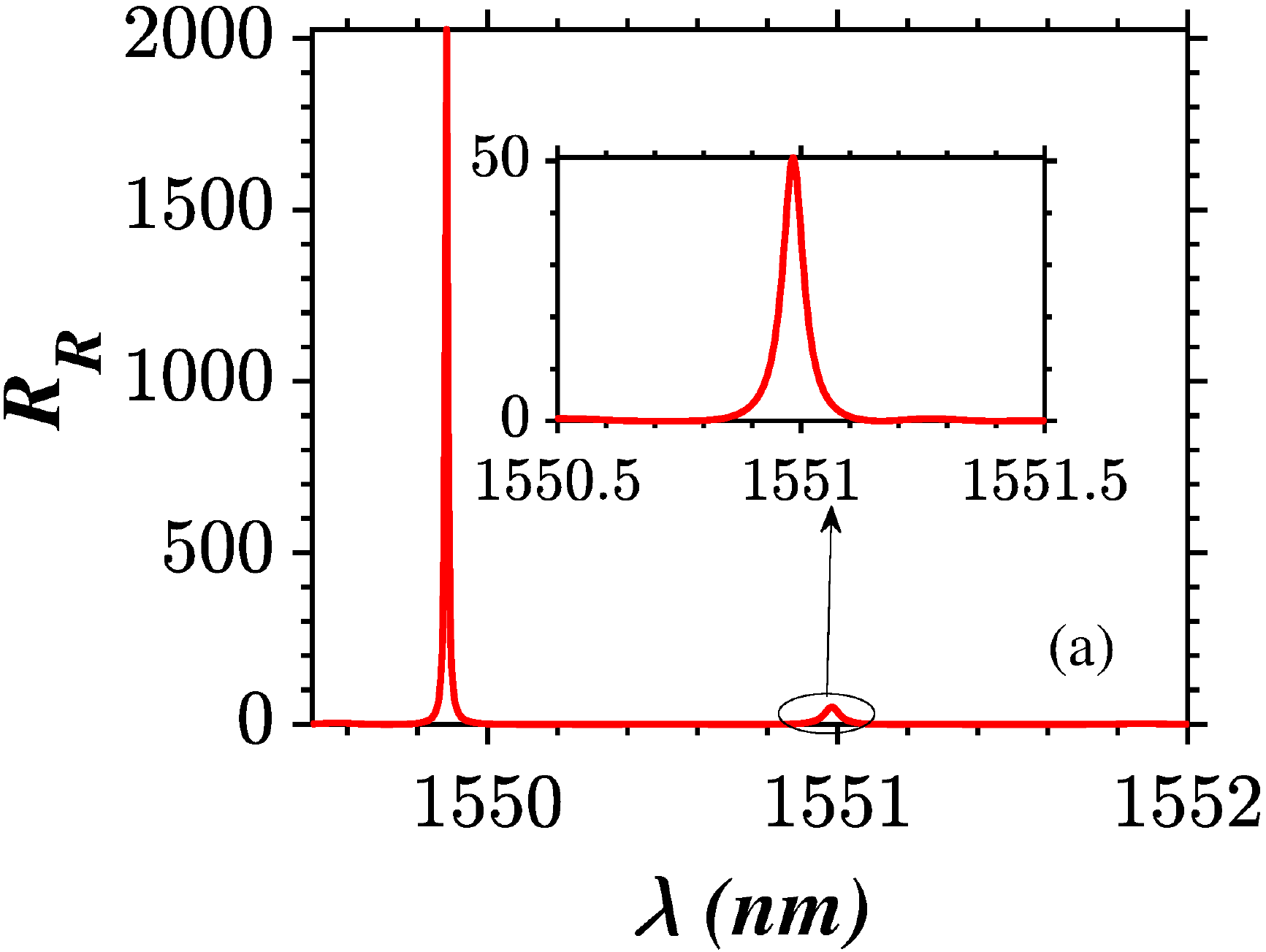}\includegraphics[width=0.5\linewidth]{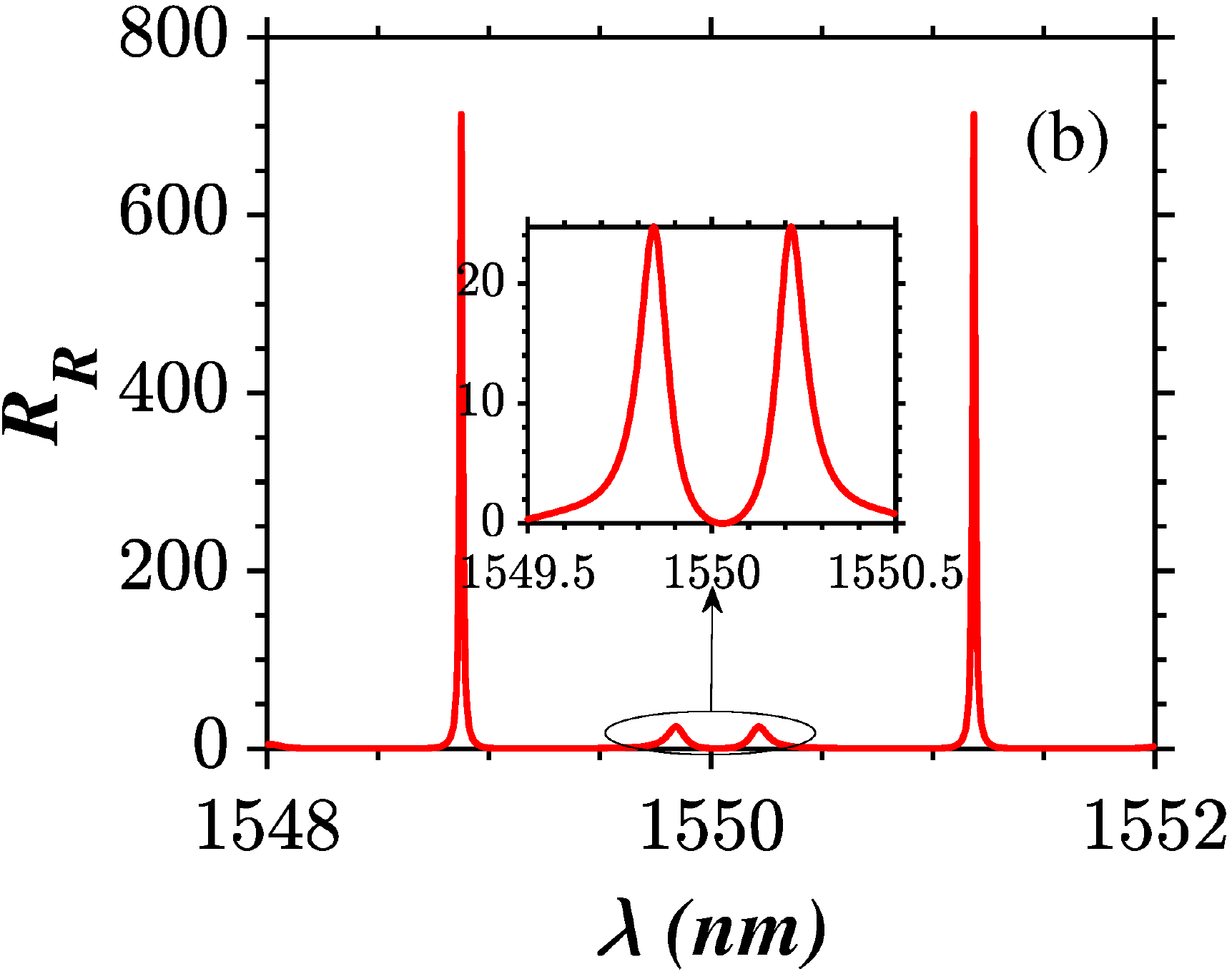}
 	\includegraphics[width=0.5\linewidth]{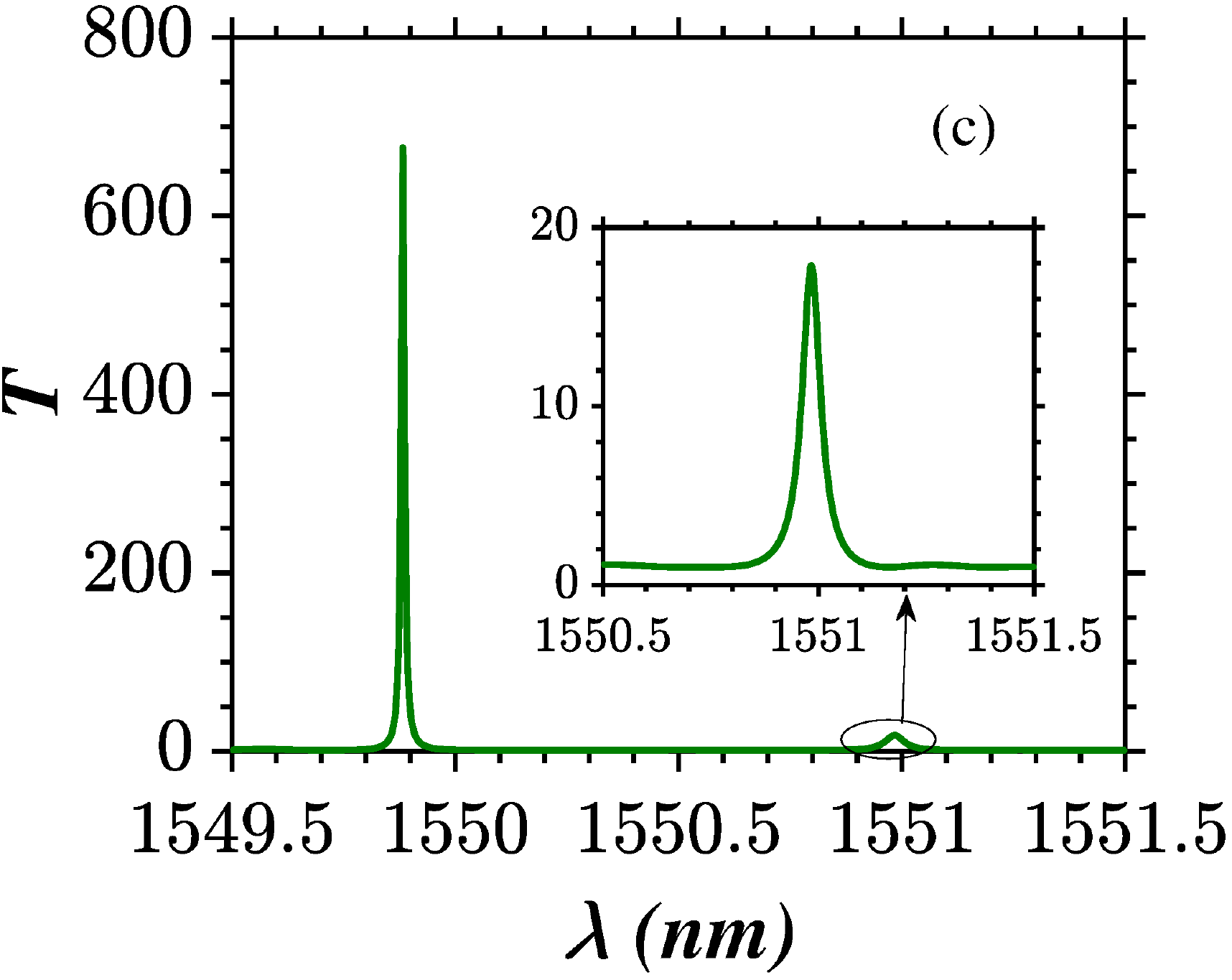}\includegraphics[width=0.5\linewidth]{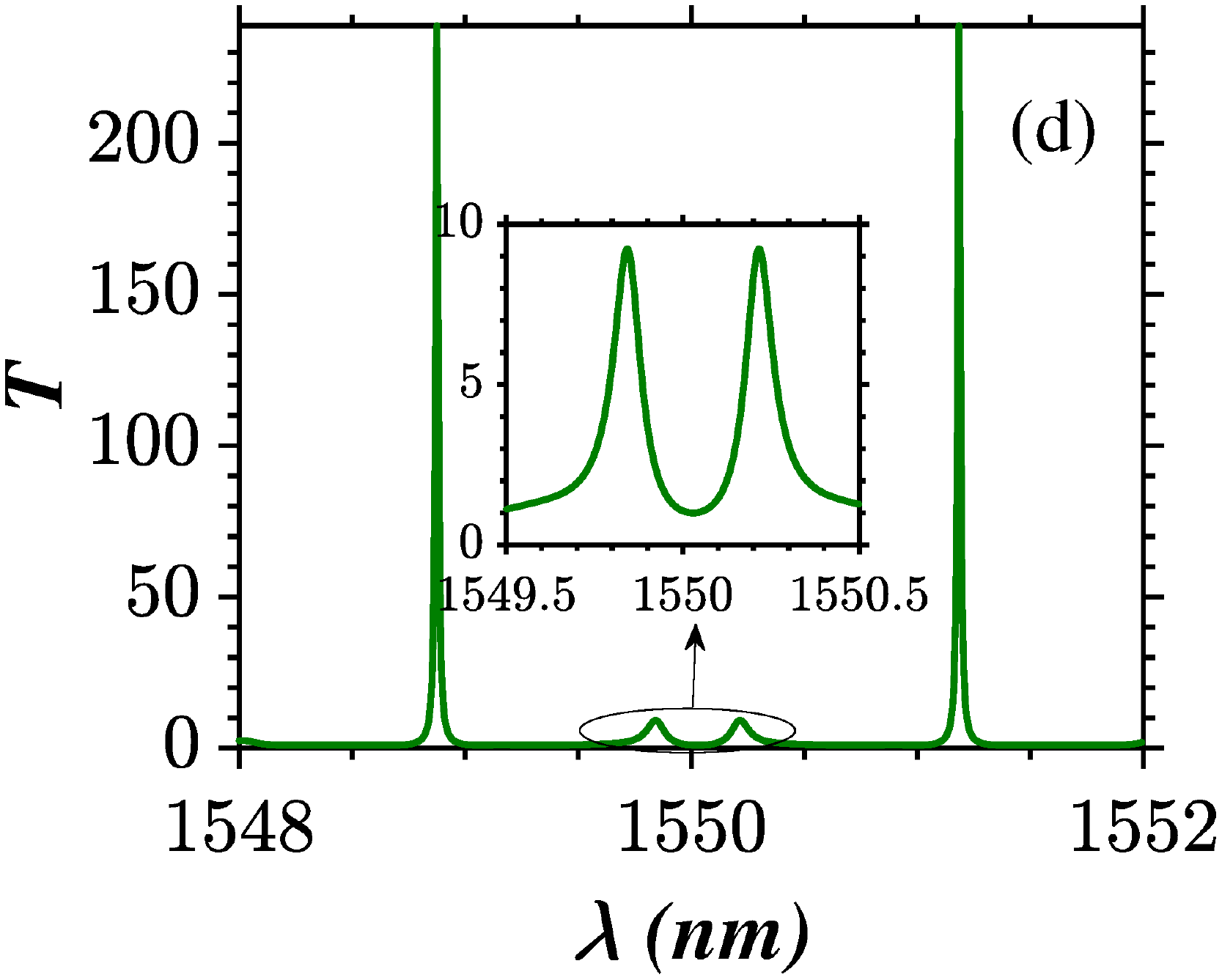}
 	\includegraphics[width=0.5\linewidth]{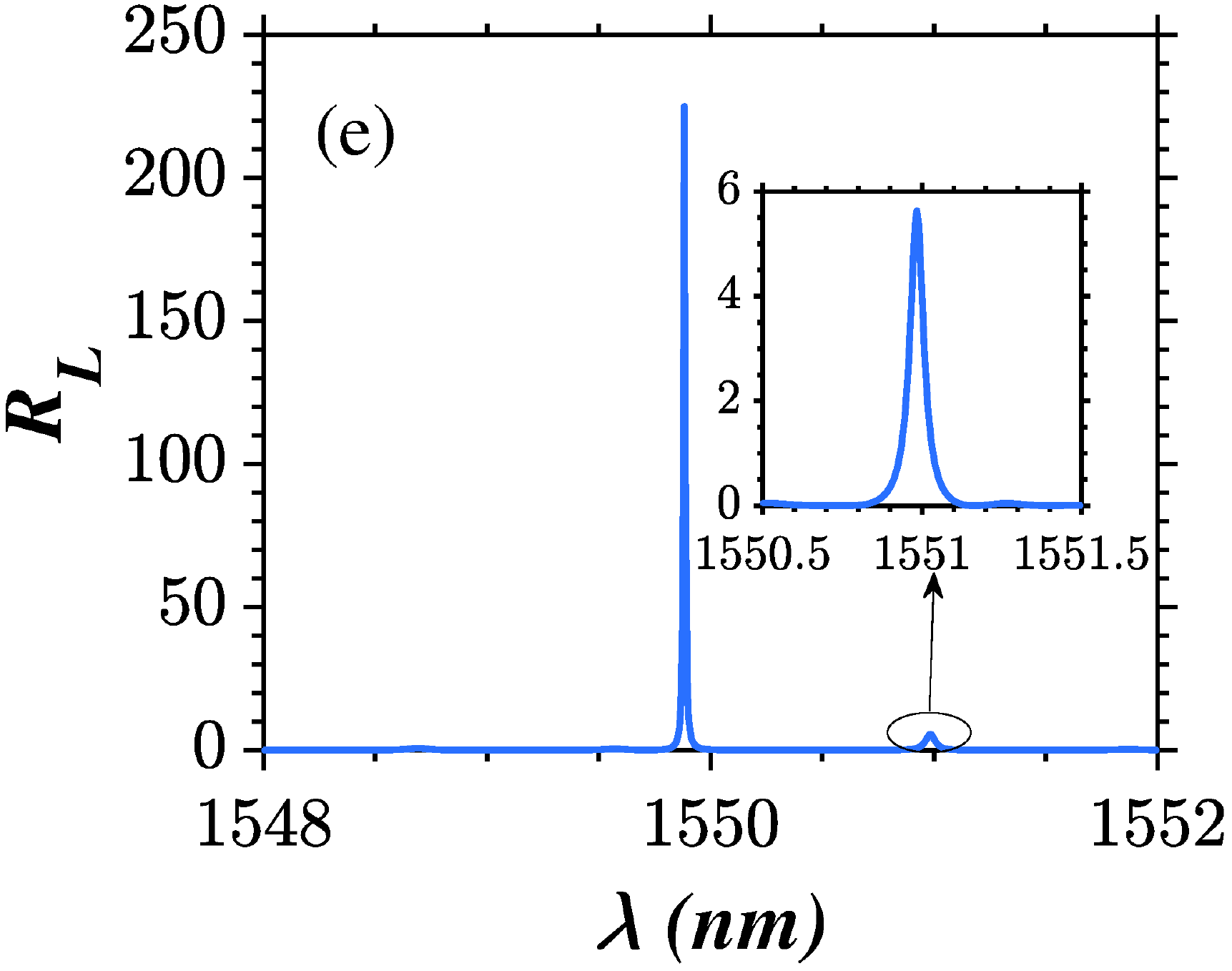}\includegraphics[width=0.5\linewidth]{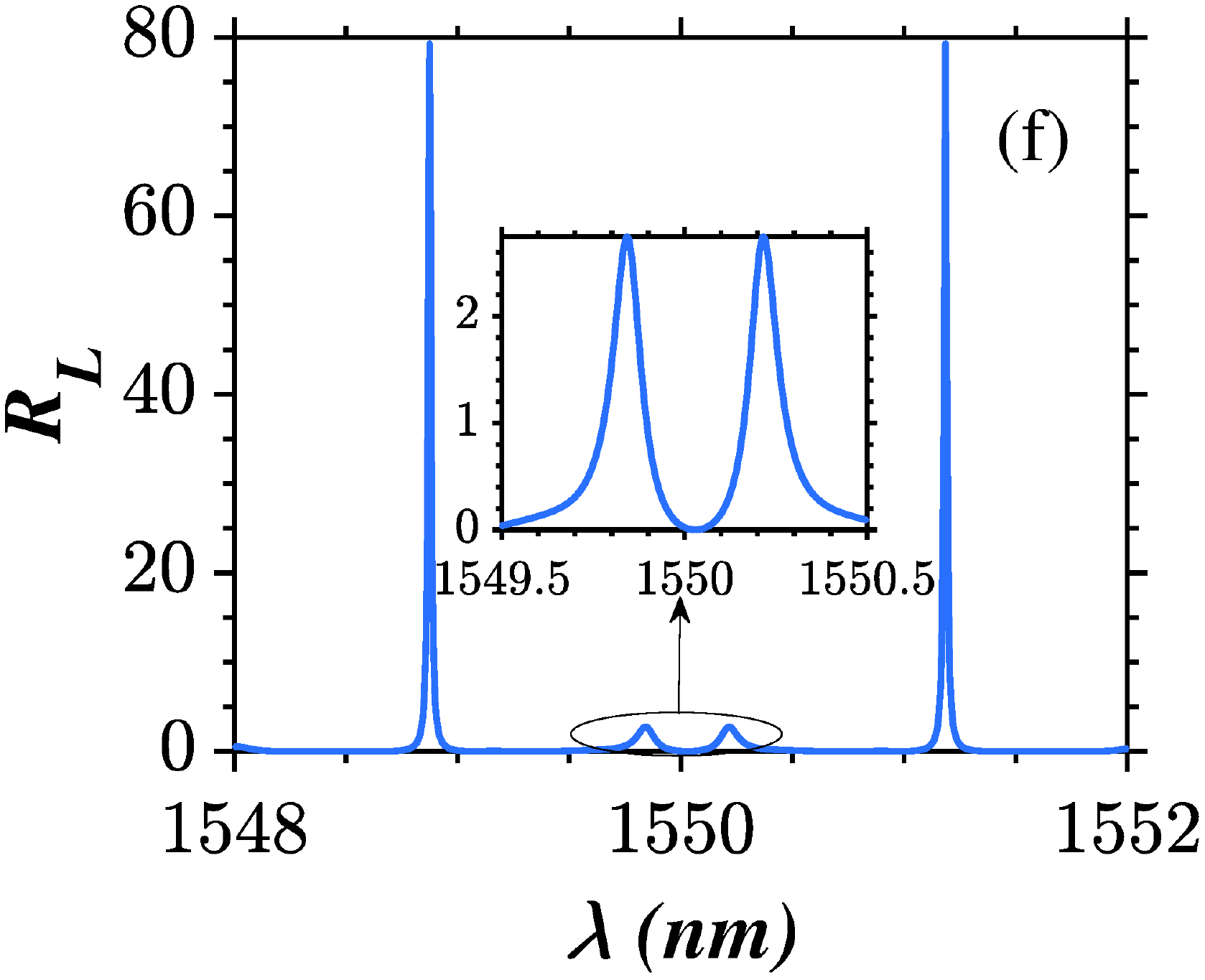}
 	\caption{Reflection and transmission spectra of a PPTFBG at the broken $\mathcal{PT}$-symmetric regime ($g = 20$ cm$^{-1}$) with phase shifts (equal) located at $-L/4$, $0$, $L/4$. The plots  in the left panel are simulated at $\phi = 90^\circ$ and the plots in the right panels are simulated at $\phi = 180^\circ$ and $270^\circ$. The top, middle and bottom panels represent the reflection right, transmission and reflection left spectra, respectively.}
 	\label{figm4}  
 \end{figure}

 Previously, we interpreted that operating the system at $\phi = 180^\circ$ leads to two symmetrical reflectivity peaks on either side of the Bragg wavelength in the presence of single phase shift region. In the case of multiple phase shifts, there are two more symmetric reflectivity peaks closer to the Bragg wavelength whose reflectivity is comparably very less to those of the peaks far away from the Bragg wavelength as shown in the right panels of Fig. \ref{figm4}. For the case $\phi = 90^\circ$, there are two reflectivity peaks observed in the left panels of Fig. \ref{figm4}, one with large reflectivity and other with very less reflectivity which is contrasting to the results obtained in the presence of single phase shift region which is characterized by single mode lasing behavior when $\phi = 90^\circ$. The results for $\phi = 270^\circ$ are simply the mirror image of plots obtained in the left panels of Fig. \ref{figm4}.

\section{conclusions}
\label{Sec:6}  
In this paper, we have analyzed the spectral characteristics of PPTFBGs in different regimes, namely the unbroken, exceptional point and the broken regimes. The spectral dynamics in the unbroken regime was quite similar to the spectra of a conventional phase shifted FBG. However, we have showed that it is possible to vary the intensity of the phase shifted spectra by varying the gain and loss parameter. Also, we made evident that the phenomenon of unidirectional reflectionless wave transport is exhibited by the proposed system. It was also shown that the device is not fully reflective within the stopband like other PTFBG systems. However, in the presence of phase it possesses symmetrical or asymmetrical spectra about one particular wavelength, where the reflection intensity is zero inside the stopband. The broken PPTFBG was found to exhibit single mode lasing behavior for some range of phase shift values and dual mode lasing behaviors for the others. Also, the broken $\mathcal{PT}$-symmetric FBG system described in this work has length of just 4 mm and thus making it becomes compact. Moreover, it involves less discrete components to perform filtering, amplification, and side lobe suppression. We have also considered the multiple phase-shifts introduced in various locations of PPTFBGs and found that they aid in controlling the number peaks in the reflection and transmission spectra. Hence it can be used to realize a narrow band single mode laser.  The results presented here give a conclusive evidence that it is possible to realize a multifunctional device which can operate as a demultiplexer and mode selective laser  from the same system configuration without a need to design a specific system for a particular application and thus simplifying the manufacturing process. We strongly believe that the fabrication of such a $\mathcal{PT}$-device(s) in future is not too far away thanks to the current optical integration methodologies.

\section*{Acknowledgments}
SVR is indebted to a financial assistantship provided
by Anna University through an Anna Centenary Research Fellowship (CFR/ACRF-2018/AR1/24). AG and ML acknowledge the support of DST-SERB for providing a Distinguished Fellowship (Grant No. SB/DF/04/2017) to ML in which AG was a Visiting Scientist. AG is now supported by  University Grants Commission (UGC), Government of India, through a Dr. D. S. Kothari Postdoctoral Fellowship (Grant No. F.4-2/2006 (BSR)/PH/19-20/0025).

\end{document}